\documentclass[acmsmall]{acmart}

\usepackage{listings}

\usepackage{amsmath,amssymb,amsfonts}
\usepackage{graphicx}
\usepackage{xcolor}
\usepackage{subfigure}
\usepackage{booktabs}
\usepackage{multirow}
\usepackage{makecell}
\usepackage{threeparttable}
\usepackage{enumitem} 
\usepackage{url}
\usepackage{bbding}
\usepackage{diagbox}
\usepackage[ruled, linesnumbered]{algorithm2e}
\usepackage[breakable]{tcolorbox}

\newtheorem{example}{Example}[section]
\newtheorem{criterion}{Criterion}
\newtheorem{principle}{Principle}
\newtheorem{assumption}{assumption}

\lstdefinelanguage{QSharp}
{morekeywords={namespace, open, fail, operation, function, use, Int, Double, Bool, true, false, if, else, while, repeat, until, body, Unit, Pauli, Result, let, set, mutable, for, in, return, Controlled, controlled, Adjoint, adjoint, Adj, Ctl, is, Qubit, auto, Length, H, X, Y, Z, I, T, S, R1, Rx, Ry, Rz, CNOT, SWAP, PauliI, PauliX, PauliY, PauliZ, M, Zero, One, within, apply},
sensitive=true,
morecomment=[l]{//},
}

\AtBeginDocument{%
	\providecommand\BibTeX{{%
			\normalfont B\kern-0.5em{\scshape i\kern-0.25em b}\kern-0.8em\TeX}}}
		
\setcopyright{acmcopyright}
\copyrightyear{2023}
\acmYear{2023}
\acmDOI{10.1145/1122445.1122456}

\begin{document}
	
\title{Testing Multi-Subroutine Quantum Programs: From Unit Testing to Integration Testing}

\author{Peixun Long}
\email{longpx@ios.ac.cn}
\affiliation{
	\institution{State Key Laboratory of Computer Science, Institute of Software, Chinese Academy of Science; University of Chinese Academy of Science}
	\country{China}
}

\author{Jianjun Zhao}
\email{zhao@ait.kyushu-u.ac.jp}
\affiliation
{
	\institution{Kyushu University}
	\country{Japan}
}

\begin{abstract}
Quantum computing has emerged as a promising field with the potential to revolutionize various domains by harnessing the principles of quantum mechanics. As quantum hardware and algorithms continue to advance, developing high-quality quantum software has become crucial. However, testing quantum programs poses unique challenges due to the distinctive characteristics of quantum systems and the complexity of multi-subroutine programs. This paper addresses the specific testing requirements of multi-subroutine quantum programs. We begin by investigating critical properties by surveying existing quantum libraries and providing insights into the challenges of testing these programs. Building upon this understanding, we focus on testing criteria and techniques based on the whole testing process perspective, spanning from unit testing to integration testing. We delve into various aspects, including IO analysis, quantum relation checking, structural testing, behavior testing, integration of subroutine pairs, and test case generation. We also introduce novel testing principles and criteria to guide the testing process. We conduct comprehensive testing on typical quantum subroutines, including diverse mutants and randomized inputs, to evaluate our proposed approach. The analysis of failures provides valuable insights into the effectiveness of our testing methodology. Additionally, we present case studies on representative multi-subroutine quantum programs, demonstrating the practical application and effectiveness of our proposed testing principles and criteria.
\end{abstract}
\keywords{Quantum computing, software testing, unit testing, integration testing}

\ccsdesc[500]{Software and its engineering~Software testing and debugging}

\maketitle

\section{Introduction}
\label{sec:introduction}

Quantum computing is a rapidly evolving field with great potential for advancements in various domains~\cite{national2019quantum}. It leverages the principles of quantum mechanics to process information and perform computational tasks. Quantum algorithms, in comparison to classical algorithms, offer the promise of accelerated solutions for specific problems~\cite{deutsch1985quantum,grover1996fast,shor1999polynomial}. As quantum hardware devices and algorithms continue to advance, developing high-quality quantum software has become increasingly crucial.

In recent years, several quantum programming languages have been proposed, including Q\#\cite{svore2018q}, Qiskit\cite{gadi_aleksandrowicz_2019_2562111}, Scaffold~\cite{abhari2012scaffold}, Quipper~\cite{green2013introduction}, and Sliq~\cite{bichsel2020silq}. These languages facilitate the creation of programs that can be executed on quantum simulators or real quantum devices. However, as quantum programs become complex and incorporate multiple subroutines, ensuring their correctness and reliability becomes increasingly critical. Testing quantum programs poses unique challenges due to the distinctive characteristics of quantum systems, such as superposition, entanglement, and quantum measurement~\cite{nielsen2002quantum}. Moreover, detecting errors in quantum programs is considered a challenging task~\cite{miranskyy2020your,huang2019statistical}.

While some testing methods have been proposed for quantum programs~\cite{ali2021assessing, honarvar2020property,wang2018quanfuzz,li2020projection,miranskyy2019testing,abreu2022metamorphic,wang2021generating,wang2021application,fortunato2022mutation,ye2023quratest}, they mainly focus on testing techniques suitable for small or fixed-scale quantum programs, neglecting the issues and challenges inherent in the whole testing process of complex practical quantum programs. As a result, there is a notable gap in the literature regarding addressing the specific testing requirements of multi-subroutine quantum programs.
%
In this paper, we explore the whole testing process for multi-subroutine quantum programs, aiming to tackle their challenges and provide solutions.

To lay the foundation for our exploration, we investigate the critical properties that significantly affect the testing tasks of multi-subroutine quantum programs through a survey of six existing quantum libraries.
Then, we review the testing process for multi-subroutine quantum programs to explore the possible challenges these properties bring and find solutions. We focus on two phases of the testing process: unit testing and integration testing.
In unit testing, we delve into various aspects, including IO analysis, general testing criteria and principles, and crucial subtasks involved in the testing process. These subtasks encompass quantum relation checking, quantum structural testing, and quantum behavior testing. Additionally, we tackle the generation of test cases, emphasizing the significance of comprehensive coverage and the incorporation of novel testing principles. 
Transitioning to integration testing, we confront the challenges inherent in integrating quantum programs and outline strategies for ensuring efficient integration testing. We address considerations such as the integration of different types of subroutine pairs. We also introduce seven novel testing principles and three innovative testing criteria tailored specifically for unit and integration testing of multi-subroutine quantum programs.

To demonstrate the effectiveness of our proposed testing approaches, we evaluate our unit testing processes using a diverse set of 17 typical quantum subroutines. To gauge the efficacy of our testing criteria, we introduce various types of mutant operators, including two novel types, and subject them to extensive testing on three distinct types of input states, encompassing tens of thousands of randomized inputs. The analysis of failures provides valuable insights into the effectiveness of our testing methodology in detecting bugs.
Furthermore, we present case studies centered around three representative multi-subroutine quantum programs. These case studies are concrete illustrations of the practical application and effectiveness of our proposed testing techniques, principles, and criteria for multi-subroutine quantum programs.
Our evaluation and case studies show the effectiveness of our proposed testing techniques, principles, and criteria for multi-subroutine quantum programs.

Our paper makes the following contributions:

\begin{itemize}
    \item \textbf{Critical properties of quantum programs:} Through a comprehensive survey of six open-source practical quantum libraries, we identify and discuss the critical properties of multi-subroutine quantum programs that significantly affect testing tasks.
    
    \item \textbf{Whole testing process perspective:} We are the first to consider testing tasks from a whole testing process perspective, spanning from unit testing to integration testing. We account for the distinctions between classical and quantum programs and provide guidance for testing multi-subroutine quantum programs.
    
    \item \textbf{Testing principles and criteria:} We propose seven novel principles and three novel testing criteria to address the specific requirements of testing multi-subroutine quantum programs. These principles and criteria serve as guidelines for designing effective tests and evaluating the quality of the testing process.

    \item \textbf{Testing methods:} Building upon the testing processes, principles, and criteria, we develop concrete methods for implementing the testing practice. We provide practical techniques and approaches that can be employed to test multi-subroutine quantum programs effectively.
    
    \item \textbf{Evaluation results:} To validate the effectiveness of our proposed testing processes and methods, we conduct experiments and case studies on unit testing design and integration testing design for multi-subroutine programs. These experiments and case studies demonstrate how our testing strategies can be applied and evaluated in different scenarios.
\end{itemize}

Overall, our contributions lay a solid foundation for testing multi-subroutine quantum programs, providing researchers and practitioners with valuable insights, guidelines, and practical techniques to ensure the correctness and reliability of quantum software systems.

The rest of the paper is organized as follows. Section~\ref{sec:related-work} provides an overview of the current work on testing quantum programs, highlighting their limitations. In Section~\ref{sec:background}, fundamental concepts of quantum computation are introduced. Section~\ref{sec:properties} focuses on the properties of multi-subroutine quantum programs and their implications on testing tasks, serving as a starting point for this paper. Detailed discussions on unit testing are provided in Section~\ref{sec:UnitTest} and integration testing in Section~\ref{sec:Integration}, including testing practices. The evaluation of our proposed testing processes and methods, along with case studies, is presented in Section~\ref{sec:evaluation} and Section~\ref{sec:casestudy}, respectively. Section~\ref{sec:conclusion} concludes the paper, offering a summary and prospects for future work.

\section{Related Work}
\label{sec:related-work}

Quantum software testing is an emerging research field still in the preliminary stage~\cite{zhao2020quantum,miranskyy2019testing}. The importance and challenges of testing quantum programs have been discussed by Miranskyy and Zhang~\cite{miranskyy2019testing}. Research progress in quantum software debugging and testing has been presented by Zhao~\cite{zhao2020quantum}, Miranskyy et al.~\cite{miranskyy2021testing}, and García de la Barrera et al.~\cite{garcia2021quantum}.

One common approach in this field is to adapt well-established testing methods for classical programs to the domain of quantum computing. Wang {\it et al.}~\cite{wang2018quanfuzz} introduced QuanFuzz, a fuzz testing method that generates test cases for quantum programs. Honarvar {\it et al.}~\cite{honarvar2020property} proposed a property-based testing approach and developed a tool named QSharpChecker to test Q\# programs. Ali {\it et al.}~\cite{ali2021assessing} defined input-output coverage criteria for quantum program testing and employed mutation analysis to evaluate their effectiveness. Wang {\it et al.}~\cite{wang2021generating} presented QuSBT, a search-based algorithm for generating test cases with a focus on capturing failure cases. Additionally, they proposed QuCAT~\cite{wang2021application}. This combinatorial testing approach explores various input variables of quantum programs by automatically generating test suites optimized for maximizing the number of failed test cases within a given test budget. Another approach, introduced by Abreu {\it et al.}~\cite{abreu2022metamorphic}, involves metamorphic testing specifically designed for oracle quantum programs. This method defines metamorphic rules tailored to quantum oracles to support their testing. Furthermore, researchers have also applied mutation testing and analysis techniques to quantum computing to test quantum programs~\cite{fortunato2022qmutpy,mendiluze2021muskit}. Moreover, to deal with quantum noise, Asmar {\it et al.}~\cite{Asmar2023Noise} have introduced a machine-learning framework to mitigate noise interference in testing quantum programs.

Assertions are crucial in checking quantum states and identifying bugs in quantum circuits and programs. One straightforward approach to implementing quantum assertions is preparing multiple copies of quantum variables and repeating the measurement process. Huang {\it et al.}~\cite{huang2019statistical} applied hypothesis testing techniques to partially reconstruct the information of quantum registers based on the measurement outcomes, enabling the implementation of assertions. However, this statistic-based method lacks support for runtime assertions. To address this limitation, Liu {\it et al.}~\cite{liu2020quantum} proposed the introduction of additional qubits to capture the information of the qubits under assertion, allowing for a minimal interruption during program execution. In another study, Li {\it et al.}~\cite{li2020projection} introduced Proq, a projection-based runtime assertion tool designed for testing and debugging quantum programs. Proq leverages projection measurement techniques to represent assertions, cleverly avoiding the destruction of quantum variables by measurements. Furthermore, Liu {\it et al.}~\cite{liu2020quantum} extended this idea to support assertions on mixed states and approximate values.

The development of benchmarks is essential to assess the effectiveness of different testing methods. QBugs, proposed by Campos and Souto~\cite{campos2021qbugs}, is a collection of reproducible bugs in quantum algorithms that supports controlled experiments for quantum software debugging and testing. However, detailed information on the usability of QBugs is not currently available. Zhao et al.~\cite{zhao2021bugs4q} introduced Bugs4Q, an open-source benchmark consisting of 36 real, validated bugs in practical Qiskit programs, accompanied by test cases for reproducing the buggy behaviors, thus supporting quantum program testing.

These existing works, however, have primarily focused on testing techniques for small or fixed-scale quantum programs, overlooking the challenges posed by practical quantum programs involving multiple subroutines and classical-quantum mixed input and output as discussed in Section~\ref{sec:properties}. Consequently, there is a gap in the literature regarding addressing the testing needs of multi-subroutine quantum programs. This paper aims to bridge this gap by presenting a comprehensive testing process specifically tailored for multi-subroutine quantum programs, covering both unit and integration testing. By carefully considering the unique properties of quantum programs, we aim to provide a robust testing process that supports practical development scenarios. We anticipate that this process will not only assist users in effectively conducting testing tasks but also serve as a valuable resource for researchers to identify and explore further research problems in the field of quantum software testing.

\section{Background}
\label{sec:background}
In this section, we provide a brief introduction to fundamental concepts in quantum computation. For a comprehensive understanding of quantum computation, readers are referred to~\cite{nielsen2002quantum} for further details.


\vspace{2mm}
\noindent $\bullet$ \textbf{Qubit and Quantum State.}\hspace*{1mm}
A quantum bit, or \textit{qubit} for short, is the fundamental unit of quantum computation. While a classical bit can be either 0 or 1, a qubit can exist in two \textit{orthogonal} states, conventionally denoted as $\left|0\right>$ and $\left|1\right>$. However, unlike classical bits, qubits can also exist in a \textit{superposition} of both states, with the general state being a linear combination of $\left|0\right>$ and $\left|1\right>$: $a\left|0\right>+b\left|1\right>$, where $a$ and $b$ are complex numbers known as \textit{amplitudes} that satisfy $|a|^2+|b|^2=1$. The amplitudes describe the probability of measuring the qubit to be in state $\left|0\right>$ or $\left|1\right>$, with the probabilities given by $|a|^2$ and $|b|^2$, respectively. Superposition is one of the fundamental aspects of quantum computing. It enables an operation to simultaneously affect all amplitudes, a phenomenon known as \textit{quantum parallelism}. For multiple qubits, the basis states are analogous to binary strings. For example, a two-qubit system has four basis states: $\left|00\right>$, $\left|01\right>$, $\left|10\right>$, and $\left|11\right>$, and the general state is a linear combination of these states with corresponding amplitudes $a_{00}$, $a_{01}$, $a_{10}$, and $a_{11}$, that is, $a_{00}\left|00\right>+a_{01}\left|01\right>+a_{10}\left|10\right>+a_{11}\left|11\right>$, where $|a_{00}|^2+|a_{01}|^2+|a_{10}|^2+|a_{11}|^2=1$. The state of a multi-qubit system can also be conveniently represented as a column vector $[a_{00}, a_{01}, a_{10}, a_{11}]^T$.

\vspace{2mm}
\noindent $\bullet$ \textbf{Hilbert Space.}\hspace*{1mm}
In quantum mechanics, a quantum state $\left|\psi\right>$ can be represented as a vector in a mathematical construct called a \textit{Hilbert space} $\mathcal{H}$, which has a dimension of $d$ for an $n$-qubit state, where $d=2^n$. A vector in the Hilbert space $\mathcal{H}$ can be expressed as a linear combination of a set of orthonormal basis vectors. For instance, the set of orthonormal basis vectors $\{\left|00\right>$, $\left|01\right>$, $\left|10\right>$, $\left|11\right>\}$ is known as the \textit{computational basis} and forms a basis for a 2-qubit (4-dimensional) Hilbert space. Any general state in the Hilbert space can be expressed as a linear combination of these basis vectors. Apart from the computational basis, there exist other orthonormal bases that can be used, such as $\left|+\right>=\frac{1}{\sqrt{2}}(\left|0\right>+\left|1\right>)$ and $\left|-\right>=\frac{1}{\sqrt{2}}(\left|0\right>-\left|1\right>)$, which form a basis for a 1-qubit (2-dimensional) Hilbert space.

\vspace{2mm}
\noindent $\bullet$ \textbf{Pure and Mixed States.}\hspace*{1mm}
The states mentioned above are known as \textit{pure states}, which are not probabilistic in nature. However, a quantum system may sometimes have a probability distribution over several pure states, known as a \textit{mixed state}. Let us consider a quantum system in state $\left|\psi_i\right>$ with probability $p_i$. The completeness of probability dictates that $\sum_i p_i = 1$. We can represent the mixed state using an \textit{ensemble representation} as ${(p_i, \left|\psi_i\right>)}$.

\vspace{2mm}
\noindent $\bullet$ \textbf{Density Matrix.}\hspace*{1mm}
Besides state vectors, the \textit{density matrix} or \textit{density operator} is another way to express a quantum state. It is particularly convenient for representing mixed states. Suppose a mixed state has an ensemble representation ${(p_i, \left|\psi_i\right>)}$, the density matrix of this state is given by $\rho = \sum_i p_i \left|\psi_i\right>\left<\psi_i\right|$, where $\left<\psi_i\right|$ denotes the conjugate transpose of $\left|\psi_i\right>$ (thus, it is a row vector and $\left|\psi_i\right>\left<\psi_i\right|$ is a $d\times d$ matrix). It is evident that the density matrix of a pure state $\left|\phi\right>$ is simply $\left|\phi\right>\left<\phi\right|$. In general, a density matrix $\rho$ is related to an ensemble ${(p_i, \left|\psi_i\right>)}$ if and only if it satisfies two conditions: (1) $tr(\rho) = 1$ and (2) $\rho$ is a positive operator. Furthermore, $\rho$ represents a pure state if and only if $tr(\rho^2) = 1$~\cite{nielsen2002quantum}.

\vspace{2mm}
\noindent
$\bullet$ \textbf{Partial Trace and Reduced Density Matrix.} Consider a composite quantum system consisting of two subsystems, denoted as $A$ and $B$. Let $\rho^{AB}$ represent the density matrix of the whole system. The \textit{reduced density matrix} for subsystem $A$, denoted as $\rho^A$, is defined as the \textit{partial trace} $tr_B$ over subsystem $B$ of $\rho^{AB}$, i.e., $\rho^A = tr_B(\rho^{AB})$. A similar definition applies to $\rho^B$. Intuitively, the partial trace operation $tr_B$ effectively removes the contribution of subsystem $B$ from the composite density matrix $\rho^{AB}$, isolating the state of subsystem $A$.

\vspace{2mm}
\noindent $\bullet$ \textbf{Entanglement.}\hspace*{1mm}
When two qubits are entangled, they cannot be treated as independent from each other, and measuring one qubit can interfere with the other, which makes entanglement a quintessential feature of quantum computation. The \textit{Bell state} $\left|\beta_{00}\right>=\frac{1}{\sqrt 2}(\left|00\right>+\left|11\right>)$ is a typical example of an entangled state between the first and second qubits. The ingenious application of entanglement has led to the development of many powerful quantum algorithms~\cite{shor1999polynomial,grover1996fast}.

\vspace{2mm}
\noindent $\bullet$ \textbf{Phase.}\hspace*{1mm}
The state $e^{i\theta}\left|\phi\right>$ is equivalent to $\left|\phi\right>$ up to a \textit{global phase factor} $e^{i\theta}$. This global phase cannot be distinguished by measurement. However, the \textit{relative phase} between components of a superposition can be distinguished by measurement. For example, the relative phase between $a$ and $b$ in the states $a\left|0\right>+b\left|1\right>$ and $a\left|0\right>+be^{i\theta}\left|1\right>$ can be determined through measurement.

\vspace{2mm}
\noindent $\bullet$ \textbf{Quantum Gate.}\hspace*{1mm}
A quantum gate is a fundamental operation in quantum computing. It is represented by a unitary matrix $G$ of size $2^n \times 2^n$ for an $n$-qubit gate. Applying the gate $G$ to a quantum state $\left|\psi\right>$ produces a new state $G\left|\psi\right>$. Several basic single-qubit gates are shown as follows.

\begin{small} 
\begin{equation}
\notag
X=\left[\begin{array}{cc}
0 & 1\\
1 & 0
\end{array}
\right], \hspace*{1mm}
Y=\left[\begin{array}{cc}
0 & -i\\
i & 0
\end{array}
\right], \hspace*{1mm}
Z=\left[\begin{array}{cc}
1 & 0\\
0 & -1
\end{array}
\right], \hspace*{1mm}
S=\left[\begin{array}{cc}
1 & 0\\
0 & i
\end{array}
\right], \hspace*{1mm}
T=\left[\begin{array}{cc}
1 & 0\\
0 & e^{i\pi / 4}
\end{array}
\right], \hspace*{1mm}
H=\frac{1}{\sqrt{2}} \left[\begin{array}{cc}
1 & 1\\
1 & -1
\end{array}
\right]
\end{equation}
\end{small}

\noindent 
Note that $G$ is a unitary matrix, meaning that it has an inverse matrix $G^{-1}$, and this inverse is given by the \textit{conjugate transpose} of $G$, denoted by $G^\dagger$. Hence, every gate has its corresponding \textit{inverse gate}, and its matrix is given by the conjugate transpose. For example, the inverses of $H$ and $S$ are:

\begin{equation}
\notag
H^{-1} = H^{\dagger} = \frac{1}{\sqrt{2}} \left[\begin{array}{cc}
1 & 1\\
1 & -1
\end{array}
\right] = H, \hspace*{5mm}
S^{-1} = S^{\dagger} = \left[\begin{array}{cc}
1 & 0\\
0 & -i
\end{array}
\right]
\end{equation}

\noindent 
Quantum gates may also have parameters. For example, $R_1(\theta)$ gate has a parameter of rotation angle $\theta$, as shown in the following:

\begin{equation}
\notag
R_1(\theta) = \left[\begin{array}{cc}
1 & 0\\
0 & e^{i\theta}
\end{array}
\right]
\end{equation}

\vspace*{1mm}
\noindent
The \textit{controlled gate} is an essential multi-qubit gate in quantum computing. For example, the gate $X$ flips the target qubit $q_1$, which means that when $X$ is applied to $\left|0\right>$, it changes to $\left|1\right>$, and when applied to $\left|1\right>$, it changes to $\left|0\right>$. By introducing another qubit, $q_2$, we can create a Controlled-X gate, also known as the CNOT gate, which operates on these two qubits. The CNOT gate flips $q_1$ only when $q_2$ is in the state $\left|1\right>$. The matrix of CNOT is:

\begin{equation}
\notag
\mathrm{CNOT} = \left[\begin{array}{cccc}
1 & 0 & 0 & 0\\
0 & 1 & 0 & 0\\
0 & 0 & 0 & 1\\
0 & 0 & 1 & 0
\end{array}
\right] = \left[\begin{array}{cc}
I & 0\\
0 & X
\end{array}
\right]
\end{equation}

\vspace*{1mm}
\noindent Similarly, other gates also have their corresponding \textit{controlled} versions.

\begin{figure}[h]
\begin{minipage}[b]{0.4\linewidth}
	\centering
	\includegraphics[scale=0.6]{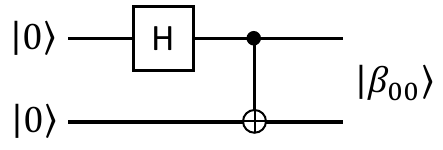}
	\caption{A quantum circuit for preparing the Bell state.}
	\label{fig:bell}
\end{minipage}
\begin{minipage}[b]{0.55\linewidth}
	\centering
	\includegraphics[scale=0.6]{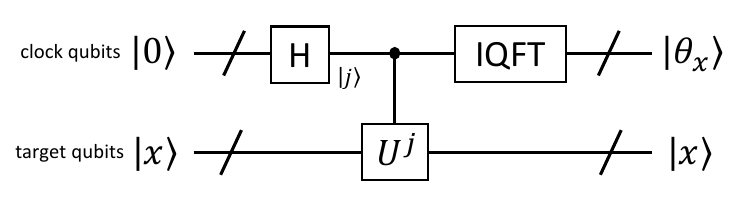}
	\caption{The quantum circuit for QPE algorithm.}
	\vspace{5mm}
	\label{fig:qpe}
\end{minipage}
\end{figure}

\vspace{2mm}
\noindent $\bullet$ \textbf{Quantum Circuit.}\hspace*{1mm}
A quantum circuit is a common model used to express the computational process in quantum computing~\cite{nielsen2002quantum}. Each line in the circuit corresponds to a qubit, and a sequence of quantum operations is applied from left to right. For example, Figure~\ref{fig:bell} illustrates a quantum circuit that prepares the Bell state $\left|\beta_{00}\right> = \frac{1}{\sqrt 2}(\left|00\right>+\left|11\right>)$, which consists of an $H$ gate and a CNOT gate.

\vspace{2mm}
\noindent $\bullet$ \textbf{Measurement.}\hspace*{1mm}
Qubit information can only be obtained in quantum devices through \textit{measurement}. Measuring a quantum system yields a classical value with a probability corresponding to the amplitude. The state of the quantum system then collapses into a basis state according to the obtained value. For example, when measuring a qubit in state $a\left|0\right>+b\left|1\right>$, the result 0 is obtained with probability $|a|^2$, and the state collapses into $\left|0\right>$; the result 1 is obtained with probability $|b|^2$, and the state collapses into $\left|1\right>$. This property introduces uncertainty and affects the testing of quantum programs.

\vspace{2mm}
\noindent $\bullet$ \textbf{Quantum Operation.}\hspace*{1mm}
A series of quantum gates can be represented as a unitary transform. However, measurement operations can disrupt the unitarity of the system. In quantum computing, a quantum operation is a mathematical model used to describe the general evolution of a quantum system, including both quantum gates and measurements, of a quantum system~\cite{nielsen2002quantum}. Given an input density matrix $\rho$ (a $d \times d$ matrix), the quantum operation $\mathcal{E}$ transforms it to $\mathcal{E}(\rho)$. This transformation can be represented by the operator-sum representation~\cite{nielsen2002quantum}, which is given by $\mathcal{E}(\rho) = \sum_i E_i \rho E_i^{\dagger}$, where $E_i$ is a $d \times d$ matrix. If $\sum_i E_i^{\dagger} E_i = I$, then $\mathcal{E}$ is called a \textit{trace-preserving} quantum operation.

\vspace{2mm}
\noindent $\bullet$ \textbf{Quantum Algorithm.}\hspace*{1mm}
Quantum algorithms leverage the principles of quantum mechanics to tackle computational problems. In Appendix~\ref{appendix:algorithms}, we provide an overview of seven algorithms that will be explored further in the subsequent sections.

\vspace{2mm}
\noindent $\bullet$ \textbf{Quantum Noise and Decoherence.}\hspace*{1mm}
Quantum computing is now in the so-called \textit{NISQ} (Noisy Intermediate-Scale Quantum) era~\cite{preskill2018quantum}. It means the scales of real quantum devices are limited ($\sim 10^2$ qubits), and the executing noise is not negligible. \textit{Quantum noise} will interfere with the quantum variables during the execution of the quantum program, making the running results deviate from the expected values. \textit{Decoherence} means quantum states can only be maintained for a short time. It limits the complexity of runnable programs.

\section{The Properties of Multi-Subroutine Quantum Programs}
\label{sec:properties}

To gain an overview of testing for multi-subroutine quantum programs, it is crucial to understand the important properties in these programs that affect testing designs. This section addresses the following research question (RQ):

\begin{itemize}
	\item \textbf{RQ1:} What are the critical properties of multi-subroutine quantum programs?
\end{itemize}

To address RQ1, we identify some critical properties that influence testing designs for multi-subroutine quantum programs. Notably, these properties have not been adequately considered in current testing methods for quantum programs. In addition to deriving the properties from the background of quantum computation, we also identify these properties through two practical surveys: (1) investigating the supported features of quantum programming languages and (2) examining existing quantum software projects. The background of quantum computation has been introduced in Section~\ref{sec:background}. Next, we will discuss each of these two surveys in detail.

\subsection{Surveys on Quantum Programming Languages and Quantum Software Projects}
\label{subsec:surveys}

\subsubsection{Survey on Quantum Programming Languages}
\label{subsec:LangugeSurvey}
A number of quantum programming languages have become available in recent years, such as QCL~\cite{omer2003structured}, Scaffold~\cite{abhari2012scaffold}, Cirq~\cite{cirq2018google}, Quipper~\cite{green2013introduction}, Qiskit~\cite{gadi_aleksandrowicz_2019_2562111}, Q\#~\cite{svore2018q}, Silq~\cite{bichsel2020silq}, isQ~\cite{Guo2022isQ}, Ket~\cite{da2021ket}, and Qrisp~\cite{seidel2022qrisp}.
These languages differ in their design and capabilities. Some, like the Python-based Qiskit, extend classical programming languages, while others, like Q\#, are independent of their classical counterparts. Some languages, such as Cirq, are lower-level and suited for describing quantum circuits, while others, such as Q\#, are higher-level and capable of implementing complex function calls. Additionally, almost all quantum programming languages provide simulators that enable programmers to run quantum programs on a classical computer. Some programming languages even offer interfaces to real quantum hardware. For example, Qiskit has an interface to IBM's quantum hardware.

We first conducted a survey on these quantum programming languages. We selected ten languages as shown in Table ~\ref{table:QLanFeature}. Our selection criteria cover two categories of languages~\cite{garhwal2021quantum,ferreira2022exploratory}: (1) widely used languages such as QCL, Scaffold, Cirq, Quipper, Qiskit, and Q\#; and (2) emerging languages that support advanced features such as Silq, isQ, Ket, and Qrisp. Table~\ref{table:QLanFeature} presents a brief summary of the features of ten quantum programming languages that offer quantum development kits (QDKs) for building executable quantum programs. In this paper, we aim to test quantum programs with multiple subroutines, focusing on language features related to program structure and organization. We use the Q\# as our primary language of choice in this paper since it supports a broad range of quantum program structures and has extensive user documentation. Although our methods primarily utilize Q\#, they are also adaptable to other quantum programming languages. 

\begin{table*}
\centering
\caption{The features of 10 quantum programming languages}
\label{table:QLanFeature}

\begin{scriptsize}
\begin{threeparttable}

\begin{tabular}{c|c|cccccccccc}
\toprule
\makecell[c]{\textbf{Feature} \\ \textbf{Type}} & \diagbox{\textbf{Feature}\tnote{2}}{\textbf{Language}} & Scaffold & Cirq & Quipper & Qiskit & Q\# & Silq & isQ & Ket & QCL & Qrisp \\
\midrule

Language & Host language\tnote{1} & \texttt{C} & \texttt{Py} & \texttt{Ha} & \texttt{Py} & SA & SA & SA & \texttt{Py} & \texttt{C} & \texttt{Py}\\
\midrule

\multirow{4}{*}{\makecell[c]{Program \\ Structure}} & Type system & \checkmark &  & \checkmark &  & \checkmark & \checkmark & \checkmark & & \checkmark & \checkmark\\
\cmidrule{2-12}
& Classical control flow & H & H & H & H & \checkmark & \checkmark & \checkmark & H & H & H \\
\cmidrule{2-12}
& Classical-quantum mix & H & H & H & H & \checkmark & \checkmark & \checkmark & H & H & H\\
\midrule

\multirow{2}{*}{Subroutine} & Subroutine calling & H & H & H & H & \checkmark & \checkmark & \checkmark & H & H & H\\
\cmidrule{2-12}
& Subroutine as parameters &  &  & H & H & \checkmark & \checkmark & \checkmark & \checkmark &  & \\
\midrule

\multirow{5}{*}{\makecell[c]{Quantum- \\ related}}& Qubit array & \checkmark & \checkmark & \checkmark & \checkmark & \checkmark & \checkmark & \checkmark & \checkmark & \checkmark & \checkmark\\
\cmidrule{2-12}
& Endian modes of qubits &&&&& \checkmark & &\\
\cmidrule{2-12}
& Auto-generate gate variants & & & & & \checkmark & \checkmark & \checkmark & \checkmark & &\\
\cmidrule{2-12}
& Gate definition by matrix/list & \checkmark & \checkmark &  & \checkmark &  &  & \checkmark & &  \checkmark &\\
\bottomrule
\end{tabular}

\begin{tablenotes}
\item[1] '\texttt{C}'/'\texttt{Py}'/'\texttt{Ha}' refers to C/Python/Haskell as the host language, while '\texttt{SA}' indicates that the quantum programming language is a standalone language.

\item[2]  '\checkmark' signifies that the quantum programming language provides support for the feature, 'H' indicates that the host language supports the feature, and a blank space means that the quantum programming language does not support the feature.
\end{tablenotes}
\end{threeparttable}
\end{scriptsize}
\end{table*}

\subsubsection{Survey on Quantum Software Projects}
\label{subsubsec:ProgramSurvey}
We next conduct a survey on examining current quantum software projects on GitHub. However, we found very few open-source projects that included multi-subroutine quantum programs besides the built-in libraries within quantum software stacks. To overcome this limitation, we selected two quantum programming languages, Q\# and Qiskit, and investigated the libraries they provide. We focus on libraries relevant to practical applications and algorithms of quantum computing rather than underlying support. For Q\#, we selected three libraries~\texttt{Numerical}, \texttt{Chemistry}, and \texttt{MachineLearning}, while for Qiskit, we selected \texttt{Aqua Chemistry}, \texttt{Finance}, and \texttt{MachineLearning}. Table~\ref{table:QProjects} shows the 16 properties we identified, divided into four groups, and the number of subroutines that satisfied each property in each project. We will discuss the details of each property in the rest of this section. From the survey process and the results shown in Table~\ref{table:QProjects}, several key observations can be made:

\begin{itemize}
\item[(1)] Practical quantum algorithms often incorporate classical subroutines for necessary pre- and post-processing steps. This hybrid approach leverages the strengths of both classical and quantum computation.

\item[(2)] Quantum algorithms implemented in the application layer frequently operate on variable qubit sizes, adapting to the specific requirements of the problem.

\item[(3)] Many subroutines within quantum algorithms rely on other subroutines to accomplish their intended tasks, indicating a modular and interconnected nature of quantum program design.

\item[(4)] Notably, distinctions exist in the overall program structures between Q\# and Qiskit, attributed to the specific features of these two programming languages. Qiskit, being hosted by Python, often employs object-oriented programming techniques. In contrast, Q\# enforces stricter type checking and has limitations on variables in its grammar, which can help prevent potential programming errors. Additionally, Q\# supports the auto-generation of variant subroutines, resulting in its projects containing more variants than those of Qiskit.
\end{itemize}

These findings provide valuable insights into the characteristics and design considerations of quantum software systems.

\begin{table*}
\centering
\caption{The number of subroutines in six quantum projects in Q\# and Qiskit which meet specific properties}
\label{table:QProjects}

\begin{small}
\begin{threeparttable}

\begin{tabular}{c|l||c|c|c||c|c|c}
\toprule
& \multirow{2}{*}{ \diagbox{\textbf{Property}}{\textbf{Project}} } & \multicolumn{3}{c||}{Q\#\quad\quad} & \multicolumn{3}{c}{Qiskit} \\
&& \texttt{\scriptsize{Numerics}} & \texttt{\scriptsize{Chemistry}} & \scriptsize{ \makecell{\texttt{Machine-}\\ \texttt{Learning}} } & \scriptsize{ \makecell{\texttt{Aqua}\\ \texttt{Chemistry}} } & \texttt{\scriptsize{Finance}} & \scriptsize{ \makecell{\texttt{Machine-}\\ \texttt{Learning}} } \\
\midrule
\multirow{4}{*}{General} & 1.1 Total count & 98 & 131 & 54 & 557 & 86 & 421 \\
& 1.2 Classical & 9 & 46 & 27 & 537 & 79 & 414 \\
& 1.3 Fixed qubits size & 0 & 0 & 0 & 0 & 0 & 0 \\
& 1.4 Variable qubits size & 89 & 84 & 27 & 10 & 7 & 7 \\

\midrule
\multirow{6}{*}{\makecell{Program \\ structure}} & 2.1 Pure structure\tnote{1} & 54 & 66 & 29 & 343 & 49 & 257\\
& 2.2 Has "if-then" block & 16 & 46 & 11 & 183 & 22 & 150 \\
& 2.3 Has "for" loop & 15 & 37 & 25 & 127 & 14 & 46 \\
& 2.4 Has "try" block & 0 & 0 & 0 & 6 & 9 & 16 \\
& 2.5 Has "while" loop\tnote{2} & 0 & 1 & 0 & 5 & 0 & 2 \\
& 2.6 Has "within-apply" & 17 & 0 & 0 & 0 & 0 & 0 \\

\midrule
\multirow{3}{*}{\makecell{Subroutine \\ calling}} & 3.1 No calling & 0 & 13 & 8 & 145 & 37 & 149 \\
& 3.2 Has calling & 98 & 118 & 46 & 411 & 39 & 269 \\
& 3.3 As input parameter & 0 & 25 & 10 & 7 & 0 & 4 \\
\midrule

\multirow{3}{*}{Variants} & 4.1 Original & 36 & 71 & 44 & 557 & 85 & 421 \\
& 4.2 Inverse & 44 & 40 & 13 & 0 & 1 & 0 \\
& 4.3 Controlled & 42 & 39 & 12 & 0 & 0 & 0 \\
\bottomrule
\end{tabular}

\begin{tablenotes}
	\item[1] "Pure structure" means the subroutine is the sequential execution of statements without any special structure.
	\item[2] Including \textit{repeat-until-success} structure.
\end{tablenotes}

\end{threeparttable}
\end{small}

\end{table*}

\subsection{The Structure of Quantum Programs}
\label{subsec:ProgramStruct}

\subsubsection{Quantum Circuits and Quantum Programs}
\begin{figure}
	\centering
	\includegraphics[scale=0.6]{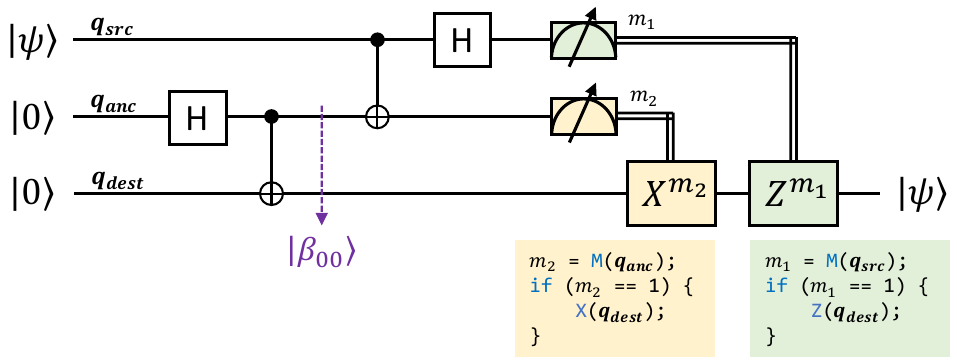}
	\caption{The quantum circuit for quantum teleportation.}
	\label{fig:teleport}
	
\begin{lstlisting}[
	xleftmargin=4mm,
	%xrightmargin=2mm,
	caption={The Q\# code for teleportation program},
	label={list:teleport}
	]
operation TeleportQubit(qsrc : Qubit, qanc : Qubit, qdest : Qubit) : Unit {
    H(qanc);
    CNOT(qanc, qdest);
    CNOT(qsrc, qanc);
    H(qsrc);
    let m2 = M(qanc);
    if (m2 == One) { X(qdest); }
    let m1 = M(qsrc);
    if (m1 == One) { Z(qdest); }
}
\end{lstlisting}
\end{figure}

The quantum circuit model, which represents a linear sequence of quantum gates applied from left to right, is commonly used to describe quantum algorithms, as shown in Figure~\ref{fig:qpe} for the QPE algorithm. 
Current research on quantum testing mainly focuses on the circuit model~\cite{wang2018quanfuzz,wang2021application,wang2021generating,wang2022mutation,abreu2022metamorphic}. However, to describe algorithms that include gates controlled by classical values from measurements, an improvement to the circuit model, called \textit{dynamic circuit model}~\cite{Hua2023CaQR,Corcoles2021DynamicCirc}, has been proposed. Figure~\ref{fig:teleport} shows a typical dynamic circuit implementing quantum teleportation, which contains two gates controlled by measurement results. Listing~\ref{list:teleport} shows the Q\# code implementation of this circuit, where gates controlled by measurement results are implemented using \texttt{if} statements (lines $7\sim 10$). However, quantum circuits are not always equivalent to quantum programs, as the circuit model is hard to describe \texttt{while} loops and quantum-classical-hybrid algorithms.

According to property 1.2 in Table~\ref{table:QProjects}, practical quantum algorithms and projects frequently incorporate both classical and quantum components. In some cases, the algorithm's input contains both classical and quantum variables, such as \textit{parameterized quantum circuit} (PQC)~\cite{marcello2019PQC}, widely used in quantum machine learning. A PQC can be denoted as $U_{\phi(\vec{x})}$, where vector $\vec{x}$ represents the classical data and $\phi$ is the encoding function. Different values of $\vec{x}$ result in different circuits; thus, $\vec{x}$ is the classical input of the PQC. In other cases, the entire program consists of both classical and quantum subroutines, as seen in quantum factoring (Algorithm~\ref{alg:Shor}), where the unique quantum subroutine is quantum order finding (QOF) and the other subroutines are classical.

\begin{figure}
	\centering
	\subfigure[$n=1$]{\includegraphics[scale=0.25]{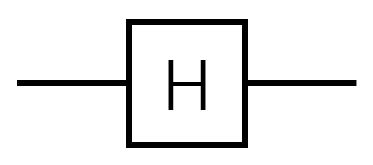}}
	\subfigure[$n=2$]{\includegraphics[scale=0.5]{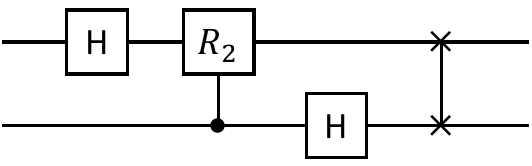}}
	\subfigure[$n=3$]{\includegraphics[scale=0.5]{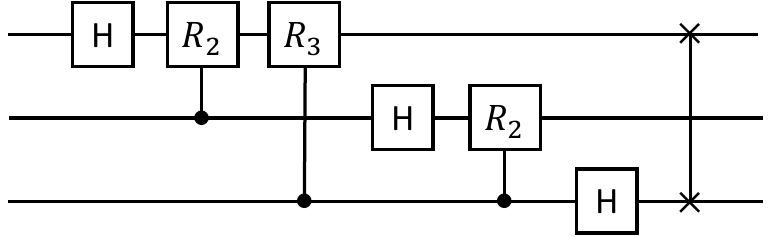}}
	\\
	\subfigure[$n=4$]{\includegraphics[scale=0.5]{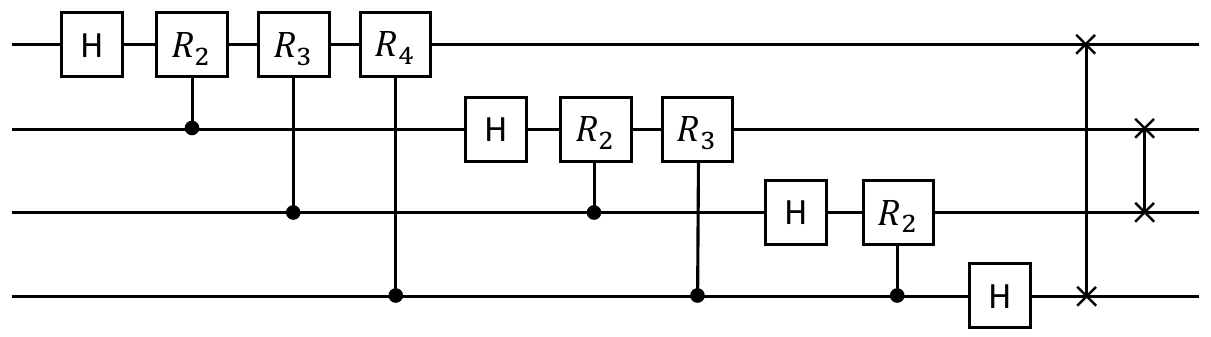}}
	\caption{The corresponding circuits of QFT program with $n=$ 1, 2, 3, 4 respectively ($n$ is the number of qubits).}
	\label{fig:qft}
\end{figure}

Unlike a fixed-size quantum circuit, a general quantum program corresponds to a family of quantum circuits rather than a specific circuit. For example, the quantum Fourier transform (QFT) program (Figure~\ref{fig:qft}) corresponds to different circuits based on the number $n$ of qubits used. It is similar to its classical counterpart, the Boolean circuit model~\cite{arora2009computational}, an essential computational model in computational complexity theory. Classical algorithms typically correspond to a uniform family of Boolean circuits rather than just one circuit.

This paper focuses on testing quantum programs containing both quantum and classical code rather than just quantum circuits. 

\subsubsection{Control Flows in Quantum Programs}
\label{subsubsec:ControlFlow}

{As property 2.1$\sim$2.6 in Table~\ref{table:QProjects},} classical control flows in quantum programs are also crucial for programming and designing practical quantum algorithms. Similar to classical programs, quantum programs frequently utilize \texttt{if} statements, \texttt{for}/\texttt{while} loops, and other control flow structures to achieve their desired outcomes. For instance, Listing~\ref{list:teleport} employs \texttt{if} statements to implement gates controlled by measurement results. Moreover, Listing~\ref{list:qft} presents an implementation of the QFT program using \texttt{for} loops to iterate over qubits of varying sizes.

Quantum programs also employ \texttt{while} loops, which have a loop condition controlled by a variable instead of an iteration index. In quantum programming, the condition variable may depend on quantum measurement results. For example, the \textit{repeat-until-success} (RUS) structure is often used in quantum programming. This structure relies on a success condition associated with measurement results. The loop continues to iterate until the success condition is satisfied. Listing~\ref{list:random} illustrates the implementation of a quantum program generating random integers 0, 1, and 2 with equal probability using a RUS structure.

Additionally, some quantum programs incorporate classical and quantum computing into one subroutine, which is referred to as a quantum-classical mixed (C-Q mixed) program. For example, \texttt{QFT} (Listing~\ref{list:qft}) contains a subroutine \texttt{CRk}, which takes classical expressions (line 3) to calculate an intermediate parameter "\texttt{theta}." This combination of classical and quantum computing is often required in developing practical quantum algorithms, such as Grover search and quantum machine learning.

\begin{figure}
\begin{lstlisting}[
	xleftmargin=4mm,
	%xrightmargin=2mm,
	caption={The Q\# code for generating random numbers 0, 1, and 2},
	label={list:random}
	]
operation Random_0to2() : Int {
    use qs = Qubit[2];
    mutable m = 0;
    repeat {
        ResetAll(qs);
        H(qs[0]);
        H(qs[1]);     //Generate uniform superposition state
        set m = ResultArrayAsInt(MultiM(qs));
    } until (m < 3);    //Success condition
    ResetAll(qs);
    return m;
}
\end{lstlisting}
\end{figure}

\subsection{Subroutines}
\label{subsec:Subroutines}

\subsubsection{The Organization of Subroutines}
In software development, programmers often use "\textit{structured programming},"~\cite{dahl1972structured} where a large program is decomposed into several subroutines and organized by~\textit{function calls}. This programming paradigm improves the readability and testability of code and allows the reuse of commonly used subroutines. As shown in Listing~\ref{list:qft}, \texttt{CRk} and \texttt{Reverse} are two subroutines called by the upper-level program \texttt{QFT}.

Besides direct calls, another way to organize subroutines is by treating the subroutine as an input parameter, as indicated by property 3.3 in Table~\ref{table:QProjects}. An example is Grover Search~\cite{grover1996fast}, which relies on an oracle subroutine to identify the solution and treats the oracle subroutine as a black box. The oracle can be implemented as an input subroutine. Listing~\ref{list:qpe} shows another example \texttt{QPE}, which requires the target quantum operation as an input parameter (\texttt{Upower} in lines 1 and 7). In general, direct calling implies a high degree of coupling between the two subroutines, whereas calling as parameters implies a low degree of coupling.

\begin{figure}
\begin{lstlisting}[
	xleftmargin=4mm,
	%xrightmargin=2mm,
	caption={The Q\# code for QFT program},
	label={list:qft}
	]
operation CRk(qctrl: Qubit, k: Int, qtarget: Qubit) : Unit is Adj {
    body(...) {
        let theta = 2.0 * Math.PI() / Convert.IntAsDouble(2 ^ k);
        Controlled R1([qctrl], (theta, qtarget));
    }
    adjoint auto;
}

operation Reverse(qs : Qubit[]) : Unit is Adj {
    body(...) {
        let n = Length(qs);
        for i in 0 .. n / 2 - 1
        { SWAP(qs[i], qs[n - i - 1]); }
    }
    adjoint auto;
}

operation QFT(qs : Qubit[]) : Unit is Adj {
    body(...) {
        let n = Length(qs);
        for i in 0 .. n - 2 {
            H(qs[i]);
            for j in i + 1 .. n - 1 {
                let k = j - i + 1;
                CRk(qs[j], k, qs[i]);   // Call "CRk"
            }
        }
        H(qs[n - 1]);
        Reverse(qs);    // Call "Reverse"
    }
    adjoint auto;
}
\end{lstlisting}

\begin{lstlisting}[
	xleftmargin=4mm,
	%xrightmargin=2mm,
	caption={The Q\# code for QPE program},
	label={list:qpe}
	]
operation QPE(Upower : (Int, Qubit[]) => Unit is Adj + Ctl,
           qsclock : Qubit[], qstarget : Qubit[]) : Unit is Adj {
    body (...) {     //Original operation
        let n = Length(qsclock);
        MultiH(qsclock);
        for i in 0 .. n - 1 {
            Controlled Upower([qsclock[n - i - 1]], (2 ^ i, qstarget)); 
        }
        Adjoint QFT(qsclock);
    }
    adjoint auto;  //Generate adjoint version of QPE by Q#.
}
\end{lstlisting}
\end{figure}

\subsubsection{Three Variants of a Subroutine}
In quantum programs, there are three important variations of subroutines: the \textit{inverse}, \textit{controlled}, and \textit{power} variants, which can also be combined. Formally, for a unitary operation $U$, these three variants implement the unitary transform $U^{-1}$, Controlled-$U$ and $U^k$, respectively. These variants play a crucial role in quantum software development. For example, in Listing~\ref{list:qpe}, the \texttt{QPE} program calls the \texttt{Adjoint QFT} (line 9), the inverse of \texttt{QFT}, and the \texttt{Controlled Upower} (line 7), which is the controlled version of \texttt{Upower}. It is important to note that these variants represent distinct subroutines compared to their original programs.

Some high-level quantum programming languages, such as Q\#~\cite{svore2018q} and isQ~\cite{Guo2022isQ}, support the generation and management of the inverse and controlled variants, as indicated by property 4.2 and property 4.3 in Table~\ref{table:QProjects}. As an example, Listing~\ref{list:hpow} gives an implementation of $H^n$, the power of the $H$ gate, which uses these variants. The power $n$ is implemented as an extra \texttt{Int} parameter \texttt{power}. Based on the result of $H^2=I$, we can obtain that $H^n = H$ if $n$ is odd; otherwise, $H = I$ (line 4). The inverse, controlled, and inverse-and-controlled variants can be generated automatically by Q\# (lines 6-8), which uses the keyword "\texttt{adjoint}" to represent inverse operation. 
However, this mechanism of automatically generating variants via programming languages is not always available. Sometimes we still need to write variants manually, which may lead to bugs. Therefore, it is necessary to propose specific techniques to test variants. We will discuss this issue in Section~\ref{subsubsec:SubroutineRelation}.

\begin{figure}[h]
\begin{lstlisting}[
	xleftmargin=4mm,
	%xrightmargin=2mm,
	caption={The implementation of the power variant of the H gate. The adjoint, controlled, and adjoint controlled versions are generated by Q\# automatically.},
	label={list:hpow}
	]
operation Hpower(power : Int, q : Qubit) : Unit is Adj + Ctl {
    body (...) {
        //If pow is odd, apply H, otherwise I
        if (power % 2 == 1) { H(q); }
    }
    adjoint auto;
    controlled auto;
    adjoint controlled auto;
}
\end{lstlisting}
\end{figure}

\subsubsection{$U^{-1}VU$ Structure}
\label{subsubsec:WithinApply}

There is a common structure in quantum programs, which can be formally represented as follows:

\begin{center}
$U^{-1}VU$
\end{center}

\begin{figure}
	\centering
	\includegraphics[scale=0.6]{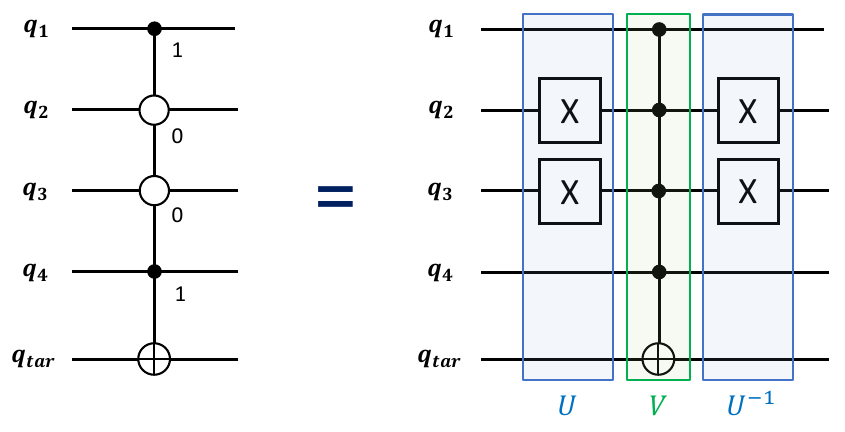}
	\caption{The decomposition of the Controlled-X gate is controlled by 1001. On the left is the note in the quantum circuit, where 0-controlling qubits are with hollow circles. It can be decomposed to the right with a standard controlled-X gate between the two \texttt{X} gates on the 0-controlling qubits.}
	\label{fig:CnCnCCNOT}
	
\begin{lstlisting}[
	xleftmargin=4mm,
	%xrightmargin=2mm,
	caption={The corresponding Q\# code of Figure~\ref{fig:CnCnCCNOT}.},
	label={list:CnCnCCNOT}
	]
operation C_nC_nC_C_X(q1 : Qubit, q2 : Qubit, q3 : Qubit,
                      q4 : Qubit, qtar : Qubit) : Unit is Adj + Ctl {
    body (...) {
        within {
            X(q2);
            X(q3);
        }
        apply {
            Controlled X([q1, q2, q3, q4], qtar);
        }
    }
    adjoint auto;
    controlled auto;
    adjoint controlled auto;
}
\end{lstlisting}
\end{figure}

\noindent Here, the execution order is from right to left: the subroutine $U$ creates an environment, and the subroutine $V$ operates on that environment. Once $V$ has been executed, the environment must be recovered to its original state, which is accomplished by applying $U^{-1}$. Some quantum programming languages, such as Q\#, provide syntactic sugar in the form of keywords \texttt{within} and \texttt{apply} to support it.
Figure~\ref{fig:CnCnCCNOT} provides an example of $U^{-1}VU$ structure, which shows the decomposition of a Controlled-$X$ gate controlled by the binary string "1001". The target gate is applied by default when the controlling qubit is 1; however, if we want it to be controlled by 0, an $X$ gate must be added before the default controlling gate. In addition, we must also recover the controlling qubits to their original state, which is achieved by applying an $X^{-1}$ ($=X$) gate after the default controlling gate. This $U^{-1}VU$ structure can be implemented in Q\# using the code shown in Listing~\ref{list:CnCnCCNOT}, which utilizes the "\texttt{within $U$ apply $V$}" statement to implement $U^{-1}VU$, with $U^{-1}$ being calculated automatically. Another example is the \texttt{HHL} program, as shown in Figure~\ref{fig:hhlcircuit}. In this case, $U$ is \texttt{QPE}, and $V$ is the controlled rotation \texttt{CRot}.

\subsection{Input and Output}
\label{subsec:IO}

\subsubsection{Input, Output, and IO Types of Quantum Programs}
\label{subsubsection:IOtype}
Testing design is typically based on the input and output of programs under test. Previous research~\cite{ali2021assessing,wang2021generating, wang2021application} has defined the input and output of a quantum program as a subset of all used qubits. However, this definition only considers qubit variables and neglects other possible types of input and output variables. In practice, many quantum subroutines contain other types of input or output variables, such as classical parameters or parameters of subroutines. For example, Listing~\ref{list:qft} shows that subroutine \texttt{CRk} contains a classical integer input parameter \texttt{k}, while Listing~\ref{list:qpe} requires \texttt{Upower}, an input of another subroutine, to be estimated. Therefore, it is valuable to classify quantum programs based on their input and output. Since quantum variables will influence the testing design, based on whether their input or output contains quantum variables, we can classify quantum subroutines into four \textit{IO types} as follows.

\begin{itemize}
\item \textsf{Type 1 - Classical:} The subroutine has no input and output quantum variables.
\item \textsf{Type 2 - Generate-Quantum:} The subroutine has input quantum variables but no output quantum variables.
\item \textsf{Type 3 - Detect-Quantum:} The subroutine has output quantum variables but no input quantum variables.
\item \textsf{Type 4 - Transform:} The subroutine has both input and output quantum variables.
\end{itemize}

\begin{figure*}
	\centering
	\subfigure[Classical]{\includegraphics[scale=0.4]{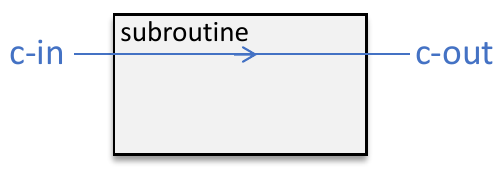}}
	\subfigure[Generate-Quantum]{\includegraphics[scale=0.4]{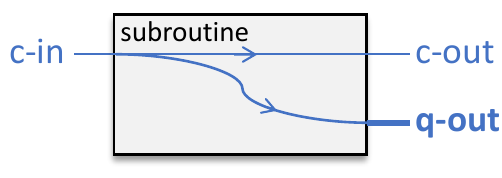}}
	\subfigure[Detect-Quantum]{\includegraphics[scale=0.4]{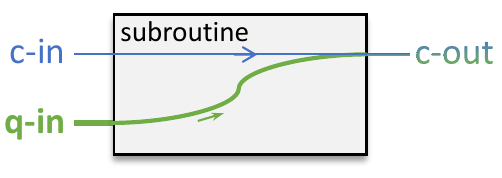}}
	\subfigure[Transform]{\includegraphics[scale=0.4]{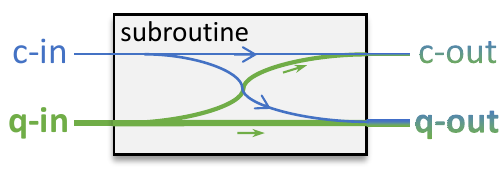}}
	\caption{The data flow of four types of quantum subroutines.}
	\label{fig:4typesdataflow}
\end{figure*}

Here, the term "quantum variable" refers not only to qubits or qubit arrays but also to subroutines that take qubits or qubit arrays as inputs or outputs. For example, in Listing~\ref{list:qpe}, the qubit arrays \texttt{qsclock} and \texttt{qstarget} in line 2 are quantum input variables. Additionally, the \texttt{Upower} subroutine in line 1, which takes a qubit array as input, is also a quantum input variable. Figure~\ref{fig:4typesdataflow} illustrates the data flow for the four IO types, showing the dependencies of classical input (c-in), quantum input (q-in), classical output (c-out), and quantum output (q-out).

\subsubsection{Logical Meaning of Quantum Variables}
It is also crucial to consider the logical meaning of quantum variables. For instance, consider the quantum adder program \texttt{QAdd}:

\begin{equation}
\label{equ:QAdd}
\texttt{QAdd}: \left|x\right>\left|y\right>\rightarrow\left|x\right>\left|x+y\right>
\end{equation}

\noindent In \texttt{QAdd}, the contents of quantum registers represent integers and add the first integer to the second. The quantum registers of $\left|x\right>$ and $\left|y\right>$ logically correspond to two sets of qubits. Therefore, the program's input should contain two quantum variables $\left|x\right>$ and $\left|y\right>$ instead of taking all qubits as one input variable. Correctly identifying the inputs and outputs of all program subroutines is a critical prerequisite for successfully finishing the testing task of a target program under test. We provide a more detailed discussion in Section~\ref{subsec:IOanalysis}.

\subsubsection{Entanglements Among Quantum Variables}
An essential difference between classical and quantum variables is the potential for entanglement among different quantum variables. Some quantum algorithms rely on these entanglements. For instance, as depicted in Figure~\ref{fig:teleport}, quantum teleportation exemplifies this concept. It operates based on the entanglement between variables $q_{anc}$ and $q_{dest}$, wherein a Bell state $\left|\beta_{00}\right>$ is generated over them. The existence of entanglements means that different quantum variables may not be independent. Operations on one variable can interfere with others. Additionally, if these quantum variables belong to different subroutines, these subroutines may interact with each other.

\subsubsection{Endian Mode of Quantum Variables}
\label{subsubsec:Endian}
Endianness describes the arrangement of bytes in computer memory. It comes in two types: \textit{big-endian} and \textit{little-endian}. Big-endian signifies that the most significant value in the sequence is stored at the lowest memory address, while little-endian means that the least significant value in the sequence is stored first.

\begin{figure*}
	\centering
	\subfigure[Big-endian input and big-endian output]{\includegraphics[scale=0.5]{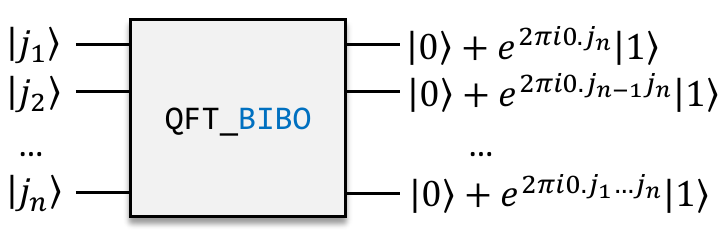}}
	\quad
	\subfigure[Big-endian input and little-endian output]{\includegraphics[scale=0.5]{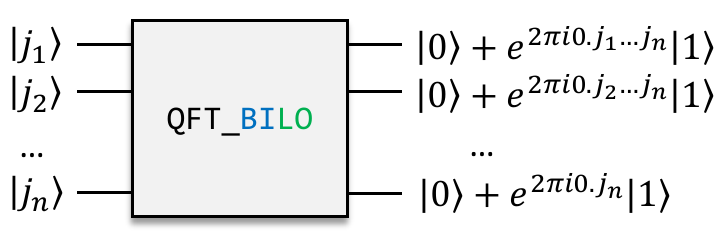}}
	\\
	\subfigure[Little-endian input and big-endian output]{\includegraphics[scale=0.5]{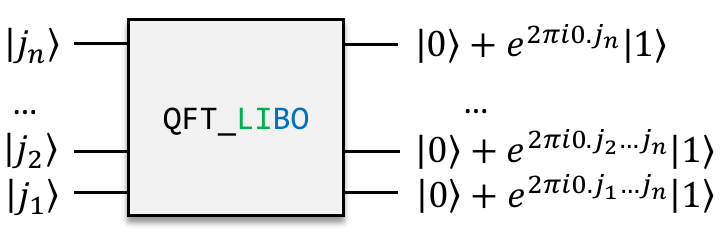}}
	\quad
	\subfigure[Little-endian input and little-endian output]{\includegraphics[scale=0.5]{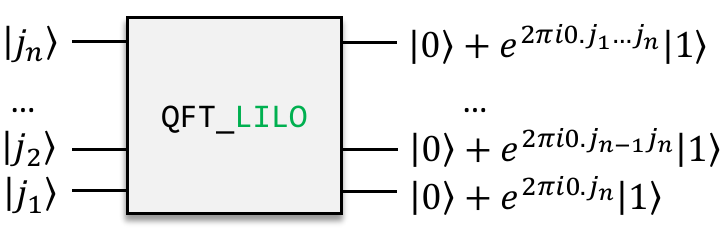}}
	\caption{Four possible IO endian modes for QFT program. Input integer $j = j_1 j_2 \cdots j_n$, i.e., $j_1$ is the highest bit of $j$. The index of qubit increases from top to bottom in each subroutine.}
	\label{fig:endians}
\end{figure*}

In classical computing, the computer architecture guarantees the endian mode of an integer, and programmers do not need to consider it while dealing with integers at the \textit{word} level. However, current quantum computing still operates on the \textit{qubit} level and requires programmers to consider the order (endian mode) of the qubits when programming quantum variables containing more than one qubit, such as quantum integers. Mismatched endian modes of two quantum subroutines can lead to errors. Figure~\ref{fig:endians} illustrates four possible qubit orders for the QFT program - input with big-endian (BI) or little-endian (LI) and output with big-endian (BO) or little-endian (LO). The QFT implementation in Listing~\ref{list:qft} is in BIBO mode, the typical implementation. However, if the calling of \texttt{Reverse} (line 29) is missed, it will result in BILO mode.

To manage the endian of qubit arrays, some programming languages such as Q\#, provide a preliminary mechanism. In the Q\# standard library, types \texttt{Bigendian} and \texttt{Littleendian} are two different encapsulations of the \textit{raw qubit array} \texttt{Qubit[]}. The \texttt{QFT()} function in the standard library requires a \texttt{Bigendian} parameter and is implemented in BIBO mode, while another function \texttt{QFTLE()} takes a \texttt{Littleendian} parameter and is implemented in LILO mode. Proper management of the endian of qubit arrays can help programmers avoid errors and ensure the correct implementation of quantum programs.

\subsection{Program Specification}
\label{subsec:PS}

Program specification gives the expected behavior of the program and is the foundation of testing execution. The purpose of testing is to check whether a program converts the given input into a specific output according to the program specification. 

\subsubsection{Probability-based Program Specification}
Previous research~\cite{wang2021generating, ali2021assessing, wang2021application} defined the program specification as the expected probability distribution of the output values under given input values. It is so-called \textit{probability-based program specification}. For example, the program in Listing~\ref{list:random} has no input variable, and we expect the output to be a uniform distribution on $\{0,1,2\}$. So the program specification can be written as follows:

\begin{center}
    \textsf{Input}: none. \qquad \textsf{Output}: (0 with $p_0=\frac{1}{3}$), (1 with $p_1=\frac{1}{3}$), (2 with $p_2=\frac{1}{3}$)
\end{center}

\subsubsection{Formula-based Program Specification}
Probability-based program specification implies that we must have a measurement at the end of each subroutine to convert quantum states into probability distributions. This restriction is not suitable for testing multi-subroutine programs. In a multi-subroutine program, its subroutine often transforms a quantum state into another state without measurement at its end. It is, therefore, better to consider the specification of quantum states rather than a probability distribution. Generally, for some given inputs, the program specification gives the expected output for each input. This relationship can be represented as a formula in most cases, and we call it \textit{formula-based program specification}. For example, Formula (\ref{equ:QAdd}) is an example of the description of \texttt{QAdd} program, and \texttt{QFT} program has the following formula description:

\begin{equation}
\label{equ:qftform}
	\texttt{QFT}: \left|j\right> \rightarrow \frac{1}{\sqrt{2^n}}\sum_{k=0}^{2^n-1}e^{2\pi i j k / 2^n}\left|k\right>
\end{equation}

\noindent where $\left|j\right>$ is a classical state with integer $j$. Formula (\ref{equ:qftform}) only gives the transform under classical state inputs. Fortunately, QFT is a unitary transform, so the expected output under general input can be deduced by linearity.

More precisely, the formula (\ref{equ:qftform}) can be rewritten in bitwise form, which is related to the endian mode. Suppose we adopt BIBO mode (Figure~\ref{fig:endians}(a)), then

\begin{equation}
\label{equ:qftqubitform}
\texttt{QFT\_BIBO}:
\begin{array}{c}
	\left|j_1\right> \rightarrow \frac{1}{\sqrt 2}(\left|0\right>+e^{2\pi i0.j_n}\left|1\right>) \\
	\cdots \\
	\left|j_n\right> \rightarrow \frac{1}{\sqrt 2} (\left|0\right>+e^{2\pi i0.j_1j_2 \ldots j_n}\left|1\right>)
\end{array}
\end{equation}

\noindent where $n$ is the length of $j$ and $j=j_1j_2 \ldots j_n$. Formula (\ref{equ:qftqubitform}) gives the output state on each qubit under the input $\left|j_1\right>\left|j_2\right> \ldots \left|j_n\right>$.

\subsection{Quantum State Generation and Detection}
\label{subsec:GenerationDetection}

Testing quantum programs relies heavily on the handling of quantum variables. However, generating and detecting quantum states have more significant challenges than their classical counterparts due to the properties of quantum states.

In classical computing, input parameters can easily be fed into a computer. A natural language description of the input, such as a block of text or a decimal number, can be converted into a binary string according to predetermined coding rules. This binary string is then converted into the internal state of a computer's components (e.g., high and low voltages represent 1 and 0, respectively). However, quantum variables not only have classical states like binary strings, but they also have superposition states or mixed states. Some states that can be easily described in natural language may be difficult to prepare on quantum devices. For example, for general superposition states, the algorithm proposed by~\cite{Lov2002Creating} can be used to generate the superposition state $\sum_i \sqrt{p_i}\left|i\right>$ if we know the probability distribution $\{p_i\}$ of each amplitude. However, this algorithm may be computationally expensive if the target states do not have specific properties. In fact, it has been proven that preparing some quantum states requires exponential costs~\cite{knill1995approximation}.

Once a quantum program has been executed with a given input, the next step is to compare the actual output with the expected output. While reading classical memory is relatively straightforward, quantum computing presents a challenge regarding reading output containing quantum variables. Measurements must be made to extract information about these variables, but they may cause the state of the quantum variables to collapse and interfere with the program's execution. Consequently, many classical testing methods that depend on intermediate variables cannot be directly applied to test quantum programs.

In the field of quantum information, various methods have been proposed to solve problems concerning quantum state detection, such as quantum distance estimation~\cite{Flammia2011FewPauli,Cerezo2020VariationalQF}, quantum discrimination~\cite{Stephen2008Discrimination,zhang2006300discri_mixed,Zhang2007discri_puremix}, quantum tomography~\cite{tomography1989,Chuang1996statetomo}, and the Swap Test~\cite{buhrman2001quantum,ekert2002direct}. Unfortunately, almost all methods require multiple copies of the target quantum state, and the results are not entirely accurate. This is particularly challenging in testing practice, as it often requires repeating the execution of the target program. 

\begin{tcolorbox}[leftrule=0.5mm, rightrule=0.5mm, toprule=0mm, bottomrule=0mm, colframe=purple, colback=white]
\textbf{Answer to RQ1: } 
In this survey, we present a summary of the critical properties of multi-subroutine quantum programs, which are as follows:

\begin{enumerate}
\item Quantum programs, distinct from quantum circuits, allow for variable scaling and incorporate classical computation and control flows.

\item Quantum subroutines can exhibit variations, with `reverse,' `controlled,' and `power' being three typical variants.

\item Quantum programs can encompass both quantum and classical inputs and outputs. They can also be categorized based on whether the inputs or outputs involve quantum variables.

\item Current quantum computing operates at the qubit level, and therefore handling endian mode remains the responsibility of programmers.

\item Due to the fundamental principles of quantum mechanics, generating and detecting quantum states pose substantial challenges.

\end{enumerate}
\end{tcolorbox}




\section{Unit Testing and its Subtasks}
\label{sec:UnitTest}

With an understanding of the properties of multi-subroutine quantum programs, our focus now shifts to testing execution. Our research centers around the following research question:

\begin{itemize}
	\item \textbf{RQ2:} How do the properties of multi-subroutine quantum programs affect the testing processes?
\end{itemize}

\noindent To address RQ2, we will delve into the testing processes and techniques that are suitable for multi-subroutine quantum programs, taking into account their unique properties.


Testing a multi-subroutine quantum program involves both unit testing for each subroutine and integration testing for the entire program structure. This section focuses on unit testing, while integration testing will be discussed in Section~\ref{sec:Integration}. We first discuss the steps of IO analysis in Section~\ref{subsec:IOanalysis}. Then, we explore general testing criteria in Section~\ref{subsec:GenericCriteria}. Following that, we prepare quantum states related to the testing process, as elaborated in Section~\ref{subsec:stategen}. Sections~\ref{subsec:subtasks}$\sim$\ref{subsec:PSCheck} provides a detailed exploration of these subtasks. Additionally, we discuss the generation and execution of each test case in Section~\ref{subsec:TestCases} for a comprehensive understanding of the unit testing process.

\subsection{IO Analysis}
\label{subsec:IOanalysis}

IO analysis involves identifying the input and output variables of a subroutine. As Section~\ref{subsec:IO} mentions, this analysis is the foundation for unit testing. In order to streamline the test design process, it is important to focus on the minimal subset of variables relevant to the test design. Specifically, for a concrete subroutine, only the variables that users need to assign are considered inputs, while only the variables that users are interested in after execution are considered outputs.
For example, in many quantum algorithms, some qubits should always be initialized with an \textit{all-zero state} $\left|0\ldots0\right>$. These qubits should not be regarded as input. Another typical case is a qubit array, such as \texttt{qs} in Listing~\ref{list:qft} (line 18). Like an array of integers, a qubit array contains both length and data. Therefore, when identifying the input and output of a subroutine, a qubit array should be regarded as two variables: length and quantum state.

To represent the input and output of a quantum program, we introduce the \textit{IO mark}, which can help developers design test cases. The general form of the IO mark is:

\vspace*{1mm}
\begin{center}
\textit{program} : (\textit{input variables}) $\rightarrow$ (\textit{output variables'})
\end{center}
\vspace*{1mm}

\noindent
We use \underline{underline} to denote the input parameters of subroutine type and \textbf{bold} font to denote quantum variables. These two marks can be combined to represent subroutine type with quantum variables. The idea to represent quantum-related content in bold font comes from the Q-UML modeling language~\cite{Perez-Delgado2020quantum}. We add an "apostrophe" (') on each output variable to distinguish input and output. If a variable~\textit{var} is both input and output, we denote the input and output as the same name, i.e., \textit{var} is the input, and \textit{var'} is the output. Sometimes, we need to specify the endian mode for multi-qubit quantum variables. In this case, we add the variables with "BE" (big-endian) or "LE" (little-endian) superscript, such as \textbf{qvar}$^{\mathrm{BE}}$.

According to Section~\ref{subsec:IO}, quantum subroutines can be classified into four types, and different types of quantum programs have different testing strategies. Having the IO mark, we are able to know which type the target program is. Example~\ref{example:QPEIO} shows how to perform IO analysis and use IO marks.

\begin{figure}[h]
	\centering
	\includegraphics[scale=0.7]{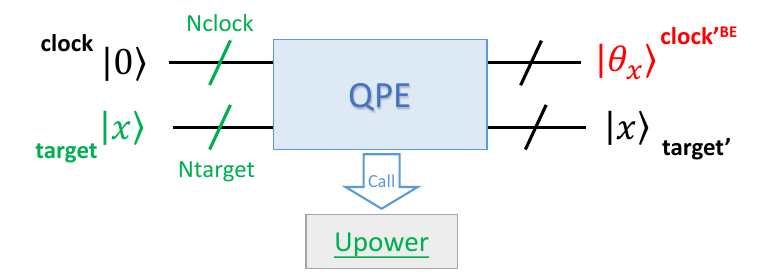}
	\caption{The diagram illustrates the input and output of the \texttt{QPE} program, with the input variables represented in green and the output variable denoted in red.}
	\label{fig:qpeio}
\end{figure}

\begin{example}
\label{example:QPEIO}
IO analysis for \texttt{QPE} program.

The \texttt{QPE} program in Listing~\ref{list:qpe} has three parameters: \texttt{Upower}, \texttt{qsclock}, and \texttt{qstarget}, while the latter two are qubit-array types. On the input side, \texttt{qsclock} contains two variables: the length "Nclock" and the data "\textbf{clock}." It is similar to \texttt{qstarget}, which contains two variables as well: the length "Ntarget" and the data "\textbf{target}." On the output side, they should be denoted as "\textbf{clock'}" and "\textbf{target'.}" \texttt{Upower} is a parameter of quantum subroutine type and should be denoted as "\underline{\textbf{\texttt{Upower}}}." Figure~\ref{fig:qpeio} shows the structure, input, and output of the \texttt{QPE} program.

In the program, the quantum state \textbf{clock} is always initialized with an all-zero state, so it is not an input. In most applications of the \texttt{QPE} program, we do not care about the post-state \textbf{target'}, so it is not output. In Figure~\ref{fig:qpeio}, the input variables are shown in green, while the output variable is shown in red. The endian mode of the \textbf{target} is not important, but that of the \textbf{clock'} is important since the \textbf{clock'} stores the binary representation of $\theta_x$. Subroutine \texttt{Adjoint QFT} is of BIBO endian mode, so the output state \textbf{clock'} is of big-endian (BE) mode. The IO mark can be written as:

\vspace*{2mm}
\begin{center}
	\texttt{QPE} : (Nclock, Ntarget, \underline{\textbf{\texttt{Upower}}}, \textbf{target}) $\rightarrow$ (\textbf{clock'}$^{\mathrm{BE}}$)
\end{center}

\vspace*{2mm}
\noindent
Obviously, \texttt{QPE} is a quantum program of \textit{transform} type.
\end{example}

\subsection{Generic Testing Principles and Criteria}
\label{subsec:GenericCriteria}

After finishing the IO analysis of a subroutine, the next step is performing unit testing by executing some subtasks. Before that, we need to discuss the generic testing criteria that can guide our testing tasks.

A testing criterion is a set of rules used to help determine whether a program is adequately tested by a test suite and guides the testing design~\cite{ammann2016introduction}. Some coverage criteria for quantum programs have been proposed, such as Quito (quantum input-output coverage)~\cite{ali2021assessing}, suitable for small-scale and fixed-scale quantum programs. However, there is still a lack of testing criteria for multi-subroutine quantum programs in previous work. In this section, we propose some quantum-related testing criteria and principles, ranging from coarse-grained to fine-grained, based on the properties of quantum variables and quantum programs. We will narrate each criterion or principle and give the reason or theoretical basis for adopting it.

\subsubsection{Basic Testing Principles}
\label{subsubsec:BasicPrinciples}

Quantum programs require special attention when selecting quantum inputs due to the distinct properties of quantum states compared to classical programs. As a result, we suggest the following principle for carefully considering the selection of quantum inputs:

\begin{principle}
\label{crit:Principle}
The fundamental principle of quantum input selection:

\begin{itemize}[leftmargin=3.5em]
\item[(1)] Quantum input states selected should be representative of the testing objectives;
\item[(2)] Selected input quantum states should be preparable using well-defined methods;
\item[(3)] Corresponding output quantum states should also be verifiable using well-defined methods.
\end{itemize}
\end{principle}

In Principle~\ref{crit:Principle}, (1) guarantees the validity of the test task, which is a similar requirement in testing classical programs. (2) and (3) are based on the fact that generating and checking general quantum states may require complex processes (see Section~\ref{subsec:GenerationDetection}), which may easily lead to errors. In the following sections, we will discuss some representative input states that are valuable for detecting errors and are relatively easy to generate. We will also discuss some representative methods for checking output states.

\subsubsection{Partition Quantum Input by State Types}
\label{subsubsec:PartByType}

In a testing task, we need to select some appropriate inputs to run the target program. Equivalence class partitioning is an important testing method for classical programs~\cite{ammann2016introduction}. The basic idea is to partition the input space into several "equivalence classes" in a logical meaning, where all cases in one class have the same effect in finding bugs. Such an idea can also be applied to quantum program testing. For quantum programs with quantum input variables, we can consider the partition of quantum input variables. A natural, coarse-grained partitioning scheme is by type of quantum state. In general, there are three types of quantum states: \textit{classical}, \textit{superposition}, and \textit{mixed states} (see Section~\ref{sec:background}). Thus, we can define a "classical-superposition-hybrid partition."

\begin{principle}
\label{crit:CSMP}
Classical-superposition-mixed partition (CSMP).

For each input variable of each quantum state type, partition it into classical state input, superposition state input, and mixed state input.
\end{principle}

However, since a mixed state can be considered as a probability distribution of several pure states unless the program specification has special regulations about the mixed state, we can omit the coverage of the mixed state, which can be simplified to the "classical superposition partition" as follows.

\begin{principle}
\label{crit:CSP}
Classical-superposition partition (CSP).

For each input variable of each quantum state type, partition it into classical state input and superposition state input.
\end{principle}

Which principle should be used during testing is determined by whether the program specification has special regulations for mixed states. If it does, we can use CSMP. Otherwise, CSP is enough.

Note that whether using CSP or CSMP, both classical and superposition states should be covered. There are two reasons for this. First, the program specification often gives the expected output states of the program under a specific input state, and the given input states are usually classical (such as formula (\ref{equ:qftform}) in Section~\ref{subsec:PS}), so covering classical states is to check the specification directly. Second, superposition is the essential difference between classical and quantum variables, and testing classical input alone cannot ensure the program behavior in superposition input. We will discuss the necessity of covering superposition states in Section~\ref{sec:evaluation}, where the result will be shown in Section~\ref{subsec:unitNecessity}.

The CSP and CSMP principles mentioned above are coarse-grained and represent minimal partitioning requirements. However, for a specific quantum program, a more fine-grained partitioning based on its logical meaning is necessary. In Section~\ref{subsubsec:EqClassPart}, we will further discuss equivalent partitioning.

\subsubsection{Criteria for Selecting Input States Derived from Density Matrix Decomposition}
\label{subsubsec:InputSelect}

While the choice of a specific input quantum state typically depends on the properties of the target program and its specifications, analyzing the structure of the quantum input can provide general selection strategies. Unlike classical variables, the input space of a quantum variable is continuous, even though it contains only limited qubits. Fortunately, we can find a set of discrete representatives for the continuous input space.


In quantum programming, transforming the quantum input state into an output state is a fundamental operation. Notably, a quantum program containing \textit{if} and \textit{while-loop} statements—where the conditions may involve measurements of qubits—can be represented as a quantum operation~\cite{ying2016foundationQP}. Consequently, we can model a quantum program's behavior as a quantum operation, mapping the quantum input density matrix space into the quantum output density matrix space (treating classical variables as a special case of quantum variables). Intriguingly, a quantum operation is a linear map from the input density matrix space to the output density matrix space, and its behavior is determined by a set of bases. For an $n$-qubit quantum system, the density matrix is of size $2^n \times 2^n$, resulting in a density matrix space of dimension $4^n$. Hence, any group of bases contains $4^n$ matrices. Fortunately, we only need to choose a small set of input states in testing tasks. The challenge lies in a base matrix potentially not being a legal density matrix of any quantum state. The solution involves further decomposing the base matrix into the sum of several matrices which are the density matrix of legal pure quantum states. Thus, we establish a feasible strategy for generating input quantum states as follows:

\begin{principle}
\label{crit:InputFromBases}
We can devise a criterion for selecting input quantum states through the following three steps:

\begin{itemize}[leftmargin=3.5em]
    \item[(1)] Choose a set of bases from the density matrix space;
    \item[(2)] Decompose each base matrix into the sum of matrices representing legitimate pure states;
    \item[(3)] The criterion involves selecting input states from these pure states.
\end{itemize}
\end{principle}

Since Principle~\ref{crit:InputFromBases} is based on the bases of the density matrix space, the selection criteria devised by it have \textit{completeness assurance} - assume that the target program has an error, the error will be detected under at least one input as long as we test all the input states of the entire set. Next, we will examine two specific groups of bases, each leading to its own respective selection criterion.

An obvious group of bases is $\{\left|x\right>\left<y\right|\}$, where $x,y = 0,1,\dots,2^n-1$, i.e., the matrix element is 1 on the location ($x,y$) and 0 on other locations. However, if $x \neq y$, $\left|x\right>\left<y\right|$ is not a legal quantum state because $tr(\left|x\right>\left<y\right|)=0\neq 1$. Fortunately, consider the following two states:

\begin{equation}
\notag
\left|+_{xy}\right> = \frac{1}{\sqrt 2}(\left|x\right> + \left|y\right>),\qquad \left|-_{xy}\right> = \frac{1}{\sqrt 2}(\left|x\right> + i\left|y\right>)
\end{equation}

\noindent where $i^2=-1$. Then $\left|x\right>\left<y\right|$ can be decomposed into the linear combination of four pure states~\cite{nielsen2002quantum}:

\begin{equation}
\notag
\left|x\right>\left<y\right| = \left|+_{xy}\right>\left<+_{xy}\right| + i\left|-_{xy}\right>\left<-_{xy}\right| - \frac{1+i}{2}\left|x\right>\left<x\right| - \frac{1+i}{2}\left|y\right>\left<y\right|
\end{equation}

\noindent
So for all $x,y = 0,1,\dots, 2^n-1$ and $x\neq y$, states with form $\left|x\right>$, $\left|+_{xy}\right>$, and $\left|-_{xy}\right>$ also constitute a group of bases. They are not orthonormal bases, but all elements are valid pure states, making them suitable for use as input states. Thus, we establish the following criterion for input selection:

\begin{criterion}
\label{crit:STV}
Single-and-two-value selection criterion (STV)

Select the quantum input states in the following three forms:

\begin{itemize}[leftmargin=3.5em]
\item[(1)] $\left|x\right>$: a classical state of single integer value $x$;

\item[(2)] $\frac{1}{\sqrt{2}}(\left|x\right>+\left|y\right>)$: a superposition state of two values $x$ and $y$;

\item[(3)] $\frac{1}{\sqrt{2}}(\left|x\right>+i\left|y\right>)$: a superposition state of two values $x$ and $y$, and the value of $y$ has an additional relative phase $i$, where $i^2=-1$.
\end{itemize}
\end{criterion}

Another typical group of bases for density matrix space are all $n$-qubit Pauli matrices, i.e., the tensor product single-qubit Pauli matrices: $\sigma_{\vec{v}} = \sigma_{v_1}\otimes \sigma_{v_2} \otimes \cdots \otimes \sigma_{v_n}$, where $v_i \in \{0,1,2,3\}$, $\vec{v} = (v_1,\dots,v_n)$, and single-qubit Pauli matrices are

\begin{equation}
\label{equ:paulis}
\sigma_0 = \left[
\begin{array}{cc}
	1 & 0\\
	0 & 1
\end{array}
\right],
\sigma_1 = \left[
\begin{array}{cc}
	0 & 1\\
	1 & 0
\end{array}
\right],
\sigma_2 = \left[
\begin{array}{cc}
	0 & -i\\
	i & 0
\end{array}
\right],
\sigma_3 = \left[
\begin{array}{cc}
	1 & 0\\
	0 & -1
\end{array}
\right]
\end{equation}

\noindent
However, $\sigma_{\vec{v}}$ is also not a legal quantum state, but we can decompose it into the sum of its eigenstates $\sigma_{\vec{v}} = \sum_i{\lambda_i\left|\psi_i\right>\left<\psi_i\right|}$, where $\lambda_i$ is the $i$-th eigenvalue of $\sigma_{\vec{v}}$ and $\left|\psi_i\right>$ is the corresponding eigenstate. $\left|\psi_i\right>$ is the tensor product of single-qubit Pauli eigenstate\footnote{Any single-qubit state is the eigenstate of $\sigma_0$, so we only need to consider the eigenstates of $\sigma_1$, $\sigma_2$, and $\sigma_3$.}:

\begin{equation}
\label{equ:PauliState}
\left|\psi_i\right> \in \{ \left|0\right>, \left|1\right>, \left|+\right>, \left|-\right>, \left|+_i\right>, \left|-_i\right> \} ^ {\otimes n}
\end{equation}
\vspace{2mm}

\noindent where $\left|+\right> = \frac{1}{\sqrt 2}(\left|0\right>+\left|1\right>)$, $\left|-\right> = \frac{1}{\sqrt 2}(\left|0\right>-\left|1\right>)$, $\left|+_i\right> = \frac{1}{\sqrt 2}(\left|0\right>+i\left|1\right>)$, and $\left|-_i\right> = \frac{1}{\sqrt 2}(\left|0\right>-i\left|1\right>)$, i.e., the eigenvectors of all Pauli matrices. Actually, the freedom degree of a $n$-qubit density matrix is $4^n$, which means some Pauli eigenstates are not independent. Fortunately, this is not a significant concern because, in testing, we typically sample only a small subset of Pauli eigenstates. This leads us to another criterion for input selection:

\begin{criterion}
\label{crit:Pauli}
Pauli selection criterion (PAULI)

Select the quantum input states from the $n$-qubit Pauli states:

\begin{equation}
\{ \left|0\right>, \left|1\right>, \left|+\right>, \left|-\right>, \left|+_i\right>, \left|-_i\right> \} ^ {\otimes n}
\end{equation}

\end{criterion}

STV and PAULI are particularly useful when generating random inputs in practice. Both STV and PAULI criteria have their own characteristics and strengths. Quantum input states from STV are controlled by two integer numbers, $x$ and $y$, so the STV criterion is useful in testing numerical quantum programs. The advantage of the PAULI criterion is that the quantum input states can be generated using only single-qubit gates (see Section~\ref{subsec:stategen}), so it is useful on operation-constrained quantum devices (e.g., devices that do not support remote CNOT operation).

\subsubsection{Superposition-Cover-All-Qubit Criterion}
\label{subsubsec:SCAQ}

Since superposition input states are able to expose some bugs, which can occur at any qubit, it is necessary to require that superposition states exist at every qubit. In the following, we introduce a novel criterion: the \textit{superposition-cover-all-qubit criterion} (SCAQ). We will begin by providing a formal description, followed by an intuitive illustration.

\begin{criterion}
\label{crit:SCAQ}
Superposition-cover-all-qubit criterion (SCAQ)

Consider an $n$-qubit quantum input state. Let each qubit be denoted as $q_1, q_2, \dots, q_n$, and define $Q=\{q_1, q_2, \dots, q_n\}$ as the set of qubits. The SCAQ criterion selects a set of quantum input states $A$ that satisfy the following condition: $\forall q_i \in Q$, $\exists \left|\psi\right> \in A$ such that $tr_{Q \textbackslash \{q_i\} }\left( \left|\psi\right>\left<\psi\right| \right) \neq \left|0\right>\left<0\right|$ and $tr_{Q \textbackslash \{q_i\} }\left( \left|\psi\right>\left<\psi\right| \right) \neq \left|1\right>\left<1\right|$, where $tr_{Q \textbackslash \{q_i\} }$ represents the partial trace over all qubits except $q_i$.
\end{criterion}

Intuitively, $tr_{Q \textbackslash {q_i} }\left( \left|\psi\right>\left<\psi\right| \right)$ denotes the substate of $\left|\psi\right>$ on qubit $q_i$. Consequently, the SCAQ criterion implies the selection of a set of input states that adhere to the following condition: ensuring that for each qubit in all quantum input variables, there must exist at least one state within the entire input set that exhibits superposition on that specific qubit. Example~\ref{example:SCAQ} presents two sets of inputs, one satisfying SCAQ but another not.

\begin{example}
\label{example:SCAQ}
The following input set satisfies SCAQ:

\begin{equation}
\notag
\{ \frac{1}{\sqrt{2}}(\left|000\right>+\left|100\right>), \frac{1}{\sqrt{2}}(\left|000\right>+\left|010\right>), \frac{1}{\sqrt{2}}(\left|000\right>+\left|001\right>) \}
\end{equation}

\noindent
The superposition of the first qubit occurs in state $\frac{1}{\sqrt{2}}(\left|000\right>+\left|100\right>) = \frac{1}{\sqrt{2}}(\left|0\right>+\left|1\right>)\left|00\right>$, and in this state, the second and third qubits have no superposition. The superposition of the second and third qubits occurs in the other two states.

The following input set does not satisfy SCQA since all three states have no superposition of the third qubit.

\begin{equation}
\notag
\{ \frac{1}{\sqrt{2}}(\left|000\right>+\left|100\right>), \frac{1}{\sqrt{2}}(\left|000\right>+\left|010\right>), \frac{1}{\sqrt{2}}(\left|000\right>+\left|110\right>) \}
\end{equation}

\end{example}

A typical superposition state is with the form $\frac{1}{\sqrt{2}}(\left|x\right>+\left|\bar{x}\right>)$, where $x$ is a binary representation of an integer and $\bar{x}$ is the \textit{bitwise-negation} of $x$. We call such a state \textit{complementary superposition state}, which is one of the minimum input sets that satisfy SCAQ. The generation of complementary superposition state is shown in Section~\ref{subapp:CompSup}. We will have an evaluation of the effectiveness of the SCAQ criterion in Section~\ref{sec:evaluation}, where the result will be shown in Section~\ref{subsec:evaluateSCAQ}.

In this section, we discussed several testing principles and criteria based on the properties of general quantum programs. However, there may be additional useful testing criteria yet to be discovered. For instance, specific structures common in quantum programs may benefit from new testing principles and criteria.

\subsection{Preparation of Some Related Quantum States}
\label{subsec:stategen}

In this section, we provide a brief overview of preparation algorithms for specific quantum states relevant to the criteria outlined in Section~\ref{subsec:GenericCriteria}. To facilitate the description of these algorithms, we adopt the convention that the index of any array starts from 0. We denote the $i$-th element of an array \textit{Arr} as \textit{Arr}$[i]$. Moreover, we treat an integer as an array of bits.

\subsubsection{Classical States}
\label{subapp:ClsState}

Generating classical states is a straightforward process. Given an integer $x$ with an $n$-bit binary representation, the $n$-qubit classical state $\left|x\right>$ can be generated from the all-zero state $\left|0\right>$ by applying an \texttt{X} gate on the qubits associated with the binary '1', as shown in Algorithm~\ref{alg:KetX}.

\begin{figure}
	\centering
	\subfigure[The circuit to generate GHZ state $\frac{1}{\sqrt{2}}(\left|0000 0\right>+\left|11111\right>)$]{\includegraphics[scale=0.35]{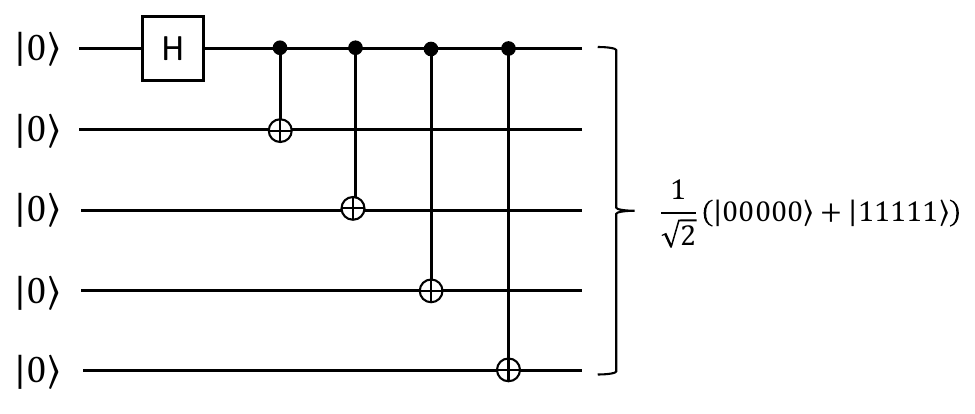}}
	\hspace{2mm}
	\subfigure[The circuit to generate state $\frac{1}{\sqrt{2}} (\left|01101\right>+e^{i\theta}\left|10010\right>)$. It is obtained by inserting several proper \texttt{X} gates (step I) and an $R_1(\theta)$ gate (step II) in the circuit of (a).]{\includegraphics[scale=0.35]{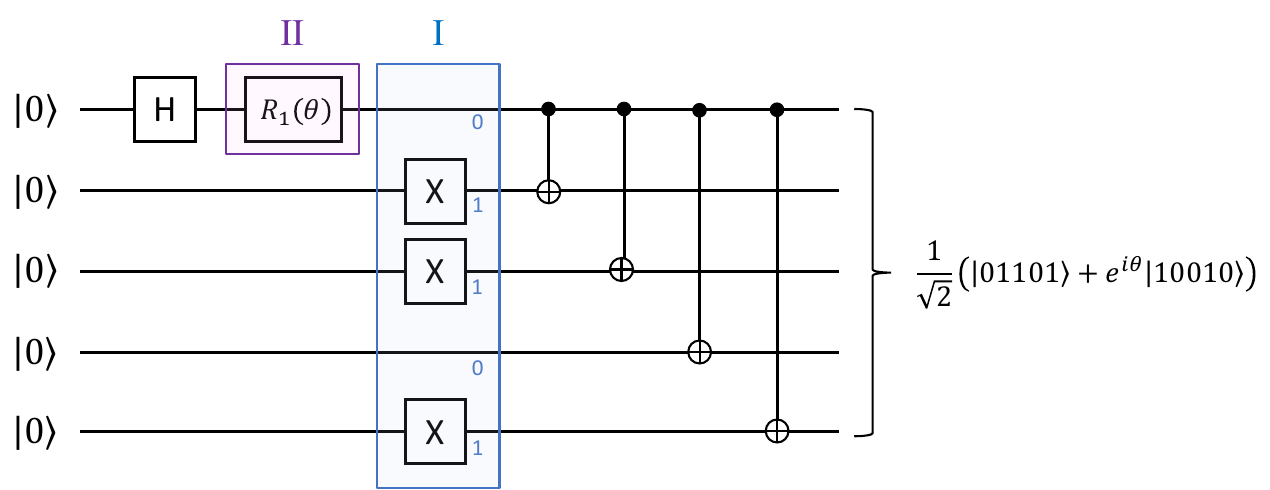}}
	\caption{Two quantum circuit examples of generating complementary superposition states.}
	\label{fig:CompSup}
\end{figure}
\begin{figure*}
\begin{minipage}{0.49\linewidth}
\begin{algorithm}[H]
\footnotesize
\caption{\footnotesize\texttt{Generate\_KetX}}
\label{alg:KetX}
\KwIn{$n$-bit integer $x$ (binary representation).}
\KwOut{$n$-qubit quantum state $\left|x\right>$}

\textbf{qs} $\leftarrow \left|0\right>^{\otimes n}$\;
\For{$i$ in $0 \dots n-1$}
{
\If{$x[i] = 1$}{\texttt{X}(\textbf{qs}$[i]$)\;}
}
\Return{\textbf{qs}}\;
\end{algorithm}
\end{minipage}\hspace{1mm}
\begin{minipage}{0.49\linewidth}
\begin{algorithm}[H]
\footnotesize
\caption{\scriptsize\texttt{Generate\_KetX\_plus\_EITheta\_KetNegX}}
\label{alg:KetXplusEIThetaKetNegX}
\KwIn{$n$-bit integer $x$ and the phase angle $\theta$.}
\KwOut{$n$-qubit state $\frac{1}{\sqrt{2}}(\left|x\right>+e^{i\theta}\left|\bar{x}\right>)$.}

\uIf{$x[0] = 0$}
{$x'\leftarrow x$\;}
\Else
{$x'\leftarrow \bar{x}$\;}
\textbf{qs} $\leftarrow \left|0\right>^{\otimes n}$;
\texttt{H}(\textbf{qs}$[0]$)\;
$R_1$($\theta$, \textbf{qs}$[0]$)\;
\For{$i$ in $1 \dots n-1$}
{
\If{$x'[i] = 1$}{\texttt{X}(\textbf{qs}$[i]$)\;}
\texttt{CNOT}(\textbf{qs}$[0]$, \textbf{qs}$[i]$)\;
}
\Return{\textbf{qs}}\;
\end{algorithm}
\end{minipage}
\\
\begin{minipage}{0.49\linewidth}
\begin{algorithm}[H]
\footnotesize
\caption{\footnotesize\texttt{Generate\_KetX\_plus\_EITheta\_KetY}}
\label{alg:KetXplusEIThetaKetY}
\KwIn{Two $n$-bit integers $x$ and $y$; phase angle $\theta$.}
\KwOut{$n$-qubit quantum state $\frac{1}{\sqrt{2}}(\left|x\right>+e^{i\theta}\left|y\right>)$.}

\textbf{qs} $\leftarrow \left|0\right>^{\otimes n}$\;
$S\leftarrow$ the indexes of same bits of $x$ and $y$\;
$D\leftarrow$ the indexes of different bits of $x$ and $y$\;
$s\leftarrow x[S]$;
$d\leftarrow x[D]$\;
\textbf{qs}$[S] \leftarrow$ \texttt{Generate\_KetX}($|S|$, $s$)\;
\textbf{qs}$[D] \leftarrow$ \texttt{Generate\_KetX\_plus\_EITheta\_KetNegX}($|D|$, $d$, $\theta$)\;
\Return{\textbf{qs}}\;
\end{algorithm}
\end{minipage}\hspace{1mm}
\begin{minipage}{0.49\linewidth}
\begin{algorithm}[H]
\footnotesize
\caption{\footnotesize\texttt{Generate\_Single\_Pauli}}
\label{alg:SinglePauli}
\KwIn{The index $i$ of single Pauli state.}
\KwOut{Single qubit Pauli state $\left|\Phi_i\right>$}

\textbf{q} $\leftarrow \left|0\right>$\;
\Switch{the value of $i$}{
\lCase{$i=2$}{\texttt{X}(\textbf{q})}
\lCase{$i=3$}{\texttt{H}(\textbf{q})}
\lCase{$i=4$}{\texttt{X}(\textbf{q}); \texttt{H}(\textbf{q})}
\lCase{$i=5$}{\texttt{H}(\textbf{q}); \texttt{S}(\textbf{q})}
\lCase{$i=6$}{\texttt{H}(\textbf{q}); \texttt{S}$^\dagger$(\textbf{q})}
}
\Return{\textbf{q}}\;
\end{algorithm}
\end{minipage}
\\
\begin{minipage}{0.49\linewidth}
\begin{algorithm}[H]
\footnotesize
\caption{\footnotesize\texttt{Generate\_Pauli\_State}}
\label{alg:PauliState}
\KwIn{The number of qubits $n$; the array $A$ of Pauli indexes (with length $n$).}
\KwOut{Pauli state associated with $A$.}

\textbf{qs} $\leftarrow \left|0\right>^{\otimes n}$\;
\For{$i$ in $0 \dots n-1$}
{\textbf{qs}$[i] \leftarrow$ \texttt{Generate\_Single\_Pauli}($A[i]$)\;}
\Return{\textbf{qs}}\;
\end{algorithm}
\end{minipage}\hspace{1mm}
\begin{minipage}{0.49\linewidth}
\begin{algorithm}[H]
\footnotesize
\caption{\footnotesize\texttt{Generate\_Mixed\_State}}
\label{alg:MixedState}
\KwIn{The ensemble representation $\{(p_i, \left|\psi_i\right>)\}$ and the generate operation $U_i$ for each element state $\left|\psi_i\right>$.}
\KwOut{Target mixed state.}
$r\leftarrow$ Random integer with probability distribution \{$Pr[X=i] = p_i$\}\;
\textbf{qs} = $\left|0\right>$; $U_r$(\textbf{qs})\;
\Return{\textbf{qs}}\;
\end{algorithm}
\end{minipage}
\end{figure*}

\subsubsection{Complementary Superposition States}
\label{subapp:CompSup}

In discussing SCAQ (Criterion~\ref{crit:SCAQ}), the complementary superposition state $\frac{1}{\sqrt{2}}(\left|x\right>+\left|\bar{x}\right>)$ is a typical state that satisfies the SCAQ criterion, where $\bar{x}$ is the bitwise negation of $x$. In this section, we present a generation algorithm for a more general state $\frac{1}{\sqrt{2}}(\left|x\right>+e^{i\theta}\left|\bar{x}\right>)$ with a relative phase $e^{i\theta}$.

To generate this state, we can start by applying an \texttt{H} gate on the first qubit, followed by a series of \texttt{CNOT} gates between the first qubit and the others to generate the \textit{GHZ state} $\frac{1}{\sqrt{2}}(\left|0\dots 0\right>+\left|1\dots 1\right>)$ (see Figure~\ref{fig:CompSup} (a)). Then, we can insert \texttt{X} gates at the appropriate locations according to the binary representation of $x$ or $\bar{x}$ (depending on whose first bit is 0) to obtain the state $\frac{1}{\sqrt{2}}(\left|x\right>+\left|\bar{x}\right>)$ (see step I in Figure~\ref{fig:CompSup} (b)). Finally, we can generate the phase factor $e^{i\theta}$ using an extra $R_1$ gate (see step II in Figure~\ref{fig:CompSup} (b)). Algorithm~\ref{alg:KetXplusEIThetaKetNegX} provides a detailed generation process.

\subsubsection{Two Value Superposition States}
\label{subapp:TwoValue}

This section outlines a generation algorithm for general two-value superposition states $\frac{1}{\sqrt{2}}(\left|x\right>+e^{i\theta}\left|y\right>)$. The process begins by comparing the binary representations of $x$ and $y$ and identifying the same and different bits, naturally dividing them into two groups. For the same bits, we apply Algorithm~\ref{alg:KetX} to the corresponding qubits. For the different bits, we use Algorithm~\ref{alg:KetXplusEIThetaKetNegX}, incorporating a phase $e^{i\theta}$ on the corresponding qubits. The overall generation process is presented in Algorithm~\ref{alg:KetXplusEIThetaKetY}. To streamline the algorithm, we denote the sub-array of an original array \textit{Arr} from the index set A as \textit{Arr}[A].

\subsubsection{Pauli States}
\label{subapp:Pauli}

As Formula (\ref{equ:PauliState}) shows, multi-qubit Pauli states are represented as tensor products of several single-qubit Pauli states. There are six single-qubit Pauli states: $\left|\Phi_1\right> = \left|0\right>$;
$\left|\Phi_2\right> = \left|1\right>$;
$\left|\Phi_3\right> = \left|+\right> = \frac{1}{\sqrt{2}}(\left|0\right> + \left|1\right>)$;
$\left|\Phi_4\right> = \left|-\right> = \frac{1}{\sqrt{2}}(\left|0\right> - \left|1\right>)$;
$\left|\Phi_5\right> = \left|+_i\right> = \frac{1}{\sqrt{2}}(\left|0\right> + i\left|1\right>)$;
$\left|\Phi_6\right> = \left|-_i\right> = \frac{1}{\sqrt{2}}(\left|0\right> - i\left|1\right>)$.

To represent a single Pauli state, we use an integer $i$ ranging from 1 to 6, where each integer corresponds to a unique state $\left|\Phi_i\right>$. An $n$-qubit Pauli state can then be represented as an integer array with $n$ elements. The process to generate a single Pauli state $\left|\Phi_i\right>$ is described in Algorithm~\ref{alg:SinglePauli}. To generate a multi-qubit Pauli state from a given integer array $A$, Algorithm~\ref{alg:PauliState} can be used.

\subsubsection{Mixed State}
\label{subapp:Mixed}

If we know the ensemble representation ${(p_i, \left|\psi_i\right>)}$ of the target state $\rho$, and also the generate operation $U_i$ for each element state $\left|\psi_i\right>$, i.e., $U_i \left|0\right> = \left|\psi_i\right>$, we can use Algorithm~\ref{alg:MixedState} to execute $U_i$ with probability $p_i$ on $\left|0\right>$, which generates $\rho$ in the statistical sense. The output state of a single execution of Algorithm~\ref{alg:MixedState} is one of the states $\left|\psi_i\right>$, but the statistical distribution of multiple executions of the algorithm yields the ensemble ${(p_i, \left|\psi_i\right>)}$, which represents the mixed state $\rho$.

Another scenario is when we only have access to the density matrix of $\rho$. In such cases, we can utilize \textit{spectrum decomposition} to identify an ensemble representation~\cite{nielsen2002quantum}, which is given by $\rho = \sum_i {\lambda_i}\left|i\right>\left<i\right|$. Assuming that we can identify the generation operation $U_i$ for each state $\left|i\right>$, Algorithm~\ref{alg:MixedState} can operate on this ensemble to generate the mixed state $\rho$.

\subsection{Subtasks in Unit Testing and Their Relations}
\label{subsec:subtasks}


In classical program testing, two main subtasks are structural (white-box) testing and behavioral (black-box) testing. Sometimes, indirect methods like metamorphic testing~\cite{chen1998metamorphic} are used when constructing testing oracles is challenging. Metamorphic testing verifies the correctness of the target program indirectly through predefined relations (metamorphic relations) of the target program.

\begin{figure*}
\centering
\subfigure[]{\includegraphics[scale=0.56]{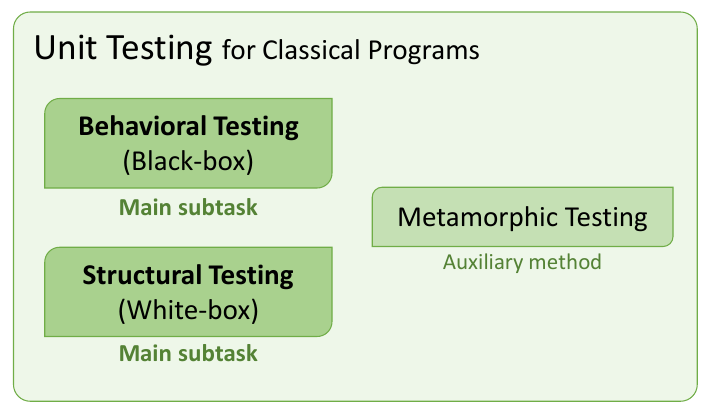}}
\subfigure[]{\includegraphics[scale=0.56]{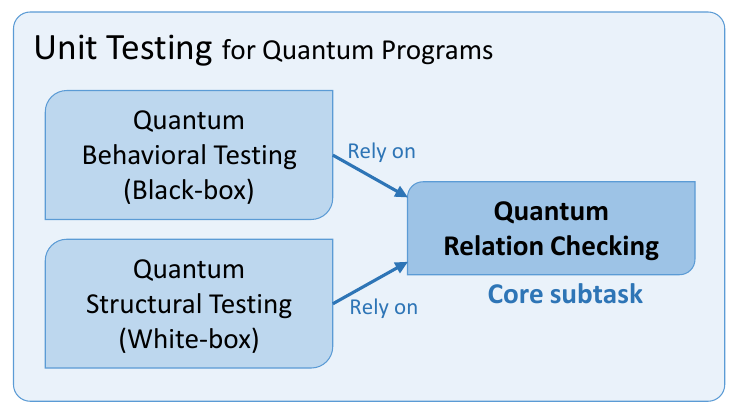}}
\caption{Unit testing in both classical and quantum programs involves three subtasks and their interrelationships.}
\label{fig:Subtasks}
\end{figure*}

In testing quantum programs, structural testing and behavioral testing are also important, along with a subtask called "quantum relation checking," which is the quantum counterpart of metamorphic testing, as indicated in Table~\ref{tab:subtasks}. Unlike metamorphic testing in classical programs, quantum relation checking plays a more significant role due to the complexities associated with generating and detecting quantum states (Section~\ref{subsec:GenerationDetection}). As depicted in Figure~\ref{fig:Subtasks} (b), in testing quantum programs, both behavioral and structural testing may depend on quantum relation checking, which serves as a core subtask (Figure~\ref{fig:Subtasks} (b)).

Indeed, there may be additional subtasks, but these three are particularly crucial in testing scenarios. Therefore, in Sections~\ref{subsec:relationcheck} to \ref{subsec:PSCheck}, we will delve into each of these key subtasks.

\begin{table*}
\centering
\caption{Three subtasks in unit testing quantum programs and their corresponding classical counterparts}
\label{tab:subtasks}

\begin{small}
\begin{tabular}{l|c|c}
\toprule
\makecell[c]{\textbf{Subtask}} & \textbf{Classical Counterpart} & \textbf{Mainly Adopted Method} \\
\midrule
\textbf{A.} Quantum relation checking & Metamorphic testing & Black-box\\
\midrule
\textbf{B.} Quantum structural testing & Structural testing & White-box \\
\midrule
\textbf{C.} Quantum behavioral testing & Behavioral testing & Black-box \\
\bottomrule
\end{tabular}
\end{small}
\end{table*}

\subsection{Subtask A: Quantum Relation Checking}
\label{subsec:relationcheck}

Given the challenges associated with generating and detecting quantum states, the indirect method of quantum relation checking plays a crucial role in testing quantum programs with multiple subroutines. As illustrated in Figure~\ref{fig:Subtasks}, the other two subtasks rely on this method. Quantum relations can be categorized into three types: (1) relations that pertain to statistical results; (2) relations that involve checking quantum states; and (3) relations that involve checking subroutines.

\subsubsection{The Relations about Statistical Results}

One simple approach to testing quantum programs involves converting the target output quantum states into specific statistical results. The most commonly used statistical result is the probability distribution of the measurements. To illustrate, let us consider the output state $\left|+\right> = \frac{1}{\sqrt 2}(\left|0\right>+\left|1\right>)$. By applying a measurement operation to this state, we can convert it into a Bernoulli distribution:

\begin{equation}
\notag
Pr[X = 0] = \frac{1}{2}, \hspace*{2mm} Pr[X = 1] = \frac{1}{2}
\end{equation}

\noindent
We can repeat the execution of the target program many times, collect the measurement results, and then check whether the output results fit this distribution (i.e., the number of outputs '0's and '1's are nearly equal). Typically, the statistical results are obtained by repeatedly running the target programs, and existing statistical methods can then be used to analyze the results. 
Huang and Martonosi~\cite{huang2019statistical} proposed statistical assertion methods for classical states, superposition states, and entanglements using hypothesis tests. This work can be integrated into the checking of relations concerning statistical measurement results.

Apart from analyzing the original probability distribution of measurements, estimating specific parameters related to the output quantum states using appropriate methods offers an alternative approach for checking the output. A common example involves comparing two quantum states $\rho$ and $\sigma$, where the parameter $tr(\rho\sigma)$ yields valuable insights. Specifically, when $\rho=\sigma$, the parameter $tr(\rho^2)$ determines whether the output state $\rho$ is a pure state. Moreover, when $\rho=\left|\alpha\right>\left<\alpha\right|$ and $\sigma=\left|\beta\right>\left<\beta\right|$ represent two pure states, $tr(\rho\sigma)=|\left<\alpha|\beta\right>|^2$, where $\left<\alpha|\beta\right>$ denotes the inner product of the states $\left|\alpha\right>$ and $\left|\beta\right>$. This calculation provides insight into the degree of overlap between the two states. Several methods, such as the SWAP test~\cite{buhrman2001quantum,barenco1997stabilization} and Pauli measurements~\cite{cai2016optimal,gross2010quantum}, have been proposed for estimating $tr(\rho\sigma)$, allowing for the extraction of valuable information from the output states.

\subsubsection{The Relations about Quantum States}
\label{subsubsec:RelationQS}

The states of quantum variables can be represented as state vectors or density matrices on paper. However, for quantum variables, there is a gap between the mathematical representation on paper and the testing process in quantum programs. Due to the challenges involved in generating and detecting quantum states (see Section~\ref{subsec:GenerationDetection}), it is often necessary to convert the testing process into verifying specific relations about quantum states. The measurement probability distribution is one such relation that has already been discussed in the preceding section. In this section, we will discuss the relation between the quantum states themselves.

To illustrate this, let us consider the previous example of the output state $\left|+\right> = \frac{1}{\sqrt 2}(\left|0\right>+\left|1\right>)$. As shown in Equation (\ref{equ:HPlusEq0}), it is noteworthy that applying the $H$ gate to this state yields $\left|0\right>$:

\begin{equation}
\label{equ:HPlusEq0}
H\left|+\right> = \left|0\right>
\end{equation}

\noindent
Consequently, when measuring $\left|0\right>$, the outcome is always 0. Hence, if a measurement result of 1 occurs, it indicates that the output state is not $\left|+\right>$.
In fact, several \textit{quantum runtime assertion} methods~\cite{li2020projection,liu2020quantum,DBLP:conf/hpca/LiuZ21} are built upon this idea of deterministic-in-one-side relations, which involves transforming the target state into a simple classical state, such as $\left|0\right>$. In this paper, we refer to such a method as  \textit{transform-based checking} for easy description. Actually, transform-based checking related to $\left|0\right>$ is equivalent to measuring the target state on another basis.

Generally, transform-based checking involves finding relations between the target quantum state and states that are easy to check. Here, we discuss the relations between pure states. Suppose that the expected output state is a pure state $\left|\psi_{eo}\right>$, which can be obtained by applying a simple unitary operation on an all-zero state, i.e., there exists a unitary operation $U_{eo}$ such that $\left|\psi_{eo}\right> = U_{eo}\left|0\ldots0\right>$. If the practical output is $\left|\psi_{eo}\right>$, applying $U_{eo}^{-1}$ on it will result in an all-zero state; otherwise will not. The measurement result for the all-zero state is always 0, and the state will not be changed. From a testing perspective, $U_{eo}^{-1}$ is the inverse variant of $U_{eo}$ and can be easily implemented as long as $U_{eo}$ can be easily implemented. Compared to converting the output into a probability distribution, the most significant advantage of this method is that since it uses the deterministic-in-one-side property, it needs only a few runs to obtain correct detection results with high probability. However, the disadvantage is that it is feasible only when a simple $U_{eo}^{-1}$ of the expected state exists. If the implementation of $U_{eo}^{-1}$ is complex, this method may also be impractical. Example~\ref{example:qftoutput} shows how to check the output of the QFT program under classical input states.

\begin{example}
\label{example:qftoutput}
Output checking for the QFT program under classical input states.

As shown in formula (\ref{equ:qftqubitform}), the program specification of QFT (with BIBO endian mode) is as follows:

\begin{equation*}
	\begin{array}{c}
		\left|j_1\right> \rightarrow \frac{1}{\sqrt 2}(\left|0\right>+e^{2\pi i0.j_n}\left|1\right>) \\
		\cdots \\
		\left|j_n\right> \rightarrow \frac{1}{\sqrt 2} (\left|0\right>+e^{2\pi i0.j_1j_2 \ldots j_n}\left|1\right>)
	\end{array}
\end{equation*}

Given classical state input $\left|j\right> = \left|j_1\right> \dots \left|j_n\right>$, the corresponding output state is the product state of $n$ single qubit states, each of which is represented in the form $\frac{1}{\sqrt 2}(\left|0\right>+e^{i\theta_{j,k}}\left|1\right>)$, where $\theta_{j,k} = 2\pi 0.j_{n-k+1} \ldots j_{n-1}j_n$ and can be generated by applying gate $U_{j,k}$ on $\left|0\right>$, where

\begin{equation}
\label{equ:genqftqubit}
	U_{j,k}= R_1(\theta_{j,k})H = \frac{1}{\sqrt 2} \left[
	\begin{array}{cc}
		1 & 1 \\
		e^{i\theta_{j,k}} & -e^{i\theta_{j,k}}
	\end{array}\right]
\end{equation}

\noindent
To check the output, we use the transform-based method for applying $U_{j,k}^{-1}$ on the $k$-th qubit, and the overall operation is $U_{eo,j}^{-1} = U_{j,1}^{-1}\otimes \cdots \otimes U_{j,n}^{-1}$. If the output is correct, all qubits after applying $U_{j,k}^{-1}$ will become $\left|0\right>$, then the results of measured qubits will always be integer 0. Otherwise, non-zero results will be possible.
\end{example}

\subsubsection{The Relations about Subroutines}
\label{subsubsec:SubroutineRelation}

Relations among subroutines are crucial in testing quantum programs, as they provide valuable insights into their correctness. When a target relation is satisfied, it indicates that the subroutines are likely correct. Conversely, if the relation is not satisfied, it indicates the presence of errors in at least one of the subroutines.

One commonly used relation is \textit{equivalence} between two subroutines, which means that both subroutines produce the same output given the same input. Let us denote the two subroutines as $\texttt{P}_1$ and $\texttt{P}_2$, the equivalence relation can be expressed as $\texttt{P}_1 = \texttt{P}_2$. A special case of equivalence is when a subroutine \texttt{P} is equal to an \textit{identity} transform, denoted as $\texttt{P} = I$. In this case, the program specification for $I$ takes the following form:

\begin{equation} 
\label{equ:Ips} I: \left|x\right> \rightarrow \left|x\right> \end{equation}

\noindent
Here, $I$ preserves the input state $\left|x\right>$ unchanged. We consider identity especially because identity is one of the most common relations and it may have a faster algorithm than equivalence checking. Importantly, as we will discuss next, the notions of \textit{equivalence} and \textit{identity} serve as the foundation for many other useful relations in quantum program testing.

Quantum programs may consist of multiple subroutines, and in addition to the output states of these subroutines, there also exist relations among them that can be used for testing quantum programs. For example, consider the execution of several subroutines, denoted as $\texttt{P}_1$, $\texttt{P}_2$, \dots, $\texttt{P}_n$. We represent their sequential execution as the composed program $\texttt{P}_n \circ \dots \circ \texttt{P}_2 \circ \texttt{P}_1$, where $\circ$ represents the sequential execution from right to left. It executes $\texttt{P}_1$ on input $x$ and produces output $o_1$, which becomes the input for $\texttt{P}_2$, and so on until the final output is produced by $\texttt{P}_n$.

While some programming languages provide mechanisms for automatically generating and managing the three variants of an original program, these mechanisms may not always be available. Consequently, there are cases where it becomes necessary to manually implement these variants and test them. Fortunately, by leveraging the relation between the variants and the original subroutine, it is possible to avoid the need for redesigning test cases. This approach allows for a unified testing process that remains independent of the specific subroutine being tested. Suppose we have finished testing the original subroutine \texttt{P}, we denote inverse, controlled, power variants of \texttt{P} as \texttt{InvP}, \texttt{CtrlP}, and \texttt{PowP}, respectively.

For \texttt{InvP}, there is an obvious relation:

\begin{equation}
\label{equ:PInvPEqI}
\texttt{P}\circ\texttt{InvP}=I
\end{equation}

\noindent where $I$ is the identity operation. Note that no matter what \texttt{P} is, the relation (\ref{equ:PInvPEqI}) always holds, so we obtain a unified test process for any \texttt{InvP}, that is, to execute the identity checking for the sequential execution of \texttt{InvP} and \texttt{P}.

The input of \texttt{PowP} contains two parts: the power $k$ (a classical integer) and target qubits \textbf{qs}. Given $k$, the effect on \textbf{qs} is equivalent to apply \texttt{P} (if $k>0$) or \texttt{InvP} (if $k<0$) for $|k|$ times. Also, the equivalence check can be converted into an identity check with the following identity relations:

\begin{equation}
\label{equ:PowP}
	\begin{array}{cc}
		\texttt{InvP}^k \circ \texttt{PowP}(k) = I, & k>0\\

		\texttt{P}^{|k|} \circ \texttt{PowP}(k) = I, & k<0
	\end{array}
\end{equation}

\noindent
Similarly, the input of \texttt{CtrlP} contains two parts: control qubits \textbf{qctrl} and target qubits \textbf{qtar}. If \textbf{qctrl} is in an \textit{all-one state} $\left|1\ldots1\right>$, then the effect on \textbf{qtar} is \texttt{P}. If \textbf{qctrl} is in a state which is orthogonal to $\left|1\ldots1\right>$, then the effect is identity $I$. We can see that the testing of variants subroutines can be converted into the checking of \textit{identity} and \textit{equivalence}, which can be finished in unified processes.

In fact, equivalence checking and identity checking have classical counterparts, which are common relations among classical programs. However, \textit{unitarity checking} is quantum-specific and has no classical counterpart. Given a quantum subroutine \texttt{P} with IO type of transform. Unitarity checking is to check whether \texttt{P} represents a unitary transform. As we know, many typical quantum algorithms are unitary transforms, and measurement is the unique way to destroy the unitarity. Therefore, incorporating unitarity checking enables testers to identify unexpected measurement statements in target programs. 
A recent study discusses the concrete implementation of equivalence, identity, and unitarity checking in quantum software testing tasks~\cite{long2023equivalence}. This work can be integrated into the proposed testing process in this paper.

\subsection{Subtask B: Quantum Structural Testing}
\label{subsec:StructureCheck}

As discussed in Sections~\ref{subsubsec:ControlFlow} and~\ref{subsubsec:WithinApply}, quantum programs may have specific structures that require careful checking to ensure correctness. In classical program testing, we commonly use branch coverage analysis to ensure the coverage of all possible control flow paths using selected input variables. However, in testing non-linear quantum programs, the execution path cannot be determined before running due to the uncertainty of quantum programs. Even the same input can lead to different execution paths in different running rounds.
While classical random programs also exhibit inherent uncertainty in execution paths, this uncertainty can be mitigated and managed by substituting the internal random number generator with additional input parameters. Conversely, quantum programs face \textit{intrinsic uncertainty}, posing a unique challenge in the realm of quantum program testing.

In practice, the challenge of structure checking for general quantum programs remains an open problem, and further research on control flows and branch coverage in quantum programs is crucial to advancing the field. Regarding structural testing for classical programs, concepts like execution path and coverage hold significance. To address structural checking for quantum programs, exploring how to define execution path and coverage based on the properties of quantum programs will be valuable in future investigations.

In this section, our focus lies on the overall testing process rather than specific testing techniques. Thus, we briefly delve into structure checking for two typical quantum program structures: repeat-until-success (RUS, see Section~\ref{subsubsec:ControlFlow}) and $U^{-1}VU$ (see Section~\ref{subsubsec:WithinApply}), offering insights into this crucial task.

To check the structure of an RUS program, the core step is to verify the loop's termination condition. This condition is usually a predicate involving several classical variables, some of which may relate to the measurement results. Since we are only concerned with the program structure in this task, we can adopt a \textit{classical-quantum separation} approach. Note that measurement is the only way for the quantum part to influence the classical part, and the condition relies only on the classical variables\footnote{Although measurement statements are allowed in the condition, their values are essentially classical.}. Therefore, we can replace each measurement statement with a controllable classical parameter and remove all quantum parts to construct a testable substitute program that is purely classical and retains the structure of the original program. We can then perform the structure checking on this substitute program. An example of classical-quantum separation is shown in Example~\ref{example:CQsep}. We refer to the substitute program as a \texttt{test double}. It accomplishes the testing objective and is more straightforward to test compared to the original program.

If the programming language supports syntax for $U^{-1}VU$ structure (such as Q\#), the correctness of the $U^{-1}VU$ structure is ensured by the language itself, and there is no need to check it separately. However, if the language does not support, programmers need to implement three parts $U$, $V$, and $U^{-1}$, respectively. In this case, we need to ensure that the manual uncomputation is correct, i.e., $U^{-1}U = I$. This task can be transformed into an identity-checking problem mentioned in Section~\ref{subsubsec:SubroutineRelation}.

\begin{example}
\label{example:CQsep}
The classical-quantum separation for Listing~\ref{list:random}.

In Listing~\ref{list:random}, we see a program with a RUS structure. The measurement statement is on Line 8, and \texttt{m} is the corresponding classical variable. Furthermore, \texttt{m} is also part of the termination condition (Line 9). To maintain classical-quantum separation, an additional parameter, \texttt{set\_inner\_m}, is introduced. We replace the measurement statement with this parameter and remove all quantum components to create the substitute. The code for the substitute is provided in Listing~\ref{list:CQsep}.
\end{example}

\begin{figure}[h]
\begin{lstlisting}[
	xleftmargin=4mm,
	%xrightmargin=2mm,
	caption={The classical substitute program of Listing~\ref{list:random}.},
	label={list:CQsep}
	]
operation Random_0to2_Classical_Substitute(set_inner_m : Int) : Int {
    //use qs = Qubit[2];
    mutable m = 0;
    repeat {
        //Remove quantum parts
        set m = set_inner_m;    //Replace measurement statement with paremeter
    } until (m < 3);    //Success condition
    //ResetAll(qs);
    return m;
}
\end{lstlisting}
\end{figure}

\subsection{Subtask C: Quantum Behavioral Testing}
\label{subsec:PSCheck}

The task of program behavioral checking is to verify whether the behavior of the target program is consistent with its specification, which is typically achieved by running the target program with given inputs and observing whether the output matches the expected values. This subtask usually employs black-box testing, and appropriate inputs must be chosen. To accomplish this subtask, the IO mark of the target subroutine (see Section~\ref{subsec:IOanalysis}) must be obtained first, and its IO type (see Figure~\ref{fig:4typesdataflow}) will influence the design of the tests. Like testing classical programs, equivalence class partitioning and the combination of variables can be adopted to complete this subtask.

\subsubsection{Equivalence Class Partition}
\label{subsubsec:EqClassPart}

Section~\ref{subsubsec:PartByType} introduces two coarse-grained input partitioning principles, CSP and CSMP. However, for a specific target program, it is necessary to consider its logical meaning to construct a more effective partition. For instance, we can treat classical input states as classical integers and apply classical techniques like dividing by a boundary value to the partitioning process. Example~\ref{example:phaseflip} demonstrates an input partitioning and test case selection strategy for a phase-flip subroutine. Instead of simply partitioning the input space into classical and superposition states, we use the program specification to create a more elaborate partitioning scheme. The final equivalence classes are determined based on the program specification, and we also consider the detectability of the output as stated in (3) of Principle~\ref{crit:Principle}.

\begin{example}\label{example:phaseflip}
\noindent 
Input selection for the phase-flip subroutine.

Consider the conditional phase-flip subroutine \texttt{PhaseFlip} in the Grover Search program. It receives an input state \textbf{qs} = $\left|x\right>$ with $n$ qubits and outputs according to the value of $x$:

\begin{center}
\texttt{PhaseFlip} : ($n$, \textbf{qs}) $\rightarrow$ (\textbf{qs'})
\end{center}

\begin{equation} \label{GScondphaseflip}
	\texttt{PhaseFlip}: \left|x\right> \rightarrow \left\{
	\begin{array} {cc}
		-\left|x\right> & x>0\\
		\left|x\right> & x=0
	\end{array}
	\right.
\end{equation}

\vspace*{2mm}
\noindent
\texttt{PhaseFlip} keeps unchanged on all-zero input and flips phase on other classical state inputs. Note that the global phase cannot be distinguished by measurement, so classical state inputs will only get trivial results. To test whether it successfully flips the phase when $x=0$, we need to use the relative phase of the component in the superposition state. We choose input states in the form $\frac{1}{\sqrt{2}}(\left|x\right>+\left|y\right>)$, where $x$ and $y$ are chosen according to two branches in the formula (\ref{GScondphaseflip}). A feasible test strategy is to select the following four types of inputs:

\begin{itemize} [itemsep=0.32em]
    \item[(1)] $\left|0\right>$;\footnote{Here $\left|0\right>$ represents a classical state with integer 0, i.e., all-zero state, rather than a single qubit.}
    \item[(2)] $\left|x_1\right>, x_1>0$;
    \item[(3)] $\frac{1}{\sqrt{2}}(\left|0\right>+\left|x_1\right>), x_1>0$;
    \item[(4)] $\frac{1}{\sqrt{2}}(\left|x_1\right>+\left|x_2\right>), x_1,x_2>0, x_1 \neq x_2$.
\end{itemize}

\noindent
Types (1) and (2) are classical state inputs corresponding to the two branches of the formula (\ref{GScondphaseflip}); types (3) and (4) are superposition state inputs, where type (3) is the superposition of the first branch and the second branch, and type (4) is the superposition state of two states in the second branch. For input types (1), (2), and (4), the difference between input and output should not be detected (same or at most global phase differences), and it ensures that the transformation is identical except for some global phases. The important case is input type (3), in which the output state will be changed into $\frac{1}{\sqrt{2}}(\left|0\right>-\left|x_1\right>)$. The difference between the output state and the input state can be detected.
\end{example}

\subsubsection{Combination of Variables}
Since many quantum programs have multiple input variables, it is also necessary to consider the combination of these variables during the testing process. Various combination coverage criteria between variables have been proposed for testing classical programs, such as All Combination Coverage (ACoC)~\cite{cohen1997aetg}, Each Choice Coverage (ECC)~\cite{ammann1994using}, Pair-wise Coverage (PWC)~\cite{burroughs1994improved}, and Base Choice Coverage (BCC)~\cite{ammann1994using,cohen1994automatic}. Obviously, these coverage criteria can be migrated into quantum program testing. The concrete choice of coverage criteria depends on the balance of accuracy and efficiency. For example, ACoC requires all possible combinations of all input variants, and thus, it is very accurate, but the cost is also high. ECC only requires the value of each equivalence class of each input variable exists in the input, and thus, it is fast but does not include all situations. Example~\ref{example:qftpart} demonstrates how to combine classical and quantum input variables and select a feasible set of test inputs based on appropriate criteria. By applying these coverage criteria in the testing process, we can ensure that the combinations of input variables are thoroughly tested, improving the quality and reliability of the quantum program.

\begin{example}
\label{example:qftpart}
Equivalence class partitioning and combination for \texttt{QFT} program.

The IO mark of the \texttt{QFT} program is:

\begin{center}
    \texttt{QFT} : ($n$, \textbf{qs}) $\rightarrow$ (\textbf{qs'})
\end{center}

\noindent
As is shown in Listing~\ref{list:qft}, there are two loops in the \texttt{QFT} program, where the maximum index of the outer loop (line 21) is $n-2$, and the maximum index of the inner loop (line 23) is $n-1$. It is easy to see that 1 and 2 are two boundary values of $n$, so partition $n$ into three parts: $n=1$, $n=2$, and $n \geq 3$. We use the CSP criterion to partition \textbf{qs} into a classical state C and a superposition state S. To combine $n$ and \textbf{qs}, we use ACoC on these two parameters. This is because the number of equivalence classes is small and thus, the cost of ACoC is acceptable. Finally, there are six equivalence classes:

\begin{center}
	1. ($n=1$, C)\quad 2. ($n=1$, S)\quad 3. ($n=2$, C)\quad 4. ($n=2$, S)\quad 5. ($n \geq 3$, C)\quad 6. ($n \geq 3$, S)
\end{center}

We select a concrete test input for each equivalence class. For the superposition state of \textbf{qs}, we use the STV criterion to select two-value-superposition states and use the SCAQ criterion to make the superposition cover all qubits. So we select complementary superposition states for S. A feasible set of tested input variables is:

\begin{itemize}
\item[] (1) $n=1$, $\mathrm{\mathbf{qs}}=\left|0\right>$
\item[] (2) $n=1$, $\mathrm{\mathbf{qs}}=\frac{1}{\sqrt{2}}(\left|0\right>+i\left|1\right>)$
\item[] (3) $n=2$, $\mathrm{\mathbf{qs}}=\left|01\right>$
\item[] (4) $n=2$, $\mathrm{\mathbf{qs}}=\frac{1}{\sqrt{2}}(\left|00\right>+\left|11\right>)$
\item[] (5) $n=7$, $\mathrm{\mathbf{qs}}=\left|1011001\right>$)
\item[] (6) $n=6$, $\mathrm{\mathbf{qs}}=\frac{1}{\sqrt{2}}(\left|101101\right>+\left|010010\right>)$
\end{itemize}
\end{example}

\subsection{Testing Cases and Execution}
\label{subsec:TestCases}

In classical testing, a test case is typically defined as a pair of inputs and expected outputs~\cite{ammann2016introduction}, as shown in the following form:

\vspace{3mm}
\qquad \texttt{Input: \textit{in\_var\_1}, \dots, \textit{in\_var\_n}}

\qquad \texttt{Expected output: \textit{out\_var\_1}, \dots, \textit{out\_var\_m}}
\vspace{3mm}

\noindent 
Here, \texttt{\textit{in\_var\_1}}, \dots, \texttt{\textit{in\_var\_n}} are the given values assigned to all input variables, and \texttt{\textit{out\_var\_1}}, \dots, \texttt{\textit{out\_var\_m}} are the expected values for all output variables. However, in the case of quantum variables, simply specifying input and output values may not be sufficient. As discussed in Section~\ref{subsec:GenerationDetection}, preparing or checking a quantum state described by a formula is not straightforward. Therefore, to ensure the effective testing of quantum programs, it is crucial to attach concrete methods for generating and checking target quantum states to the given quantum variables in the test cases. To this end, we propose the following principle:

\begin{principle}
\label{crit:IOmethod}
In each test case, attach the generation method for each quantum input state and the checking method for each quantum output state.
\end{principle}


In practice, the specific form of a test case for a quantum program may vary depending on the complexity and nature of the program. However, by adhering to the proposed principle, developers and testers can ensure that each test case is executable, allowing for effective testing and validation of the quantum programs. The generation method should specify the steps for preparing the input state, such as initializing qubits to a particular state or applying quantum gates to a set of qubits. The checking method should specify how to measure and verify the output state, such as performing quantum state tomography or measuring the expectation value of certain observables. Specifically, a test case for a quantum program has the following form:

\vspace*{4mm}
\texttt{Input:}

\qquad \texttt{\textit{cls\_in\_1}, \dots, \textit{cls\_in\_$n_{ci}$},}

\qquad \texttt{[\textit{qs\_in\_1}, GenProc\_qs\_in\_1]},

\qquad \texttt{$\cdots$}

\qquad \texttt{[\textit{qs\_in\_$n_{qi}$}, GenProc\_qs\_in\_$n_{qi}$]}\\

\texttt{Expected Output:}

\qquad \texttt{\textit{cls\_out\_1}, \dots, \textit{cls\_out\_$n_{co}$},}

\qquad \texttt{[\textit{qs\_out\_1}, ChkProc\_qs\_out\_1]},

\qquad \texttt{$\cdots$}

\qquad \texttt{[\textit{qs\_out\_$n_{qo}$}, ChkProc\_qs\_out\_$n_{qo}$]}
\vspace*{4mm}

\noindent
Here, the values assigned to all input classical variables are denoted by \texttt{\textit{cls\_in\_1}}, \dots, \texttt{\textit{cls\_in\_$n_{ci}$}}, while the expected values for all output classical variables are denoted by \texttt{\textit{cls\_out\_1}}, \dots, \texttt{\textit{cls\_out\_$n_{co}$}}. For each quantum input item, such as \texttt{[\textit{qs\_in\_1}, GenProc\_qs\_in\_1]}, it contains the input state \texttt{\textit{qs\_in\_1}} itself and the procedure \texttt{GenProc\_qs\_in\_1} indicating how to generate the target state. Similarly, for each quantum output item, \texttt{ChkProc\_qs\_out\_1} is the checking procedure for the target expected state \texttt{\textit{qs\_out\_1}}.

The following Example~\ref{example:GenQFTcases} illustrates how to construct the set of test cases for the QFT program using the equivalence classes provided in Example~\ref{example:qftpart}.

\begin{example}
\label{example:GenQFTcases}
Test cases for the QFT program.

At the end of Example~\ref{example:qftpart}, we give a feasible set of tested input variables. In the following, we describe how to expand this set to a set of executable test cases. Table~\ref{table:qftio} shows the \textit{input} and the \textit{expected output} pairs. We denote state $\left|\psi_{j}\right>$ as

\begin{equation}
\label{equ:qftstate}
\left|\psi_{j}\right> = \frac{1}{\sqrt{2^{n_j}}}\sum_{k=0}^{2^{n_j}-1}e^{2\pi i j k / 2^{n_j}}\left|k\right>
\end{equation}

\noindent
for any integer (binary string) $j$ with length $n_j$. 

The checking for the output under the classical input has been discussed in Example~\ref{example:qftoutput}. What remains is how to check the output with the state of the form $\frac{1}{\sqrt 2}(\left|\psi_{a}\right>+\left|\psi_{b}\right>)$ (the last three items in Table~\ref{table:qftio}), where $a$ and $b$ are two binary strings. There is no intuitive way to generate this state, but it has a useful property. For any $j$, suppose $U_{eo,j}\left|0\ldots0\right> = \left|\psi_j\right>$. Then applying $U_{eo,a}^{-1}$ on the target state will get the following formula:

\begin{equation}
U_{eo,a}^{-1}[\frac{1}{\sqrt 2}(\left|\psi_{a}\right>+\left|\psi_{b}\right>)] = \frac{1}{\sqrt 2}(\left|0\ldots0\right>+U_{eo,a}^{-1}\left|\psi_{b}\right>)
\end{equation}

\noindent where $\left|0 \dots 0\right> \bot U_{eo,a}^{-1}\left|\psi_b\right>$ (since $\left|a\right> \bot \left|b\right>$ and unitary operations keep the orthogonality). Therefore, the probability of measuring this state to obtain an integer 0 is 1/2. This is similar to applying $U_{eo,b}^{-1}$. In order to check the output, we can repeat the following steps and count the number of results as 0 for the output.

\begin{itemize}
\item [] 1. Generate the input state $\frac{1}{\sqrt 2}(\left|a\right>+\left|b\right>)$;
\item [] 2. Run the target QFT program on the input state;
\item [] 3. Apply $U_{eo,a}^{-1}$ or $U_{eo,b}^{-1}$;
\item [] 4. Measure the output state.
\end{itemize}

If the number of results 0 occurs nearly half for both $U_{eo, a}^{-1}$ and $U_{eo,b}^{-1}$, we may be able to assume that the output state is as expected. In fact, due to the difficulty of checking a target output state, this is a compromise approach based on some properties of the target program.
\end{example}

\begin{table}
\centering
\caption{A set of feasible test cases for the QFT program.}
\label{table:qftio}

\begin{scriptsize}
\begin{threeparttable}
	\centering
	
 	\begin{tabular}{c||l|l||l|l}
		\toprule
		\textbf{\makecell{Equivalence \\ Class}} & \textbf{Input} & \textbf{\makecell[l]{ Input\\ Generation\\ Procedure }} & \makecell[l]{\textbf{Expected Output}}& \textbf{\makecell[l]{Output \\ Checking \\ Procedure}} \\
		\midrule
		($n=1$, C) & $\left|0\right>$ & \multirow{3}{*}{Algorithm~\ref{alg:KetX}} & $\left|\psi_{0}\right>$\tnote{1} & \multirow{3}{*}{See Example~\ref{example:qftoutput}} \\
		($n=2$, C) & $\left|01\right>$ && $\left|\psi_{01}\right>$& \\
		($n\geq 3$, C) & $\left|1011001\right>$ && $\left|\psi_{1011001}\right>$& \\
		\midrule
		($n=1$, S) & $\frac{1}{\sqrt 2}(\left|0\right>+i\left|1\right>)$ & Apply $SH$ on $\left|0\right>$\tnote{1} & $\frac{1+i}{2}\left|0\right> + \frac{1-i}{2}\left|1\right>$ & Apply $HS$ on output state\tnote{1}. \\
		\cmidrule{2-5}
		($n=2$, S) & $\frac{1}{\sqrt 2}(\left|00\right>+\left|11\right>)$ & \multirow{2}{*}{Algorithm~\ref{alg:KetXplusEIThetaKetNegX}} & $\frac{1}{\sqrt 2}(\left|\psi_{00}\right>+\left|\psi_{11}\right>)$ & \multirow{2}{*}{See Example~\ref{example:GenQFTcases}} \\
		($n\geq 3$, S) & $\frac{1}{\sqrt 2}(\left|101101\right>+\left|010010\right>)$ && $\frac{1}{\sqrt 2}(\left|\psi_{101101}\right>+\left|\psi_{010010}\right>)$& \\
		\bottomrule
	\end{tabular}
	
	\begin{tablenotes}
	    \item[1] Note that the product of two gates in quantum computing is right-associative. $SH$ denotes applying the gate $H$ first, followed by the gate $S$, while $HS$ denotes applying the gate $S$ first, followed by the gate $H$.
	\end{tablenotes}
\end{threeparttable}
\end{scriptsize}
\end{table}

In fact, checking the correctness of a quantum program solely based on output states under classical state input is not sufficient, but the behavior of the program with general input states can be complex and difficult to identify. Fortunately, if the target program is a unitary transform, its unitarity guarantees that the correctness of the classical state input implies the correctness of the general input. Therefore, a feasible approach to test a unitary program, denoted as \texttt{P}, is as follows:

\begin{itemize} 
\item[] 1. Check the output states of \texttt{P} under classical input states; 
\item[] 2. Conduct additional unitarity checking (see Section~\ref{subsubsec:SubroutineRelation}) for \texttt{P}. 
\end{itemize}

\noindent
By following these steps, the testing process incorporates the evaluation of superposition states within a unified unitarity checking framework, thereby reducing the burden on testers.

\section{Integration Testing and Testing Practices}
\label{sec:Integration}

Similar to classical programs, the correctness of individual subroutines in quantum programs does not necessarily guarantee the correctness of their composition. Bugs can manifest in the interface between two subroutines. In this section, we delve into integration testing and testing practices for quantum programs.

\subsection{Overview of Integration Testing}
\label{subsec:Integration}


Integration testing involves integrating multiple subroutines within a quantum program and identifying bugs that may occur at the interfaces of coupled subroutines. In practice, most issues encountered during the integration testing of classical programs can also arise in the integration testing of quantum programs. For instance, the \textit{integration order} is a critical concern due to the dependency relations among subroutines, and some algorithms like \textit{topological sorting}~\cite{Knuth97algorithms} have been proposed to address this. Example~\ref{example:QPEintgeration} in the following discusses the integration order of the QPE program.

\begin{figure}
	\centering
	\includegraphics[scale=0.6]{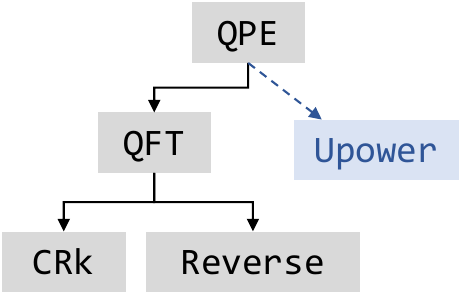}
	\caption{The subroutine dependencies of QFT and QPE programs in Listings~\ref{list:qft} and~\ref{list:qpe}. The solid-line arrows represent direct calling, and the dashed-line arrow represents calling as a parameter.}
	\label{fig:QPEdepend}
\end{figure}

\begin{example}
\label{example:QPEintgeration}
Determine the integration order of \texttt{QPE} programs.

Figure~\ref{fig:QPEdepend} illustrates the subroutine dependencies of the \texttt{QFT} and \texttt{QPE} programs, with direct callings represented by solid line arrows. Since \texttt{Upower} is an input parameter of \texttt{QPE}, it is not considered a direct calling, and we represent it as a dotted line arrow. Since there is no circular dependency and the subroutines are fixed except for \texttt{Upower}, we can employ topological sorting to determine feasible integration orders for the subroutines, excluding \texttt{Upower}. One example of a feasible integration order is

\begin{center}
(1) \texttt{CRk} $\rightarrow$ (2) \texttt{Reverse} $\rightarrow$ (3) \texttt{QFT} $\rightarrow$ (4) \texttt{QPE}
\end{center}

\noindent
In addition, \texttt{Upower} is an abstract subroutine, which is used as an input parameter. The integration strategies for it will be discussed in Section~\ref{subsec:IntQUQL}.
\end{example}

\begin{figure*}
	\centering
	\subfigure[Type CUCL]{\includegraphics[scale=0.4]{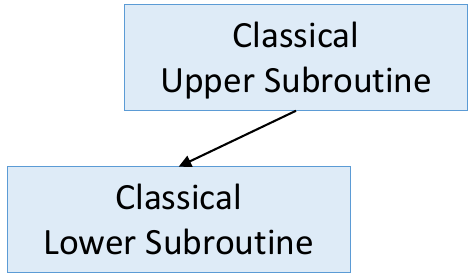}}
	\subfigure[Type QUQL]{\includegraphics[scale=0.4]{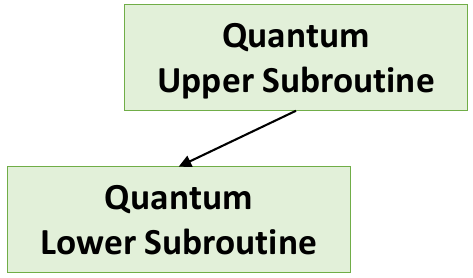}}
	\subfigure[Type QUCL]{\includegraphics[scale=0.4]{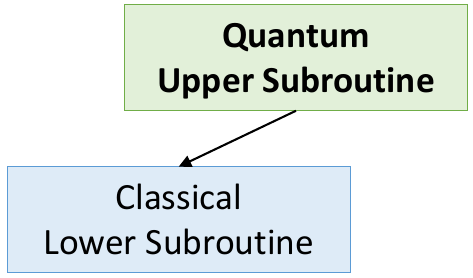}}
	\subfigure[Type CUQL]{\includegraphics[scale=0.4]{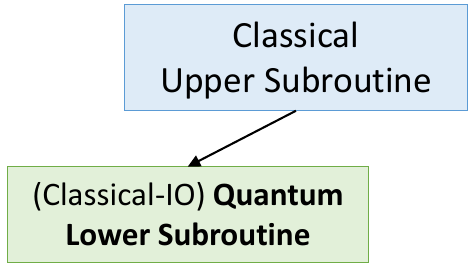}}
	\caption{Four types of coupling between the upper and lower subroutines.}
	\label{fig:4typescouple}
\end{figure*}

However, due to the unique properties of quantum programs outlined in Section~\ref{sec:properties}, there exist distinctive problems and challenges in the integration testing of quantum programs. Considering that the integration of two subroutines can be viewed as a coupling pair of subroutines, we first examine the types of coupling pairs. Similar to the classification of quantum subroutines based on their IO (Section~\ref{subsec:IO}), it is also valuable to categorize the coupling based on the types of the upper and lower subroutines. As shown in Figure~\ref{fig:4typescouple}, depending on whether the upper subroutine (caller) is classical (CU) or quantum (QU) and whether the lower subroutine (callee) is classical (CL) or quantum (QL), we can obtain four possible types of coupling: CUCL, QUQL, QUCL, and CUQL types.

Among these types, since type CUCL is essentially the integration of classical subroutines and numerous testing strategies and techniques have been devised for it, we will not consider it in this paper. In the subsequent part of this section, our primary focus will be on discussing the three remaining coupling types related to quantum. It is important to highlight the CUQL type: a quantum lower subroutine can be invoked by a classical upper subroutine only when the quantum subroutine has an IO type of "classical."

\subsection{Integration of Quantum-Quantum Coupling Pairs}
\label{subsec:IntQUQL}

Let us start by discussing the coupling of two quantum programs. As we discussed in Section~\ref{subsec:Subroutines}, the lower subroutine can be directly called or called as a parameter by the upper subroutine. When employing direct calling, a bottom-up approach can be adopted, as demonstrated in Example~\ref{example:QPEintgeration}.

In the remainder of this section, we delve into the scenario where the lower subroutine is called a parameter. In this type of coupling pair, the upper subroutine operates independently of the lower one. We can adopt a bottom-up order if the input subroutine parameters are specified in the entire program. Additionally, due to the independence of the two subroutines, a top-down order is also feasible. In this approach, when testing the upper subroutine first, we utilize additional subroutines as input parameters. These additional input subroutines, referred to as \textit{test doubles}, act as substitutes for the lower subroutine. A test double essentially replaces the original subroutine while preserving its critical properties for testing, making it easier to test. Example~\ref{example:oracle} demonstrates the adoption of a top-down order for integrating the Grover program, and Example~\ref{example:IntQPE} illustrates the use of test doubles to implement the concrete target input operation for integration testing in a quantum phase estimation program.

\begin{figure}
\centering
\includegraphics[scale=0.52]{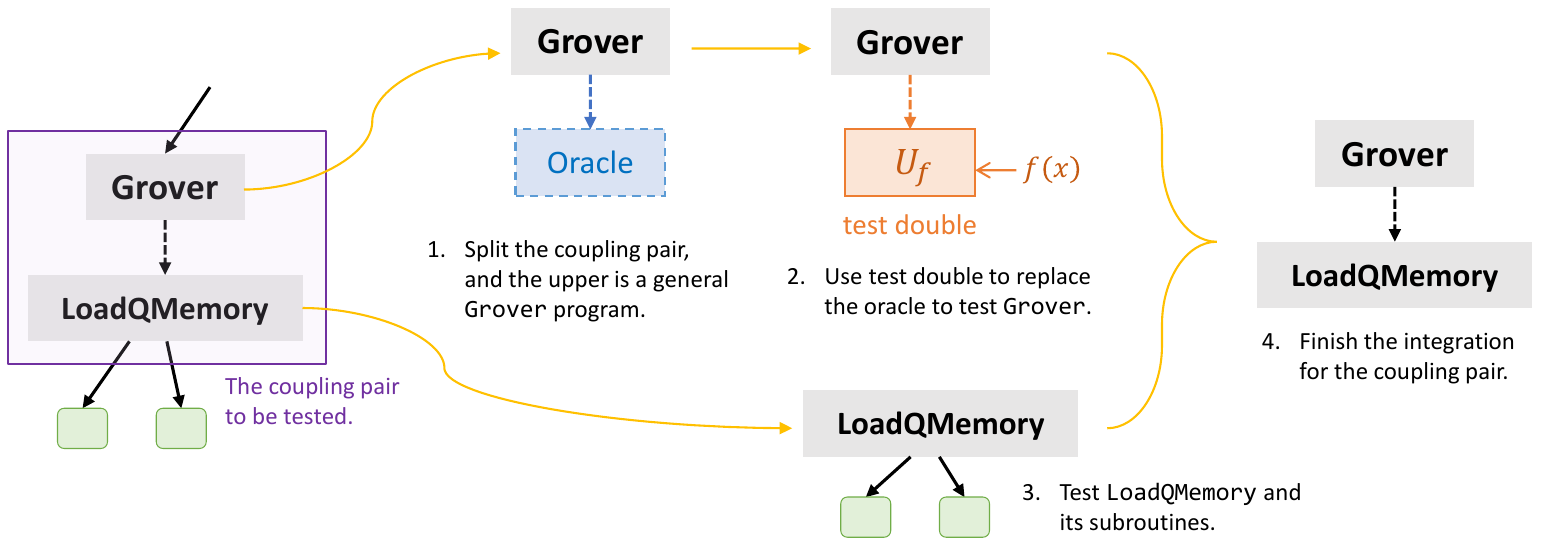}
\caption{A top-down integration strategy for the coupling pair of \texttt{Grover} and \texttt{LoadQMemory}.}
\label{fig:IntGrover}
\end{figure}

\begin{example}
\label{example:oracle}
Top-down integration order for the Grover program.

Consider a Grover program designed to search a quantum memory. As shown in Figure~\ref{fig:IntGrover}, the \texttt{Grover} program uses \texttt{LoadQMemory} as its oracle implementation. This \texttt{LoadQMemory} itself has subroutines, and some upper modules may also call the whole program. Figure~\ref{fig:IntGrover} illustrates the top-down integration of this coupling pair, showcasing the connection between \texttt{Grover} and \texttt{LoadQMemory}. Initially, we divide this pair into two parts: the upper being a general Grover program with an abstract oracle. To test \texttt{Grover}, we use 'test doubles' to substitute the abstract oracle. These 'test doubles' can be realized by a series of unitary operations denoted as $U_f$:

\begin{equation}
\notag
U_{f}: \left|x\right> \rightarrow (-1)^{f(x)}\left|x\right>
\end{equation}

\noindent where the indicator function $f$ is chosen and controlled by testers. Through selecting various indicator functions $f$ as input, the upper \texttt{Grover} can be tested without involving the lower \texttt{LoadQMemory}. Once \texttt{LoadQMemory} is successfully tested, we can integrate it into \texttt{Grover}, completing the integration of this coupling pair.
\end{example}

\begin{example}
\label{example:IntQPE}
Test doubles in the integration testing of the QPE program

We continue discussing the integration of the \texttt{QPE} program. In Example~\ref{example:QPEintgeration}, we delved into the program structure and subroutine dependencies of the \texttt{QPE} program. It is based on \texttt{QFT} and relies on an essential input subroutine called \texttt{Upower}. Our approach involves a bottom-up order, initiating with the testing of the \texttt{QFT} subroutine. The testing strategy for \texttt{QFT} has been thoroughly discussed in Example~\ref{example:GenQFTcases}.

Next, let us address the integration of \texttt{Upower}. As outlined in the IO mark (refer to Example~\ref{example:QPEIO}), \texttt{QPE} encompasses four crucial input variables: Nclock, Ntarget, \underline{\texttt{\textbf{Upower}}}, and \textbf{target}. In this task, \texttt{Upower} functions as an abstract subroutine, implying that there is no concrete implementation of \texttt{Upower} within the target program. To effectively test the \texttt{QPE} program, we employ test doubles to represent \texttt{Upower}. These test doubles are carefully chosen based on equivalence class partitioning. Owing that the program yields the estimated eigenvalue on the quantum variable \texttt{clock'} using a limited number of qubits, we categorize the target operations into two primary types:

\begin{itemize}
\item[(1)] Eigenvalues that can be precisely stored on \textbf{clock'}.
\item[(2)] Eigenvalues that need truncation to store into \textbf{clock'}.
\end{itemize}

An example of eigenvalues that can be precisely stored is the target \texttt{Upower} being \texttt{CNOT}, due to the eigenvalues of gate \texttt{CNOT} include only $+1 = e^{2\pi i 0.0}$ and $-1 = e^{2\pi i 0.1}$ (binary representation). Another example of eigenvalues that need truncation to store is the target \texttt{Upower} being $ControlledR_z(\frac{2}{3}\pi)$. $ControlledR_z(\frac{2}{3}\pi)$ has eigenvalues $e^{\pm i\frac{\pi}{3}}$ which cannot be represent as limited binary strings.

\begin{footnotesize}
\begin{equation}
\notag
\mathrm{CNOT} = \left[\begin{array}{cccc}
1 & 0 & 0 & 0\\
0 & 1 & 0 & 0\\
0 & 0 & 0 & 1\\
0 & 0 & 1 & 0
\end{array}
\right], \qquad
ControlledR_z(\frac{2}{3}\pi) = \left[\begin{array}{cccc}
1 & 0 & 0 & 0\\
0 & 1 & 0 & 0\\
0 & 0 & e^{-i\frac{\pi}{3}} & 0\\
0 & 0 & 0 & e^{i\frac{\pi}{3}}
\end{array}
\right]
\end{equation}
\end{footnotesize}

\end{example}

\subsection{Endian Mode Matching}
\label{subsec:EndianMatch}

In classical computing, the endian mode is guaranteed by the computer architecture and is transparent to programmers. However, as discussed in Section~\ref{subsubsec:Endian}, current quantum programming is still at the "bit level" rather than the "word level," which means that we need to consider the operations on each qubit, and the endian mode must be guaranteed by programmers. Thus, the endian mode mismatching of coupling subroutines is a typical interface bug. The following example shows the endian mode matching in the integration of QFT in QPE.

\begin{figure}
	\centering
	\subfigure[Adopt InvQFT with BIBO mode and the corresponding index of \texttt{qsclock} is "\texttt{n-i-1}." ]{\includegraphics[scale=0.46]{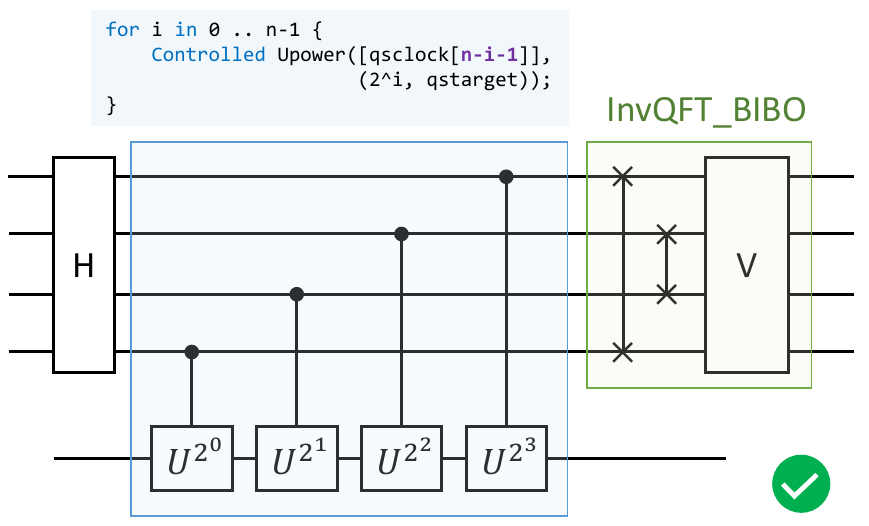}}
	\hspace{2mm}
	\subfigure[Adopt InvQFT with LIBO mode and the corresponding index of \texttt{qsclock} is "\texttt{i}." ]{\includegraphics[scale=0.46]{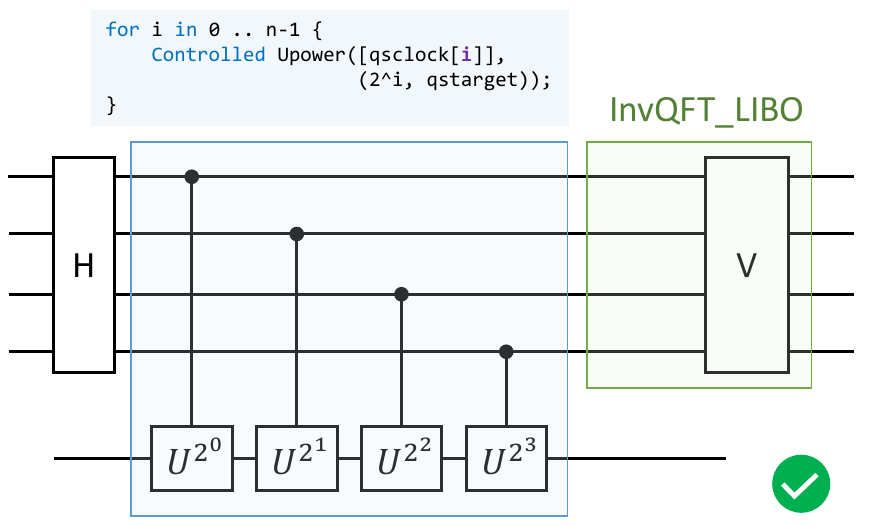}}
	\\
	\subfigure[Error: Adopt InvQFT with LIBO mode, but the corresponding index of \texttt{qsclock} is still "\texttt{n-i-1}." ]{\includegraphics[scale=0.46]{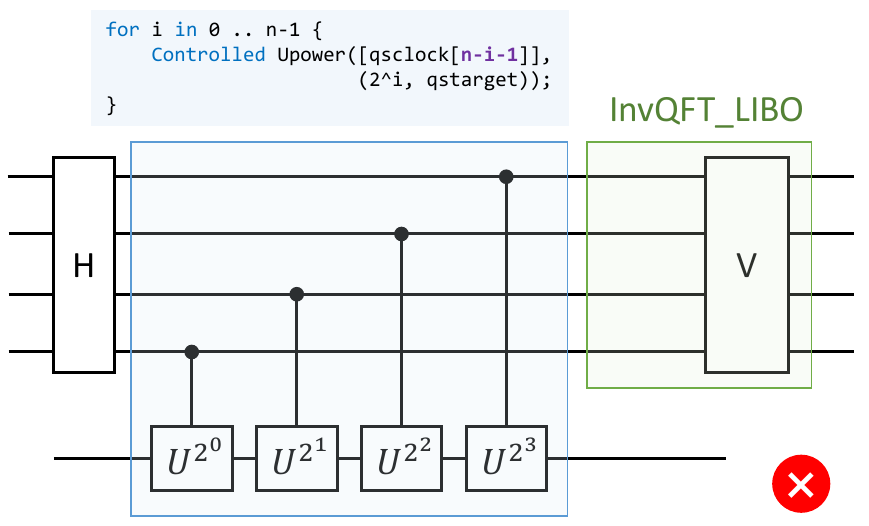}}
	\caption{The endian mode matching of the QPE program and its InvQFT subroutine.}
	\label{fig:QPEendian}
\end{figure}

\begin{example}
\label{example:EndianError}
Endian mode matching in the integration of QFT in QPE.

The Q\# code implementation of the QPE algorithm is presented in Listing~\ref{list:qpe}. In this implementation, the inverse QFT subroutine (\texttt{Adjoint QFT}) is called with BIBO endian mode at line 9. Figure~\ref{fig:QPEendian} (a) illustrates the corresponding circuit representation for the case of \texttt{Nclock}$=4$. The index of \texttt{qsclock} at line 7 is denoted as $\texttt{n-i-1}$. Since the inverse QFT (InvQFT) subroutine adopts BIBO mode, it initiates with several SWAP gates\footnote{Note that the circuit of QFT with BIBO mode ends with several SWAP gates (see Listing~\ref{list:qft} and Figure~\ref{fig:qft}), so its inverse version (InvQFT) initiates with several SWAP gates.}. Notably, these SWAP gates at the beginning of InvQFT can be omitted, and the corresponding index in line 7 needs to be modified to "\texttt{i}," as depicted in Figure~\ref{fig:QPEendian} (b). This modification avoids unnecessary SWAP gates. In this particular implementation, InvQFT is in LIBO mode. It is crucial to note that if the index in line 7 remains unaltered (Figure~\ref{fig:QPEendian} (c)), it will result in an error in the whole program.

\end{example}

Unfortunately, based on our survey in Section~\ref{subsec:surveys}, most current programming languages lack support for the endian mode, or offer only limited support. This means developers might unintentionally misconfigure the endian mode while coding. To avoid this problem, it is vital to consider endian modes not only during testing but throughout the whole development life cycle, as emphasized in the following principle.

\begin{principle}
\label{crit:endian}
Endian mode needs careful consideration throughout the whole development lifecycle and during all development activities for quantum programs.
\end{principle}

In the design stage, the designer should clearly define the chosen endian mode for each software module. During the coding stage, all programmers must adhere to the specified endian mode throughout development. In the testing stage, testers should remain vigilant about potential endian mode-related bugs. In practice, we can use the terms 'big-endian (BE)' and 'little-endian (LE)' to denote the endian mode. However, these terms may sometimes cause confusion among programmers. For instance, it may not be intuitive to discern whether the output of QFT programs is big endian or little endian (it is the decomposition of single-qubit superposition states). To mitigate this confusion, providing a specific descriptive example to clarify the concrete endian mode, as depicted in Figure~\ref{fig:endians}, is highly valuable.

\subsection{Integration of Classical-Quantum Coupling Pairs}
\label{subsec:CQcouple}

In this section, we delve into the integration of classical-quantum couplings, namely the coupling types QUCL and CUQL. Our discussion operates under a fundamental assumption:

\begin{assumption}
Testing quantum subroutines is more challenging than testing classical subroutines.
\end{assumption}

This assumption is grounded in two key factors: (1) the complexities involved in generating and detecting quantum states, as discussed in Section~\ref{subsec:GenerationDetection}; (2) classical program testing is a well-established research domain with numerous mature testing methods and techniques.

In the coupling type QUCL, a quantum subroutine calls a classical subroutine. A common scenario of QUCL is where the classical subroutine computes the essential parameters for the quantum subroutine. For instance, in quantum order finding (QOF), the program calls a continued fraction expansion subroutine after the quantum circuit to deduce the order from the circuit's measurement results. Given the dependence of the upper quantum subroutine on the lower classical subroutine, a bottom-up order is a natural choice.

Next, let us consider the coupling type CUQL. As outlined in Section~\ref{subsec:Integration}, the quantum lower subroutine has an IO type of "classical," implying it can be substituted by an equivalent classical subroutine (test double). Consequently, the coupling pair transforms into a pure classical form (type CUCL), allowing us to employ integration strategies applicable to classical programs. To illustrate, we provide an example (Example~\ref{example:IntOF}) demonstrating the use of a classical test double. However, it is crucial to verify the equivalence of the classical test double, as bugs may still emerge during its development. Thus, executing a (classical) equivalence check between the original classical-IO quantum subroutine and the classical test double is imperative.

\begin{example}
\label{example:IntOF}
Integration of quantum order finding in quantum factoring.

As demonstrated in Algorithm~\ref{alg:Shor} within Appendix~\ref{appendix:algorithms}, the crux of the quantum factoring algorithm resides in the quantum order finding subroutine (\texttt{QOF}). This subroutine, utilizing quantum mechanics, calculates the order of $x$ modulo $N$. Importantly, \texttt{QOF} has an IO type classified as "classical." Within the realm of \texttt{Quantum\_Factoring}, \texttt{QOF} stands as the sole quantum subroutine, yet it can also be implemented in a classical manner. Our testing approach entails individual tests for \texttt{QOF} and the creation of a classical test double named \texttt{ClassicalOF}, which serves as a substitute for the original \texttt{QOF} within the \texttt{Quantum\_Factoring} algorithm. This transformation renders the algorithm classical, facilitating the use of standard testing methods. Finally, we reintegrate \texttt{QOF} into the algorithm to complete the integration process. Additional details regarding the testing of the quantum factoring program will be provided in Section~\ref{subsec:CaseShor}.
\end{example}

In fact, many quantum algorithms are designed to address classical problems such as factoring, discrete logarithms, and hidden subgroup problems~\cite{nielsen2002quantum}. They blend quantum subroutines at the lower level with classical subroutines at the top level. The link between these levels is formed through a quantum subroutine with an IO type of "classical," serving as a bridge between quantum and classical components. When testing the top classical subroutines, it is feasible to substitute quantum subroutines with equivalent classical ones, effectively converting the entire target program into an equivalent classical program. This approach allows the top-down integration process to leverage existing classical testing techniques. For instance, as illustrated in Example~\ref{example:IntOF}, the quantum subroutine \texttt{QOF} can be replaced by a classical equivalent subroutine \texttt{ClassicalOF}, facilitating the implementation of top-down integration.

\subsection{Testing Practices}
\label{subsec:Tool}

In practice, testing quantum programs can be broken down into three steps: (1) writing testing scripts; (2) executing them on hardware or a simulator; and (3) analyzing the results. A testing script for a subroutine contains several testing cases, each of which involves generating input states and checking output states.

If the testing task runs on a quantum simulator, the main problem is the execution efficiency. The cost of classical computer simulating quantum computation is exponential, and thus, we can only run target programs with limited scale. In this situation, we need to design the testing scheme under a limited scale carefully. Example~\ref{example:limited} shows how to design the testing scheme on a limited number of qubits.

\begin{example}
\label{example:limited}
Testing \texttt{QPE} program on a limited number of qubits

Suppose the maximum available number of qubits is 9, and we plan to test the \texttt{QPE} program. According to Example~\ref{example:QPEIO}, the total number of qubits is $\textit{Nclock}+\textit{Ntarget}$, and it must satisfy the following condition:

$$\textit{Nclock}+\textit{Ntarget} \leq 9$$

Note that \textbf{qsclock} stores the result of phase by binary decimal. The more the number of qubits of \textbf{qsclock} has, the more accurate the result is. \textbf{qstarget} is just the workspace of the target \texttt{Upower}. A rational choice is to allocate more qubits to \textbf{qsclock} than \textbf{qstarget}, i.e., select a larger \textit{Nclock} than \textit{Ntarget}, especially for the case of eigenvalues that need truncation to store (see Example~\ref{example:IntQPE}). For example, we can choose $(\textit{Nclock}=7, \textit{Ntarget}=2)$ or $(\textit{Nclock}=6, \textit{Ntarget}=3)$.
\end{example}

If a testing task is executed on a quantum device, it is imperative to consider the influence of quantum noise and decoherence. Specific to program testing, quantum noise may lead to a problem: when a `FAIL' result occurs, it is difficult to distinguish whether the failure is caused by code errors in the program or caused by the noise. It means quantum noise will significantly influence the testing execution in the NISQ era. Therefore, when running tests on quantum devices, it is imperative to address quantum noise. Recent research, as demonstrated in~\cite{Asmar2023Noise}, has introduced machine learning techniques to filter out quantum noise during the testing of quantum programs. In future work, it is valuable to further explore techniques for mitigating quantum noise, with the aim of incorporating these techniques into the testing processes for multi-subroutine quantum programs.  Furthermore, when executing tests on a quantum device, careful consideration must be given to the influence of decoherence. Test schemes should be thoughtfully designed to ensure the execution is completed within the limitations imposed by decoherence times.

Like testing classical programs, a well-designed testing tool can help test engineers complete testing tasks more efficiently. Based on the properties of quantum programs and the testing process mentioned above, we propose four basic functional requirements that a testing tool for multi-subroutine quantum programs should provide:

\begin{principle} 
\label{crit:Tool} 
The basic functional requirements of a testing tool for multi-subroutine quantum programs are as follows:
\begin{itemize} 
\item[] 1. Repeated execution support for target programs; 
\item[] 2. Quantum information processing support; 
\item[] 3. Quantum subroutine composition support; 
\item[] 4. Results analysis support. 
\end{itemize} 
\end{principle}

Let us discuss the necessity of these requirements. Requirement 1 is crucial because checking many quantum properties requires multiple copies of target states, which are the results of target programs. Therefore, the testing tool should provide a mechanism for running the target programs repeatedly. Requirement 2 is based on the discussion in Section~\ref{subsec:TestCases}, where we highlighted the importance of quantum information processing methods in generating test cases. Requirement 3 is essential for the integration testing of multi-subroutine quantum programs. Finally, requirement 4 is critical as a testing tool should not only output the experimental results but also provide necessary modules for result analysis. These four functions represent the minimum requirements for a testing tool, and a practical testing tool is allowed to provide additional features to meet specific needs.

\subsection{Summary of Testing Processes}
\label{subsec:summary_of_test}

\begin{figure*}
	\centering
	\includegraphics[width=\textwidth]{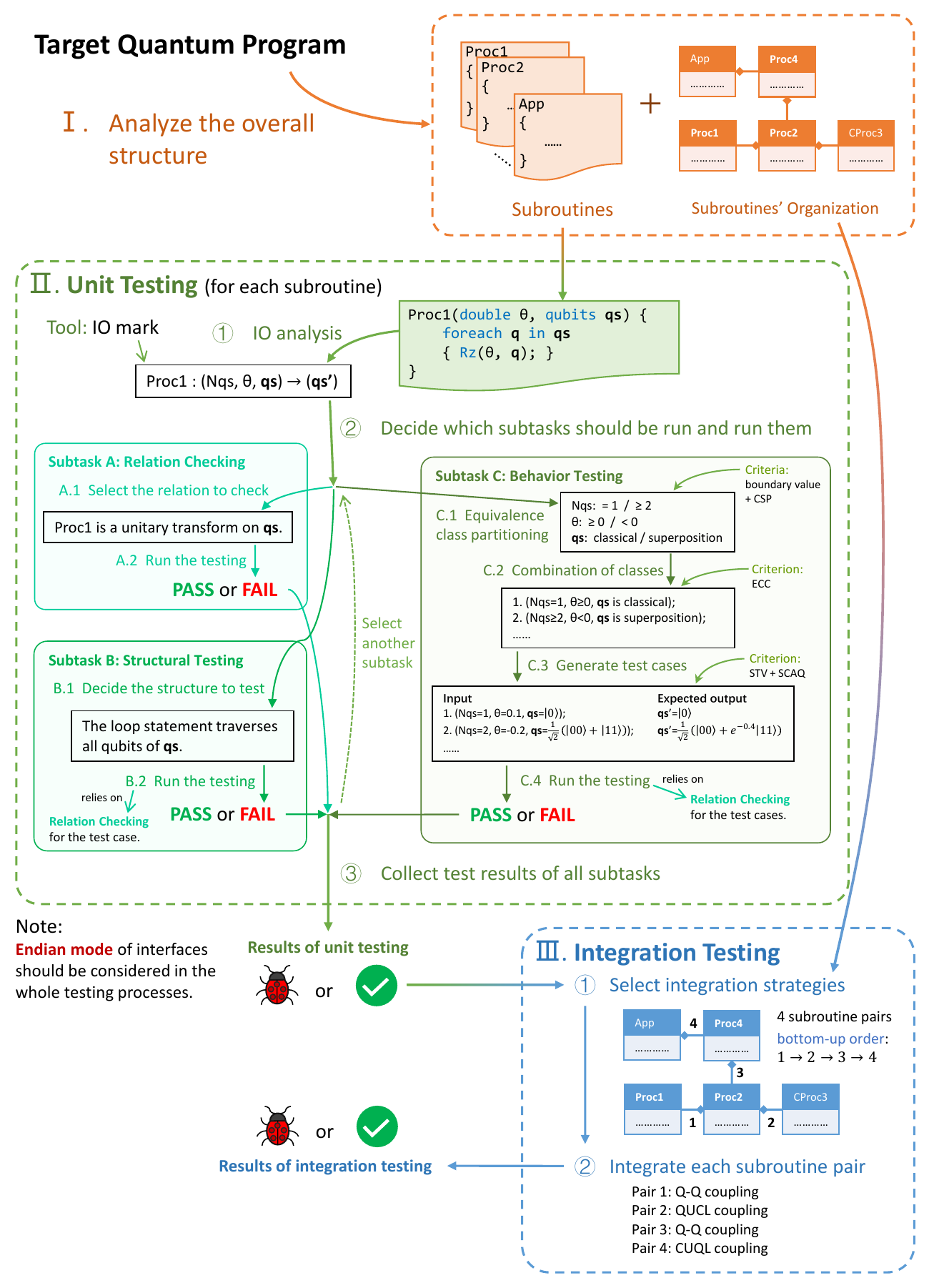}
	\caption{The overall process of testing multi-subroutine quantum programs.}
	\label{fig:overall}
\end{figure*}

We have examined the testing processes and techniques for multi-subroutine quantum programs in Sections~\ref{sec:UnitTest} and~\ref{sec:Integration}, which involve both unit testing and integration testing. In this section, we summarize the steps of testing multi-subroutine quantum programs in Figure~\ref{fig:overall}.

Before conducting unit testing on each subroutine and integration testing on the entire program, it is essential to conduct a thorough structural analysis, identify subroutines, and understand the program's organization (Step I in Figure~\ref{fig:overall}). Recent advancements have introduced several quantum modeling languages~\cite{Perez-Delgado2020quantum,ExmanS21,perez2021modelling,ali2020modeling} designed for specifying quantum software systems. In the subsequent Section~\ref{sec:casestudy}, we will utilize Q-UML~\cite{Perez-Delgado2020quantum}, an extension of the Unified Modeling Language (UML) tailored for quantum programming, to represent the structure of the target programs. This extension encompasses two types of UML diagrams: class diagrams and sequence diagrams. Q-UML adopts a distinct convention where quantum elements, such as classes, variables, functions, and more, are depicted in bold font, while classical components are represented in a standard font. In fact, we have applied this convention in the IO analysis (Section~\ref{subsec:IOanalysis}).

In Sections~\ref{sec:UnitTest} and~\ref{sec:Integration}, we introduced various testing principles and criteria aimed at addressing the challenges posed by such programs. To facilitate clarity, we compiled a summary of these testing principles and criteria in Table~\ref{table:summarize_principle}. Each principle and criterion has been linked to the specific properties of multi-subroutine quantum programs, as discussed in our responses to "Answer to RQ1."

\begin{table*}
\centering
\caption{Summary of testing principles and criteria in this paper}
\label{table:summarize_principle}
\footnotesize

\begin{tabular}{c|p{9cm}|c}
\toprule
\textbf{No.} & \makecell[c]{\textbf{Description}} & \makecell{ \textbf{Related Properties} \\ \textbf{in "Answer to RQ1"}} \\
\bottomrule
\multirow{2}{*}{Principle~\ref{crit:Principle}} & Select the input quantum states satisfying: representative, easy to prepare, and corresponding output quantum states being also easy to check. & \multirow{2}{*}{(1) (5)} \\
\hline
\multirow{2}{*}{\makecell{Principle~\ref{crit:CSMP} \\ (CSMP)}} & Partition each quantum input variable into the classical state, superposition state, and mixed state input. & \multirow{2}{*}{(5)} \\
\hline
\multirow{2}{*}{\makecell{Principle~\ref{crit:CSP} \\ (CSP)}} & Partition each quantum input variable into the classical state and superposition state. & \multirow{2}{*}{(5)} \\
\hline
\multirow{4}{*}{Principle~\ref{crit:InputFromBases}} & A criterion for selecting input quantum states can be devised by: 1. choose a set of bases from the density matrix space; 2. decompose each base matrix into the sum of matrices representing legitimate pure states; 3. select input states from these pure states. & \multirow{4}{*}{(1) (5)} \\
\hline
\multirow{2}{*}{Principle~\ref{crit:IOmethod}} & In each test case, attach the generation method for each quantum input state and the checking method for each quantum output state. & \multirow{2}{*}{(3) (5)} \\
\hline
\multirow{2}{*}{Principle~\ref{crit:endian}} & Endian mode needs careful consideration throughout the whole development lifecycle and during all development activities for quantum programs. & \multirow{2}{*}{(4)} \\
\hline
\multirow{3}{*}{Principle~\ref{crit:Tool}} & The basic functional requirements of a testing tool: 1. repeated execution support for target programs; 2. quantum information processing support; 3. quantum subroutine composition support; 4. result analysis support. & \multirow{3}{*}{(1) (5)} \\
\toprule
\multirow{2}{*}{\makecell{Criterion~\ref{crit:STV} \\ (STV)}} & Select the quantum input states in 3 forms: $\left|x\right>$; $\frac{1}{\sqrt{2}}(\left|x\right>+\left|y\right>)$; and $\frac{1}{\sqrt{2}}(\left|x\right>+i\left|y\right>)$, where $i^2=-1$. & \multirow{2}{*}{(1) (5)} \\
\hline
\multirow{2}{*}{\makecell{Criterion~\ref{crit:Pauli} \\ (PAULI)}} & Select the quantum input states from the $n$-qubit Pauli states: $\{ \left|0\right>, \left|1\right>, \left|+\right>, \left|-\right>, \left|+_i\right>, \left|-_i\right> \} ^ {\otimes n}$. & \multirow{2}{*}{(1) (5)} \\
\hline
\multirow{3}{*}{\makecell{Criterion~\ref{crit:SCAQ} \\ (SCAQ)}} & Select a set of input states that adhere to the following condition: ensuring that for each qubit in all quantum input variables, there must exist at least one state within the entire input set that exhibits superposition on that specific qubit. & \multirow{3}{*}{(5)} \\
\bottomrule
\end{tabular}
\end{table*}

\begin{tcolorbox}[leftrule=0.5mm, rightrule=0.5mm, toprule=0mm, bottomrule=0mm, colframe=purple, colback=white]
\textbf{Answer to RQ2:} Below is a summary of how the properties of multi-subroutine quantum programs influence the testing tasks:

\begin{enumerate}
\item Unlike testing quantum circuits, testing quantum programs involves considering classical variables, control flows, and varying scales of qubit numbers;
\item Due to the challenges of generating and detecting quantum states, indirect methods, such as metamorphic testing based on specific relations, play a more critical role in testing quantum programs compared to their auxiliary role in classical program testing (see Figure~\ref{fig:Subtasks}).
\item New testing principles and criteria are needed to address the unique properties of multi-subroutine quantum programs, summarized in Table~\ref{table:summarize_principle};
\item Testing variant subroutines can be accomplished through quantum relation checking;
\item Endian mode should be considered during the development processes due to current quantum programming being at the bit-level;
\item Testing on NISQ devices requires consideration of device scale limits, noise, and decoherence.
\end{enumerate}

\end{tcolorbox}

\section{Evaluation of Methods and Criteria for Unit Testing}
\label{sec:evaluation}

To demonstrate the effectiveness of our proposed testing criteria and techniques for multi-subroutine quantum programs, we will evaluate them using practical quantum programs. First, we evaluate our methods and criteria for unit testing using a set of benchmark subroutines in this section. Then we evaluate our complete testing process using three practical cases in the next section. For the evaluation of unit testing, we focus on the following research questions (RQs):

\begin{itemize}
\setlength{\itemsep}{3pt}
    \item \textbf{RQ3:} Are our unit testing design strategies effective for testing various quantum subroutines?

    \item \textbf{RQ4:} Is it necessary to cover both classical and superposition states as inputs to ensure adequate test coverage?
    
    \item \textbf{RQ5:} How effectively does the Superposition-Cover-All-Qubit (SCAQ) criterion reveal bugs in quantum programs?
\end{itemize}

\subsection{RQ3: Case Studies for Designing Unit Testing}
\label{subsec:unitCaseDesign}

To address RQ3, we conducted case studies on various types of quantum subroutines.

\subsubsection{Study Cases}
\label{subsubsec:CasesForUnit}
We have carefully selected a set of subroutines with relatively simple behavior, making them ideal candidates for unit testing. These subroutines cover all four input-output (IO) types and may have dependencies among each other. Table~\ref{table:benchmark-subroutines} gives a brief introduction to these 17 subroutines.

\begin{table*}[ht]
\centering
\caption{Benchmark subroutines (programs) for unit testing}
\label{table:benchmark-subroutines}
\renewcommand\arraystretch{1.1}
\footnotesize

\begin{tabular}{l|p{9.4cm}|c}
\toprule
\textbf{\scriptsize Subroutine Name} & \makecell[c]{\textbf{Description}} & \textbf{IO Type}  \\
\bottomrule
\texttt{QRandom} & Input an integer $n$, return a uniform random integer in $0, \dots, 2^n - 1$. & Classical \\
\bottomrule
\texttt{GenQInt} & Input a binary string $x$ with length $n$. Generate $n$-qubit state $\left|x\right>$. & \multirow{4}{*}{ \makecell{Generate-\\Quantum} } \\
\cline{1-2}
\texttt{GenXPlusY} &  Input two binary strings $x$ and $y$ with length $n$. Generate $n$-qubit state $\frac{1}{\sqrt{2}}(\left|x\right>+\left|y\right>)$. & \\
\cline{1-2}
\texttt{GenMaxSup} &  Generate $n$-qubit maximum superposition state. & \\  
\cline{1-2}
\texttt{GenMaxMix} & Generate $n$-qubit maximum mixed state. & \\  
\bottomrule
\multirow{2}{*}{\texttt{SwapTest}} & Input two subroutines which generate $n$-qubit states $\rho_1$ and $\rho_2$. Return their single-shot SWAP test result. & \multirow{6}{*}{ \makecell{Detect-\\Quantum} } \\
\cline{1-2}
\multirow{2}{*}{\texttt{Purity}} & Input a subroutine that generates $n$-qubit state $\rho$ and the number of repeat $t$. Returns whether $\rho$ is a pure state. It uses \texttt{SwapTest} as its subroutine. & \\
\cline{1-2}
\multirow{2}{*}{\texttt{InnerProduct}} & Input two subroutines that generate $n$-qubit state $\rho_1, \rho_2$ and the number of repeat $t$. Returns the estimation of $tr(\rho_1\rho_2)$. It uses \texttt{SwapTest} as its subroutine. & \\
\bottomrule
\texttt{Empty} & A quantum circuit without any gate. It is equivalent to an identity transform. & \multirow{10}{*}{Transform} \\
\cline{1-2}
\texttt{Reverse} & It reverses the order of input states. See lines $9\sim 16$ in Listing~\ref{list:qft}. & \\
\cline{1-2}
\texttt{MultiSWAP} & It swaps quantum variables. & \\
\cline{1-2}
\texttt{CRk} & Subroutine of lines $1\sim 7$ in Listing~\ref{list:qft}. & \\
\cline{1-2}
\texttt{PhaseFlip} & See Example~\ref{example:phaseflip}. & \\
\cline{1-2}
\multirow{2}{*}{\texttt{Grover}} & Given a single-target oracle (implemented as an input subroutine), it performs the Grover Search algorithm. It contains \texttt{PhaseFlip} as its subroutine. & \\
\cline{1-2}
\texttt{QFT} & It performs the Quantum Fourier transform (see lines $18\sim 32$ in Listing~\ref{list:qft}). & \\
\cline{1-2}
\texttt{QAdd} & It performs the transform: $\left|x\right>\left|y\right> \rightarrow \left|x\right>\left|x+y\right>$. & \\
\cline{1-2}
\texttt{Teleport} & See Listing~\ref{list:teleport}. & \\
\bottomrule
\end{tabular}
\end{table*}

\subsubsection{Design of Testing Plans}
We conducted case studies by designing a unit testing plan for each subroutine. Our work involves analyzing the input-output (IO) mark and program specifications of each subroutine, followed by designing a testing plan for each case. The analysis of the IO mark method is based on the discussion in Section~\ref{subsec:IOanalysis}. Designing the testing plan entails equivalence class partitioning, generating test cases, and identifying methods for input preparation and output verification, as discussed in Sections~\ref{subsec:PSCheck} and~\ref{subsec:TestCases}.

Our work results are summarized in Table~\ref{tab:unitDesign}. It is evident that regardless of the IO type, the IO mark can effectively represent the input and output variables of the target subroutine, allowing for clear differentiation between classical and quantum variables. In the testing design, we discuss equivalence class partitioning and provide corresponding test cases. Moreover, we include the generation methods for input quantum variables and the checking methods for output quantum variables. These case studies demonstrate the effectiveness of our unit testing design strategies for a variety of quantum subroutines.

\begin{table*}
\centering
\caption{IO mark, program specification, and testing design for the benchmark subroutines.}
\label{tab:unitDesign}
\begin{scriptsize}
\begin{tabular*}{\linewidth}{c|c|c}
\toprule
\textbf{Program} & \makecell{ \textbf{IO Mark and} \\ \textbf{Program Specification} } & \textbf{Testing Design} \\
\toprule
\texttt{QRandom} & \makecell{ ($n$) $\rightarrow$ (Int) \\ return a uniform random integer \\ in $[0,2^n-1]$ } & \makecell[l]{Equivalence classes: $n=0$, $n>0$ \\ Test cases: \\ \quad \texttt{[input: $n$ = 0; output: 0$\sim$1]} \\ \quad \texttt{[input: $n$ = 5; output: 0$\sim$31]} \\ Check outputs: \\ \quad Repeat running and check the output range. } \\
\toprule

\texttt{GenQInt} & \makecell{ ($n$, $x$) $\rightarrow$ (\textbf{qs}') \\ return $\left|x\right>$ ($n$ qubits) } & \makecell[l]{ Test cases: \\ \quad \texttt{[input: $n = 5, x = 0$; output: $\left|00000\right>$]} \\ \quad \texttt{[input: $n=6, x=23$; output: $\left|010111\right>$]} \\ Check output: \\ \quad Measure \textbf{qs'} and compare the result with input $x$. } \\
\midrule
\texttt{GenXPlusY} & \makecell{ ($n$, $x$, $y$) $\rightarrow$ (\textbf{qs}') \\ return $\frac{1}{\sqrt{2}}(\left|x\right>+\left|y\right>$) } & \makecell[l]{ Equivalence class: $x=y$, $x \neq y$. \\ Test cases: \\ \quad \texttt{[input: $n=5$, $x=y=20$; output: $\left|10100\right>$]} \\ \quad \texttt{[input: $n=4$, $x=5$, $y=12$; output: $\frac{1}{\sqrt{2}}(\left|0101\right>+\left|1100\right>)$]} \\ Check output: \\ \quad Measure \textbf{qs'} and compare the result with $x$ and $y$. } \\
\midrule
\texttt{GenMaxSup} & \makecell{ ($n$) $\rightarrow$ (\textbf{qs}') \\ return $\frac{1}{\sqrt{2^n}}\sum_{x \in \{0,1\}^n}\left|x\right>$ } & \makecell[l]{ Test cases: \\ \quad \texttt{[input: $n=1$; output: $\left|+\right>$]} \\ \quad \texttt{[input: $n=5$; output: $\left|+++++\right>$]} \\ Check output: \\ \quad Apply $H^{\otimes n}$ on the output \textbf{qs'} and measurement. } \\
\midrule
\texttt{GenMaxMix} & \makecell{ ($n$) $\rightarrow$ (\textbf{qs}') \\ return $\frac{1}{2^n}\sum_{x \in \{0,1\}^n}\left|x\right>\left<x\right|$ } & \makecell[l]{ Test cases: \\ \quad \texttt{[input: $n=1$; output: $\frac{1}{2}\left|0\right>\left<0\right| + \frac{1}{2}\left|1\right>\left<1\right|$]} \\ \quad \texttt{[input: $n=32$; output: $\rho_{out} = \frac{1}{2^5}I_{32}$]} \\ Check output: \\ \quad Measure \textbf{qs'} and check whether the output range is $0 \sim 2^n-1$. } \\
\toprule

\texttt{SwapTest} & \makecell{ ($n$, \underline{\textbf{GenRho1}}, \underline{\textbf{GenRho2}}) \\  $\rightarrow$ (Result) \\ return once SWAP test result \\ 0 or 1. } & \makecell[l]{ Equivalence classes: \\ \quad CSMP for $\rho_1$ and $\rho_2$; $\rho_1=\rho_2$ or $\rho_1 \neq \rho_2$. \\ Test cases: \\ \quad \texttt{[input: $\rho_1 = \left|011\right>\left<011\right|$, $\rho_2 = \left|110\right>\left<110\right|$; return 0 with $p_0=0.5$]} \\ \quad \texttt{[input: $\rho_1 = \rho_2 =  \left|1001\right>\left<1001\right|$; return 0 with $p_0=1$]} \\ \quad \texttt{[input: $\rho_1 = \left|0\right>\left<0\right|$, $\rho_2 = \left|+\right>\left<+\right|$; return 0 with $p_0=0.75$]} \\ \quad \texttt{[input: $\rho_1 = \rho_2 = \frac{1}{2}\left|0\right>\left<0\right| + \frac{1}{2}\left|1\right>\left<1\right|$; return 0 with $p_0=0.75$]} \\ \quad \texttt{[input: $\rho_1 = \left|+\right>\left<+\right|$, $\rho_2 = \frac{1}{2}\left|0\right>\left<0\right| + \frac{1}{2}\left|1\right>\left<1\right|$; $p_0=0.75$]} \\ Generate input: \\ \quad Use Algorithms~\ref{alg:KetX},~\ref{alg:KetXplusEIThetaKetY}, and~\ref{alg:MixedState}. \\ Check output: \\ \quad Repeat several times and count the proportion of result 1. } \\
\midrule
\texttt{Purity} & \makecell{ ($n$, $t$, \underline{\textbf{GenRho}}) $\rightarrow$ (isPure) \\ return TRUE if $\rho$ is pure state. } & \makecell[l]{ Equivalence classes: \\ \quad $\rho$ is classical, superposition and mixed (CSMP criterion). \\ Test cases: \\ \quad \texttt{[input: $\rho:\left|0101\right>$, $t=100$; output: True]} \\ \quad \texttt{[input: $\rho:\frac{1}{\sqrt{2}}(\left|00\right>+\left|11\right>)$, $t=100$; output: True]} \\ \quad \texttt{[input: $\rho = \frac{1}{2}\left|0\right>\left<0\right> + \frac{1}{2}\left|1\right>\left<1\right>$, $t=100$; output: False]} \\ Generate input: \\ \quad Use Algorithms~\ref{alg:KetX} and~\ref{alg:KetXplusEIThetaKetY}. \\ Check output: \\ \quad The expected output is returned with high probability. }\\
\midrule
\texttt{InnerProduct} & \makecell{ ($n$, $t$, \underline{\textbf{GenRho1}}, \underline{\textbf{GenRho2}}) \\ $\rightarrow$ (Float) \\ return estimation of $tr(\rho_1\rho_2)$ } & \makecell[l]{ Equivalence classes: \\ \quad CSMP for $\rho_1$ and $\rho_2$; $\rho_1=\rho_2$ or $\rho_1 \neq \rho_2$. \\ Test cases: \\ \quad \texttt{[input: $\rho_1 = \left|0011\right>\left<0011\right|$, $\rho_2 = \left|1010\right>\left<1010\right|$; output: $\sim 0$]} \\ \quad \texttt{[input: $\rho_1 = \rho_2 =  \left|1001\right>\left<1001\right|$; output: $\sim 1$]} \\ \quad \texttt{[input: $\rho_1 = \left|0\right>\left<0\right|$, $\rho_2 = \left|+\right>\left<+\right|$; output: $\sim 0.5$]} \\ \quad \texttt{[input: $\rho_1 = \rho_2 = \frac{1}{2}\left|0\right>\left<0\right| + \frac{1}{2}\left|1\right>\left<1\right|$; output: $\sim 0.5$]} \\ \quad \texttt{[input: $\rho_1 = \left|+\right>\left<+\right|$, $\rho_2 = \frac{1}{2}\left|0\right>\left<0\right| + \frac{1}{2}\left|1\right>\left<1\right|$; output: $\sim 0.5$]} \\ Generate input: \\ \quad Use Algorithms~\ref{alg:KetX},~\ref{alg:KetXplusEIThetaKetY}, and~\ref{alg:MixedState}. } \\
\bottomrule

\end{tabular*}
\end{scriptsize}
\end{table*}

\begin{table*}
\begin{scriptsize}
Continued from Table~\ref{tab:unitDesign}

\begin{tabular*}{\linewidth}{c|c|c}
\toprule
\textbf{Program} & \makecell{ \textbf{IO Mark and} \\ \textbf{Program Specification} } & \textbf{Testing Design} \\
\toprule
\texttt{Empty} & \makecell{ ($n$, \textbf{qs}) $\rightarrow$ (\textbf{qs'}) \\ $\left|\psi\right>$ $\rightarrow$ $\left|\psi\right>$ } & \makecell[l]{Use \textit{identity checking}. } \\
\midrule
\texttt{Reverse} & \makecell{ ($n$, \textbf{qs}) $\rightarrow$ (\textbf{qs'}) \\ $\left|j_1\right>\dots\left|j_n\right> \rightarrow \left|j_n\right>\dots\left|j_1\right>$ } & \makecell[l]{ Equivalence classes: $n=1$; $n>1$ is odd; $n$ is even. \\ Test cases: \\ \quad \texttt{[input: $\left|1\right>$, output: $\left|1\right>$]} \\ \quad [\texttt{input: $\left|11001\right>$, output: $\left|10011\right>$]} \\ \quad [\texttt{input: $\left|010011\right>$, output: $\left|110010\right>$]} \\ Generate input: \\ \quad Use Algorithm~\ref{alg:KetX}. \\ Check output: \\ \quad Use the inverse of Algorithm~\ref{alg:KetX}. \\ \quad Additional \textit{unitarity checking}. } \\
\midrule
\texttt{MultiSWAP} & \makecell{ ($n$, \textbf{qs1}, \textbf{qs2}) $\rightarrow$ (\textbf{qs1'}, \textbf{qs2'}) \\ $\left|\phi\right>\left|\psi\right>\rightarrow\left|\psi\right>\left|\phi\right>$ } & \makecell[l]{ Take random testing. \\ Generate input: \\ \quad $(U_\phi \otimes U_\psi) \left|0\right>\otimes \left|0\right>$, with randomly chosen $U_{\phi}$ and $U_{\psi}$. \\ Check output: \\ \quad Apply $U_\psi \otimes U_\phi$ on the output state and measurement. \\ \quad Compare the measurement result with 0. } \\
\midrule
\texttt{CRk} & \makecell{ ($k$, \textbf{qctrl}, \textbf{qtar}) $\rightarrow$ (\textbf{qctrl'}, \textbf{qtar'}) \\ $\left|c\right>\left|\psi\right> \rightarrow \left|c\right>\left[
	\begin{array}{cc}
	1 & 0\\
	0 & \exp\left(\frac{i\pi c}{2^k}\right)\\
	\end{array}
\right] \left|\psi\right>$ } & \makecell[l]{ Equivalence classes: $c=0$; $c=1$. \\ Test cases: \\ \quad \texttt{[input: $k=2$, $\left|0\right>\left|+\right>$; output: $\left|0\right>\left|+\right>$]} \\ \quad \texttt{[input: $k=1$, $\left|1\right>\left|+\right>$; output: $\frac{1}{\sqrt{2}}\left|1\right> (\left|0\right>+i\left|1\right>)$]} \\ Generate input and check output: \\ \quad See Example~\ref{example:qftoutput}. } \\
\midrule
\texttt{PhaseFlip} & \makecell{ ($n$, \textbf{qs}) $\rightarrow$ (\textbf{qs'}) \\ $	\left|x\right> \rightarrow \left\{
			\begin{array} {cc}
				-\left|x\right> & x>0\\
				\left|x\right> & x=0
			\end{array}
			\right.$ } & \makecell[l]{Equivalence classes: \\ \quad $n$: '$=1$' , '$>1$'; \\ \quad \textbf{qs}: see Example~\ref{example:phaseflip}. \\ Test cases: \\ \quad \texttt{[input: \textbf{qs}=$\left|0\right>$; output: $\sim$input]} \\ \quad \texttt{[input: \textbf{qs}=$\left|1\right>$; output: $\sim$input]} \\ \quad \texttt{[input: \textbf{qs}=$\left|+\right>$; output: $\neq$input]} \\ \quad \texttt{[input: \textbf{qs}=$\left|00000\right>$; output: $\sim$input]} \\ \quad \texttt{[input: \textbf{qs}=$\left|01101\right>$; output: $\sim$input]} \\ \quad \texttt{[input: \textbf{qs}=$\frac{1}{\sqrt{2}}(\left|00000\right> + \left|11001\right>)$; output: $\neq$input]} \\ \quad \texttt{[input: \textbf{qs}=$\frac{1}{\sqrt{2}}(\left|0110\right> + \left|1001\right>$); output: $\sim$input]} \\ Generate input: \\ \quad Use Algorithms~\ref{alg:KetX} and~\ref{alg:KetXplusEIThetaKetY}. \\ Check output: \\ \quad Use \textit{identity checking}. }  \\
\midrule
\texttt{Grover} & \makecell{ ($n$, \underline{\textbf{OracleK}}) $\rightarrow$ (\textbf{qs'}) \\ return $\approx \left|K\right>$ } & \makecell[l]{ Take random testing. \\ Generate input: \\ \quad Randomly choose integer $K$ and construct \underline{\textbf{OracleK}}. \\ Check output: \\ \quad Measure \textbf{qs'} and the result should be $K$ with high probability. } \\
\midrule
\texttt{QFT} & \makecell{ ($n$, \textbf{qs}) $\rightarrow$ (\textbf{qs'}) \\ $\left|j\right> \rightarrow \frac{1}{\sqrt{2^n}}\sum_{k=0}^{2^n-1}e^{2\pi i j k / 2^n}\left|k\right>$ } & \makecell[l]{ Equivalence classes: see Example~\ref{example:qftpart}. \\ Test cases: see Example~\ref{example:GenQFTcases}. \\ Generate input and checking output: \\ \quad See Example~\ref{example:qftoutput} and Example~\ref{example:GenQFTcases}. } \\
\midrule
\texttt{QAdd} & \makecell{ ($n$, \textbf{qs1}, \textbf{qs2}) $\rightarrow$ (\textbf{qs1'}, \textbf{qs2'}) \\ $\left|x\right>\left|y\right>\rightarrow\left|x\right>\left|x+y\right>$ } & \makecell[l]{ Equivalence classes: Result overflow / not overflow. \\ Test cases: \\ \quad [\texttt{input: $\left|0011\right>\left|0110\right>$, output: $\left|0011\right>\left|1001\right>$]} \\ \quad \texttt{[input: $\left|1100\right>\left|1001\right>$, output: $\left|1100\right>\left|0111\right>$]} \\ Generate input: \\ \quad Use Algorithm~\ref{alg:KetX}. \\ Check output: \\ \quad Use the inverse of Algorithm~\ref{alg:KetX}; \\ \quad Additional \textit{unitarity checking}. } \\
\midrule
\texttt{Teleport} & \makecell{ (\textbf{qsrc}) $\rightarrow$ (\textbf{qdest'}) \\ $\left|q\right> \rightarrow \left|q\right>$ } & \makecell[l]{Use \textit{identity checking} } \\
\bottomrule
\end{tabular*}
\end{scriptsize}
\end{table*}

\begin{tcolorbox}[leftrule=0.5mm, rightrule=0.5mm, toprule=0mm, bottomrule=0mm, colframe=purple, colback=white]
\textbf{Answer to RQ3: } 
The results of our case studies, as shown in Table~\ref{tab:unitDesign}, demonstrate that our proposed testing design strategies effectively conduct IO analysis, equivalence class partitioning, and test case design for various quantum subroutines.
\end{tcolorbox}

\subsection{Mutant Programs}
\label{subsec:mutations}

RQ4 and RQ5 focus on evaluating the effectiveness of testing principles or criteria for identifying erroneous programs. One common approach to simulate bugs in practical programs is through mutation analysis. For multi-subroutine quantum programs, it is valuable to consider four types of mutant operators:

\begin{itemize}
\item \textsf{Gate Mutant (GM) operators:} This type involves adding, removing, or modifying a basic quantum gate.
\item \textsf{Subroutine Mutant (SM) operators:} This type involves adding, removing, or modifying a calling statement. In the case of modification, it refers to making the calling statement invoke an erroneous version of the original subroutine.
\item \textsf{Classical Mutant (CM) operators:} This type focuses on modifying control-flow statements, classical expressions, or assignments of classical variables.
\item \textsf{Measurement Mutant (MM) operators:} This type involves adding or removing measurement statements\footnote{Actually, changing the measurement bases may also be included as a type of mutant operator. However, considering that changing the measurement bases is implemented as an additional unitary transform ahead of the measurement in practice, we regard it as an SM rather than an MM.}.
\end{itemize}

\begin{figure}
	\centering
	\includegraphics[scale=0.44]{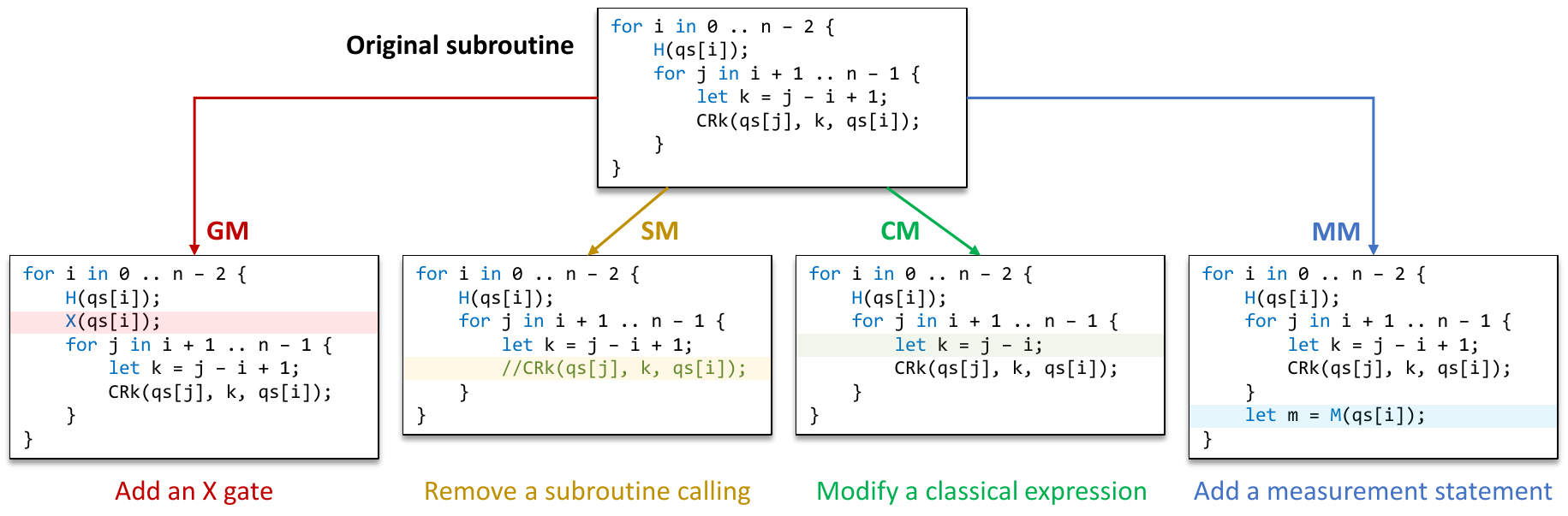}
	\caption{Examples about gate mutant (GM), subroutine mutant (SM), classical mutant (CM), and measurement mutant (MM) operators.}
	\label{fig:mutations}
\end{figure}

Figure~\ref{fig:mutations} provides specific examples of these four types of mutant operators applied to an original subroutine. Previous research on testing quantum programs~\cite{ali2021assessing,wang2021generating,wang2021application,fortunato2022qmutpy,fortunato2022mutation} and quantum platforms~\cite{wang2021qdiff} has mainly considered gate mutant (GM) and measurement mutant (MM) operators. This paper introduces two new mutant operator types - subroutine mutant (SM) and classical mutant (CM) operators, which are essential for evaluating multi-subroutine and quantum-classical-mixed programs.

\subsection{RQ4 and RQ5: Experimental Design}
\label{subsec:RQ4RQ5design}

To address RQ4 and RQ5, we have selected six subroutines from Section~\ref{subsec:unitCaseDesign} with an IO type of "transform." As shown in Table~\ref{table:RQ4RQ5}, for \texttt{Reverse}, \texttt{MultiSWAP}, and \texttt{QFT}, we consider different scales $n$. Our objective is to evaluate the necessity of covering both classical and superposition states and the effectiveness of the SCAQ criteria using mutation testing techniques.

In the row labeled "\#mutant subroutines," we have generated several mutant subroutines for each original correct subroutine by using Gate Mutant (GM), Subroutine Mutant (SM)\footnote{SM is only applicable to the subroutines that utilize other subroutines.}, Classical Mutant (CM), and Measurement Mutant (MM) operators, as discussed in Section~\ref{subsec:mutations}. Each mutant subroutine used in the evaluation is generated by using only one mutant operator, affecting one line of code, and can be successfully compiled. We have ensured that the mutants cover all parts of the original subroutine uniformly, with each type of mutant operator including addition, deletion, and replacement.

For the evaluation, we have employed random testing, where we randomly generate input test cases of specific types for each mutant subroutine and check whether the test cases can kill the mutant subroutine. The \textit{kill rate} is
defined as the proportion of the number of test cases that kill the mutant in all test cases. 
We consider three types of input states: classical input (CI), represented as $\left|x\right>$ with a random integer $x$; random two-value superposition input (RTI), represented as $\frac{1}{\sqrt{2}}(\left|x\right>+\left|y\right>)$ with two random integers $x$ and $y$; and complementary superposition input (CSI), represented as $\frac{1}{\sqrt{2}}(\left|x\right>+\left|\bar{x}\right>)$ with a random integer $x$ and its bitwise negation $\bar{x}$. To check the output, we use transform-based checking for RTI and CSI of \texttt{QFT}, and the method described in Section~\ref{subsubsec:RelationQS} for CI of \texttt{QFT} and other subroutines. For a significant comparison, we perform each transform-based checking only once and record the kill result directly. It is important to note that the kill rate observed in this evaluation may not reflect the rates encountered in practical testing tasks. In practical scenarios, multiple repetitions are often performed to enhance the accuracy of the judgment. Additionally, due to the larger input space with increasing scale $n$, we have selected the number of test cases for each mutant subroutine to be $100n$ to cover a broader range of scenarios adequately. This selection is based on empirical grounds.

We have used the Q\# language and its simulator for our experiments, as (1) our methods and benchmark programs involve variable-scale and classical-quantum-mixed scenarios, and (2) Q\# is a high-level quantum programming language that supports such program patterns. The experiments were conducted on a personal computer equipped with an Intel Core i7-1280P CPU and 16 GB RAM.

The results of the experiments are presented in Table~\ref{table:RQ4RQ5}, which consists of five subtables. Subtable (a) provides the parameters of the tested subroutines, mutant operators, and test cases. Subtables (b), (c), and (d) display the numbers of killed mutant subroutines and the kill rates under CI, RTI, and CSI, respectively. Subtable (e) presents the overall kill rate for each subroutine. In this table, the "not applicable" items are denoted as the dash '-,' such as $n$ is not applicable for single-qubit subroutines, and SM mutant operators are not applicable for subroutines without callings.

\subsection{RQ4: The Necessity of Covering Both Classical and Superposition States}
\label{subsec:unitNecessity}

According to the subtables (b), (c), and (e) in Table~\ref{table:RQ4RQ5}, we observe that for most mutants, the kill rate is higher for RTI compared to CI. Interestingly, the kill rate of MMs using classical input states is significantly lower than that of MMs using superposition input states. In Particular, for \texttt{CRk}, \texttt{Empty}, \texttt{Reverse}, and \texttt{MultiSWAP}, the classical input states fail to kill any MMs. Further investigation reveals the following findings: 
\begin{itemize}

\item[(1)] If the program consists of "essentially classical" gates (such as $X$, CNOT, and SWAP), classical input states cannot kill any MMs. This is because, under classical input states, the computation process remains the output states classical, and measurements do not alter the variable contents. 

\item[(2)] If the MM occurs at the beginning of a program, classical input states are unable to kill it since measurements do not modify the given input. 

\item[(3)] Certain "phase-only" mutants, such as adding an $S$ or $Z$ gate, cannot be killed by classical input states.
\end{itemize}

Another intriguing observation relates to the kill rate with scale $n$. For cases where $n=1$, the kill rate for each program is consistently low. This can be attributed to the absence of loop executions at $n=1$, resulting in the mutants inside loops not being killed. By examining the results of the $n=1$ test cases, we can determine whether bugs are located inside or outside the loops.

During the execution of the experiments, it is worth noting that some mutants fail due to exceptions rather than output checking, with most of these cases involving CMs. In practice, mutants in classical parameters, such as array indices, can be killed by exceptions through the exception-handling mechanisms present in programming languages.

While superposition input states demonstrate higher kill rates compared to classical input states, this does not imply that we should solely focus on testing superposition inputs. In fact, checking output under superposition input states usually relies on checking output under classical input states (as demonstrated in Example~\ref{example:GenQFTcases}). Additionally, we also observed that certain mutants killed by classical input states fail to be killed by superposition input states. This highlights the crucial role of superposition as a differentiating factor between classical and quantum computing. Therefore, it is necessary to cover both classical and superposition inputs when testing quantum programs.

\begin{tcolorbox}[leftrule=0.5mm, rightrule=0.5mm, toprule=0mm, bottomrule=0mm, colframe=purple, colback=white]
\textbf{Answer to RQ4: } 
Superposition is fundamental to quantum computing. Our experimental results, as shown in Table~\ref{table:RQ4RQ5} (b)(c)(e),  demonstrate that both classical and superposition inputs are effective in uncovering various bugs. Therefore, it is essential to cover both classical and superposition inputs when testing quantum programs.
\end{tcolorbox}

\begin{table*}
\centering
\caption{The evaluation results of RQ4 and RQ5}
\label{table:RQ4RQ5}
\begin{scriptsize}
\begin{tabular}{cl|c|c|c|c|c|c|c|c|c|c|c}
\multicolumn{13}{l}{ \textsf{\footnotesize (a) Experiments configurations }} \\

\toprule
&& \texttt{CRk} & \texttt{\tiny Teleport} & \texttt{Empty} & \multicolumn{3}{c|}{\texttt{Reverse}} & \multicolumn{2}{c|}{\texttt{MultiSWAP}} & \multicolumn{3}{c}{\texttt{QFT}} \\
\midrule
\multirow{4}{*}{\#mutant subroutine} & GM & 6 & 13 & 9 & \multicolumn{3}{c|}{16} & \multicolumn{2}{c|}{13} & \multicolumn{3}{c}{20} \\
& SM & 0 & 0 & 0 & \multicolumn{3}{c|}{0} & \multicolumn{2}{c|}{0} & \multicolumn{3}{c}{16} \\
& CM & 4 & 12 & 0 & \multicolumn{3}{c|}{12} & \multicolumn{2}{c|}{8} & \multicolumn{3}{c}{18} \\
& MM & 4 & 7 & 2 & \multicolumn{3}{c|}{10} & \multicolumn{2}{c|}{10} & \multicolumn{3}{c}{12} \\
\midrule
\multicolumn{2}{c|}{parameter $n$} & - & - & 10 & 1 & 8 & 9 & 1 & 6 & 1 & 2 & 6  \\
\midrule
\multicolumn{2}{c|}{\#total qubits} & 2 & 3 & 10 & 1 & 8 & 9 & 1 & 12 & 1 & 2 & 6  \\
\midrule
\multicolumn{2}{c|}{\makecell[c]{\#test cases for each \\ mutant subroutine}} & 200 & 100 & 1000 & 100 & 800 & 900 & 100 & 600 & 100 & 200 & 600 \\
\midrule
\multirow{4}{*}{\#test cases}& GM & 1200 & 1300 & 9000 & 1600 & 12800 & 14400 & 1300 & 7800 & 2000 & 4000 & 12000 \\
& SM & - & - & - & - & - & - & - & - & 1600 & 3200 & 9600 \\
& CM & 800 & 1200 & - & 1200 & 9600 & 10800 & 800 & 4800 & 1800 & 3600 & 10800 \\
& MM & 800 & 700 & 2000 & 1000 & 8000 & 9000 & 1000 & 6000 & 1200 & 2400 & 7200 \\
\midrule
\multicolumn{2}{c|}{\#total test cases} & 1400 & 3200 & 11000 & 3800 & 30400 & 34200 & 3100 & 18600 & 6600 & 13200 & 39600 \\
\bottomrule
\\[-3pt]

\multicolumn{13}{l}{\textsf{\footnotesize (b) The number of mutant subroutines killed by classical input (CI)} } \\

\toprule
\multirow{4}{*}{\makecell{ \#mutant subroutines \\ killed by CI }}& GM & 224 & 436 & 4999 & 348 & 9262 & 10473 & 868 & 6860 & 300 & 2068 & 9900 \\
& SM & - & - & - & - & - & - & - & - & 52 & 1530 & 7771 \\
& CM & 248 & 461 & - & 300 & 8094 & 9598 & 353 & 4442 & 154 & 1781 & 9545 \\
& MM & 0 & 56 & 0 & 0 & 0 & 0 & 0 & 0 & 240 & 1087 & 4725 \\
\midrule
\multirow{4}{*}{\makecell{Kill rate \\ by CI (\%)} }& GM & 18.7 & 33.5 & 55.6 & 21.8 & 72.4 & 72.7 & 66.8 & 87.9 & 15.0 & 51.7 & 82.5 \\
& SM & - & - & - & - & - & - & - & - & 3.3 & 47.8 & 80.9 \\
& CM & 31.0 & 38.4 & - & 25.0 & 84.3 & 88.9 & 44.1 & 92.5 & 8.6 & 49.5 & 88.4 \\
& MM & 0.0 & 8.0 & 0.0 & 0.0 & 0.0 & 0.0 & 0.0 & 0.0 & 20.0 & 45.3 & 65.6 \\
\bottomrule
\\[-3pt]

\multicolumn{13}{l}{\textsf{\footnotesize (c) The number of mutant subroutines killed by random two-value superposition input (RTI)} } \\

\toprule
\multirow{4}{*}{\makecell{ \#mutant subroutines\\ killed by RTI }}& GM & 241 & 569 & 6220 & 442 & 10912 & 12353 & 338 & 7202 & 615 & 2898 & 11625 \\
& SM & - & - & - & - & - & - & - & - & 19 & 2599 & 8954 \\
& CM & 260 & 453 & - & 300 & 8515 & 9995 & 300 & 4576 & 219 & 2183 & 10762 \\
& MM & 201 & 208 & 788 & 200 & 3287 & 3766 & 524 & 2848 & 12 & 1540 & 5424 \\
\midrule
\multirow{4}{*}{\makecell{ Kill rate \\ by RTI (\%) }}& GM & 20.1 & 43.8 & 69.1 & 27.6 & 85.3 & 85.8 & 26.0 & 92.3 & 30.8 & 72.5 & 96.9 \\
& SM & - & - & - & - & - & - & - & - & 1.2 & 81.2 & 93.3 \\
& CM & 32.5 & 37.8 & - & 25.0 & 88.7 & 92.5 & 37.5 & 95.3 & 12.2 & 60.6 & 99.6 \\
& MM & 25.1 & 29.7 & 39.4 & 20.0 & 41.1 & 41.8 & 52.4 & 47.5 & 1.0 & 64.2 & 75.3 \\
\bottomrule
\\[-3pt]

\multicolumn{13}{l}{\makecell[l]{\textsf{\footnotesize (d) The number of mutant subroutines killed by complementary superposition input (CSI)} }} \\

\toprule
\multirow{4}{*}{\makecell{ \#mutant subroutines \\ killed by CSI }}& GM & 610 & 603 & 7414 & 452 & 12249 & 13901 & 938 & 7775 & 615 & 3294 & 11611 \\
& SM & - & - & - & - & - & - & - & - & 13 & 2804 & 8865 \\
& CM & 358 & 441 & - & 300 & 8071 & 9628 & 683 & 4781 & 215 & 2310 & 10799 \\
& MM & 387 & 212 & 1014 & 211 & 4035 & 4487 & 748 & 5963 & 19 & 1786 & 5417 \\
\midrule
\multirow{4}{*}{\makecell{ Kill rate \\ by CSI (\%) }}& GM & 50.8 & 46.4 & 82.4 & 28.3 & 95.7 & 96.5 & 72.2 & 99.7 & 30.1 & 82.4 & 96.8 \\
& SM & - & - & - & - & - & - & - & - & 0.8 & 87.6 & 92.3 \\
& CM & 44.8 & 36.8 & - & 25.0 & 84.1 & 89.1 & 85.4 & 99.6 & 11.9 & 64.2 & 99.99 \\
& MM & 48.4 & 30.3 & 50.7 & 21.1 & 50.4 & 49.9 & 74.8 & 99.4 & 1.6 & 74.4 & 75.2 \\
\bottomrule
\end{tabular}
\end{scriptsize}

\vspace{5mm}
\footnotesize
\textsf{(e) The overall kill rates of six subroutines with three types of input: CI, RTI, and CSI}
\includegraphics[scale=0.73]{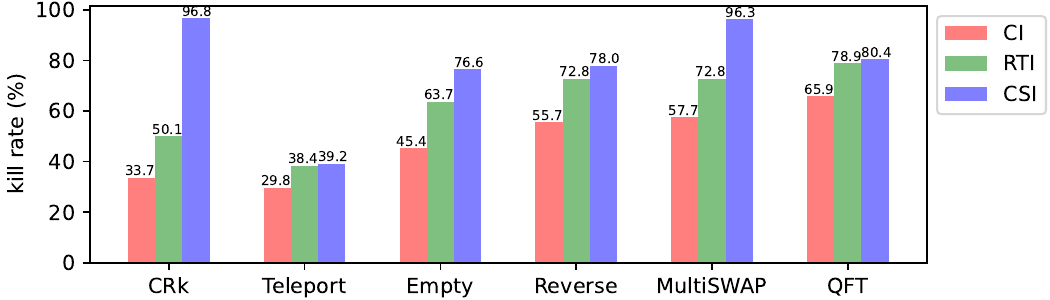}

\end{table*}

\subsection{RQ5: The Effectiveness of Superposition Covering All Qubits}
\label{subsec:evaluateSCAQ}

According to the subtables (c), (d), and (e) in Table~\ref{table:RQ4RQ5}, it can be observed that CSI has a higher kill rate than RTI, even though both are superposition input states. This indicates that superposition with a larger extent has a greater ability to reveal bugs compared to superposition with a lower extent. Actually, this result can be attributed to the occurrence possibility of mutants on every qubit, coupled with the comprehensive coverage of the input superposition facilitated by the SCAQ criterion. Furthermore, we have noticed that different subroutines exhibit varying degrees of sensitivity to kill rates with RTI and CSI. Specifically, \texttt{CRk} and \texttt{MultiSWAP} demonstrate sensitivity, while \texttt{Teleport} and \texttt{QFT} do not exhibit the same level of sensitivity. 

\begin{tcolorbox}[leftrule=0.5mm, rightrule=0.5mm, toprule=0mm, bottomrule=0mm, colframe=purple, colback=white]
\textbf{Answer to RQ5: } 
The effectiveness of superposition inputs to reveal different bugs also exists at the level of the qubit. Our experimental results, as shown in Table~\ref{table:RQ4RQ5} (c)(d)(e), demonstrate that when superpositions are applied across all input qubits, they are more effective at revealing bugs. It is worth noting that bugs can arise on any qubit, highlighting the importance of test cases that adhere to the SCAQ criterion in detecting bugs in quantum programs.
\end{tcolorbox}

\section{Case Studies for Complete Testing Process}
\label{sec:casestudy}

In this section, we evaluate the effectiveness of our overall testing process by conducting case studies on three representative multi-subroutine quantum programs. Our goal is to address the following research question:

\begin{itemize}
\item \textbf{RQ6:} Do our testing design strategies effectively test multi-subroutine quantum programs?
\end{itemize}

We have chosen three cases for examination: (1) Quantum Phase Estimation, (2) Shor's Factoring Program, and (3) Linear System Solver. Each case study will begin with an overview of the program's design and then delve into the testing design. We will provide the complete testing process details for case 1, and our focus for the other two cases will be primarily on integration testing strategies. This analysis will enable us to assess the effectiveness of our testing approach within the context of multi-subroutine quantum programs. At the end of this section, we will share insights and testing experiences derived from these case studies.

\subsection{Case 1: Quantum Phase Estimation (\texttt{QPE})}
\label{subsec:CaseQPE}

In Sections ~\ref{sec:UnitTest} and ~\ref{sec:Integration}, we have employed the QPE program as an illustrative example to provide in-depth explanations of several integration testing methods (Examples ~\ref{example:QPEIO}, ~\ref{example:QPEintgeration}, ~\ref{example:IntQPE}, ~\ref{example:EndianError}, and ~\ref{example:limited}). In this case study, we present a comprehensive testing process for the QPE program, which evaluates the effectiveness of our overall testing process.

\vspace{.7mm}
\begin{small}
\begin{itemize}[leftmargin=4em]
\setlength{\itemsep}{0.5pt}
\setlength{\baselineskip}{10pt}
\item[\textbf{Step 1:}] \textit{Analyze the target program's profile:} This analysis consists of two key aspects: the Input-Output (IO) structure and the subroutine structure. In Example~\ref{example:QPEIO} and Figure~\ref{fig:QPEdepend}, we presented detailed insights into how to analyze the IO structure and subroutine structure. Regarding the subroutine structure, we recommend adopting a bottom-up integration order. Additionally, for subroutines called as parameters, such as \texttt{Upower}, it is essential to create and utilize test doubles during implementation.

\vspace{1mm}
\item[\textbf{Step 2:}] \textit{Test the \texttt{QFT} subroutine:}
Detailed information regarding testing the \texttt{QFT} subroutine can be found in Section~\ref{sec:UnitTest}, specifically in Examples~\ref{example:qftoutput}, \ref{example:qftpart}, and \ref{example:GenQFTcases}.

\vspace{1mm}
\item[\textbf{Step 3:}] \textit{Design the test plan for the topmost module \texttt{QPE}:} This step involves unit testing for the \texttt{QPE} module. It is important to note that \texttt{QPE} includes subroutines \texttt{QFT} and \texttt{Upower}, which means the test plan should encompass the integration of both \texttt{QFT} and \texttt{Upower} into \texttt{QPE}.

\vspace{1mm}
    \begin{itemize}
	\item Input variables (see Example~\ref{example:QPEIO}): Nclock, Ntarget, \underline{\texttt{\textbf{Upower}}}, and \textbf{target}.
	\item Output variable: \textbf{clock'}.

    \item Equivalence class partition:
    \begin{itemize}
    	\item For variable Ntarget: boundary value partition.
    	\item For variables Nclock, \underline{\texttt{\textbf{Upower}}}: see Example~\ref{example:IntQPE}.
    	\item For variable \textbf{target}: consider whether the input state is the eigenvector of the target \underline{\texttt{\textbf{Upower}}}.
        \item A feasible partition:
        
        Nclock[ins, suf]\footnote{ins/suf: the number of clock qubits which is insufficient/sufficient to contain all digits of phase estimation results.}; Ntarget[=1, $\geq$2]; \underline{\texttt{\textbf{Upower}}}[fev, ifev]\footnote{fev/ifev: the operation with eigenvalue which has finite/infinite binary argument.}; \textbf{target}[es, nes]\footnote{es: the eigenstate of corresponding to \texttt{Upower}; nes: not the eigenstate.}.
    \end{itemize}
    
    \item Combine input variables:
    \begin{itemize}
    	\item Combination criterion: ECC (prefer efficiency in this testing task).
    	
    	\item A set of feasible equivalence classes:
    	
    	\begin{itemize}
    		\item (Nclock is suf, Ntarget = 1, \underline{\texttt{\textbf{Upower}}} is fev, \textbf{target} is es)
    		\item (Nclock is suf, Ntarget = 1, \underline{\texttt{\textbf{Upower}}} is fev, \textbf{target} is nes)
    		\item (Nclock is suf, Ntarget $\geq$ 2, \underline{\texttt{\textbf{Upower}}} is fev, \textbf{target} is es)
    		\item (Nclock is suf, Ntarget $\geq$ 2, \underline{\texttt{\textbf{Upower}}} is fev, \textbf{target} is nes)
    		\item (Nclock is ins, Ntarget $\geq$ 2, \underline{\texttt{\textbf{Upower}}} is ifev, \textbf{target} is es)
    	\end{itemize}
    \end{itemize}
\end{itemize}

\vspace{1mm}
\item[\textbf{Step 4:}] \textit{Construct the concrete test cases for testing module \texttt{QPE}:} According to the equivalence classes in Step 3, we constructed the following test cases:
\begin{itemize}
	\item \texttt{[input: Ntarget=1, Nclock = 3, \underline{\texttt{\textbf{Upower}}}$=X$,} \textbf{target}$=\left|-\right>$; 
	
		\texttt{output:} \textbf{clock'}$=\left|100\right>$ \texttt{]}
	
	\item \texttt{[input: Ntarget=1, Nclock = 3, \underline{\texttt{\textbf{Upower}}}$=H$,} \textbf{target}$=\left|0\right>$; 
		
		\texttt{output:} \textbf{clock'}: $\left|000\right>$ with prob $\sim 0.8536$, $\left|111\right>$ with prob $\sim 0.1464$ \texttt{]}
		
	\item \texttt{[input: Ntarget=3, Nclock = 3, \underline{\texttt{\textbf{Upower}}}$=Controlled(S\otimes S^\dagger)$,} \textbf{target}$=\left|101\right>$; 
	
		\texttt{output:} \textbf{clock'}$=\left|110\right>$ \texttt{]}
		
	\item \texttt{[input: Ntarget=3, Nclock = 3, \underline{\texttt{\textbf{Upower}}}$=Controlled(S\otimes S^\dagger)$,} \textbf{target}$=\frac{1}{\sqrt{2}}(\left|101\right> + \left|110\right>)$; 
		
		\texttt{output:} \textbf{clock'}: $\left|110\right>$ with prob $\sim 0.5$, $\left|010\right>$ with prob $\sim 0.5$ \texttt{]}
		
	\item \texttt{[input: Ntarget=2, Nclock = 7, \underline{\texttt{\textbf{Upower}}}$=Controlled R_z(\frac{2}{3}\pi)$,} \textbf{target}$=\left|11\right>$; 
	
		\texttt{output:} \textbf{clock'}$=\left|x\right>$, $x\approx \frac{1}{6}$\texttt{]} \footnote{Note that eigenvalue $e^{i\frac{1}{3}\pi} = e^{2\pi i \frac{1}{6}}$ and it needs truncation to store, so the output is approximate to $\frac{1}{6}$. }
\end{itemize}

\vspace{1mm}
\item[\textbf{Step 5:}] \textit{Construct the generation methods and the checking methods for the test cases:}

\begin{itemize}
    \item Generate the input of \underline{\textbf{\texttt{Upower}}}: built by corresponding quantum gates.
    \item Generate the input of \textbf{target}: use Algorithms~\ref{alg:KetX} and~\ref{alg:KetXplusEIThetaKetY}.
    \item Check the output of \textbf{clock'}: relation checking about statistical results.
\end{itemize}

\vspace{1mm}
\item[\textbf{Step 6:}] \textit{Execute the test cases constructed in Step 4 using the methods outlined in Step 5.} 
\end{itemize}
\end{small}

\subsection{Case 2: Shor's Factoring Program (\texttt{Factoring})}
\label{subsec:CaseShor}

\subsubsection{Program Structure}
The \texttt{Factoring} program is designed to factorize a positive integer $N$ and produce the factorization series of $N$. The program's structure is represented by the Q-UML diagram shown in Figure~\ref{fig:shoruml}(a). We have modeled the program using several Q-UML classes. The core module is the \textit{order-finding} class \texttt{QOrderFinding}. This class takes a positive integer $x$ as input and determines its \textit{order} $r$, the minimum value of $r$ satisfying the equation $x^r \equiv 1 (\mathrm{mod}\hspace{1mm}N)$. The \texttt{QOrderFinding} class consists of three subroutines: \texttt{QPhaseEstimation}, which performs quantum phase estimation, \texttt{QModularMultiply}, which implements the transform $\left|y\right> \rightarrow \left|xy(\mathrm{mod}\hspace{1mm}N)\right>$, and a classical subroutine for continued fraction calculations. As discussed in Section~\ref{appendix:algorithms}, the \texttt{Factoring} program also incorporates other number theory subroutines such as greatest common factors (\texttt{GCD}) and a check for whether $N$ is in the form of $a^b$. These subroutines are encapsulated within the \texttt{FactorTest} class. At the top level, the \texttt{Application} class provides the user interface for the program.

\subsubsection{Testing Design}

\begin{figure}
	\centering
	\subfigure[The Q-UML diagram illustrating the structure the program.]{\includegraphics[scale=0.6]{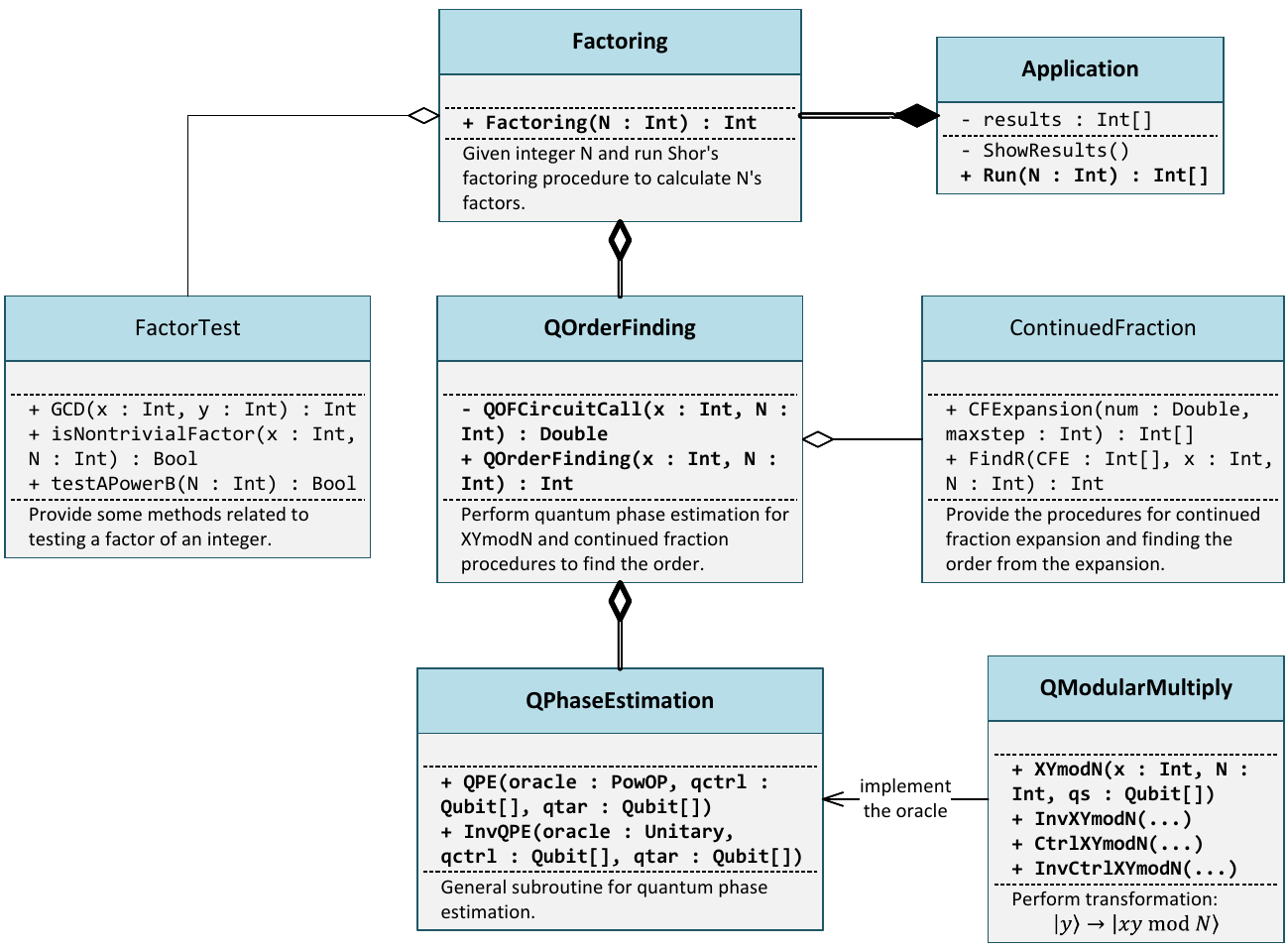}}
	\\
	\subfigure[The process of integration testing. We split the original program into two parts and use the classical test double of subroutine \texttt{QOrderFinding}. The bold numbers denote the integration order.]{\includegraphics[scale=0.5]{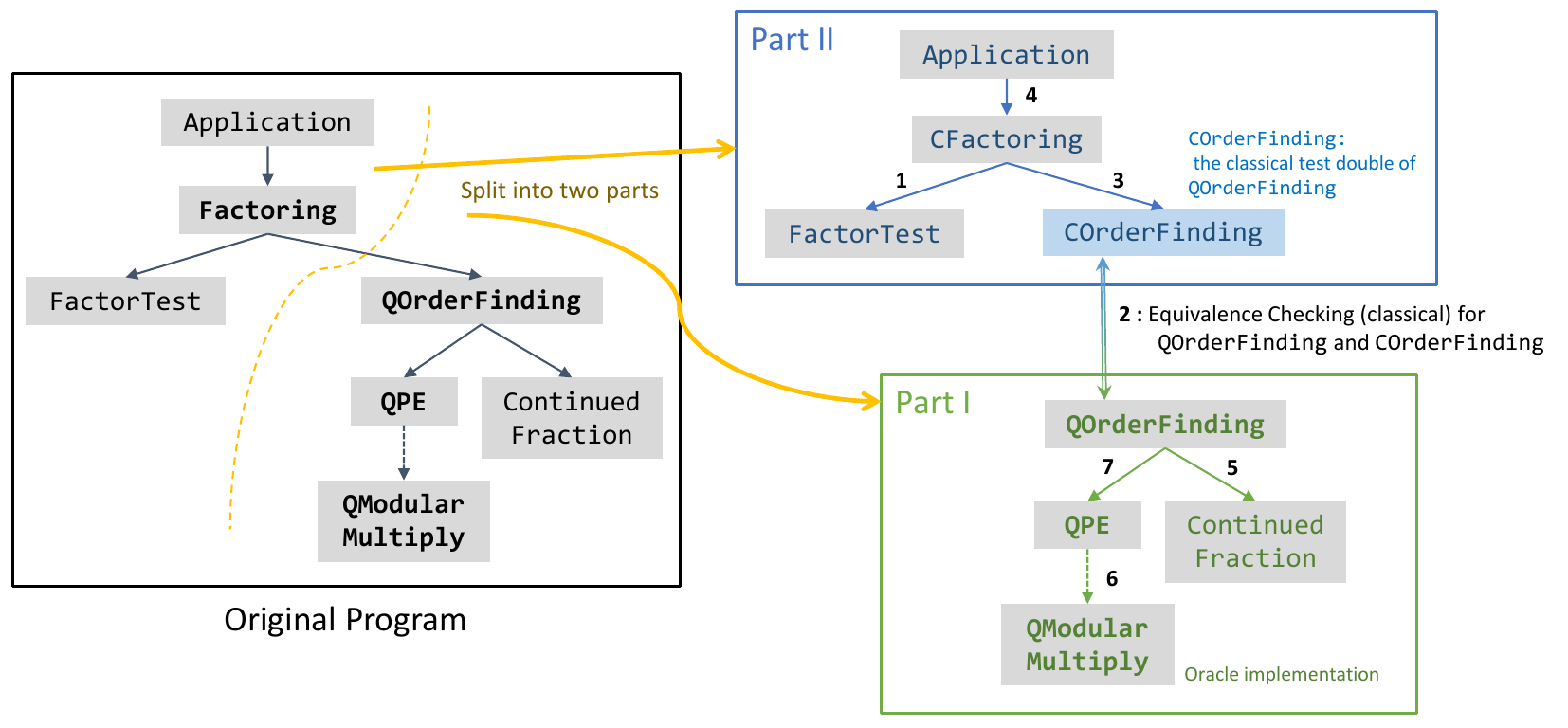}}
	\caption{The Q-UML and the testing strategies for the demo program for Shor's factoring algorithm.}
	\label{fig:shoruml}
\end{figure}

It is important to note that the subroutines upper than \texttt{QOrderFinding} are classical in nature and can also be implemented as a classical version called \texttt{COrderFinding}. By replacing \texttt{QOrderFinding} with \texttt{COrderFinding}, the program \texttt{Factoring} transforms into a classical factoring program named \texttt{CFactoring} that relies on classical order finding. Therefore, test double technology is suitable for the subroutines above the order-finding process. In practice, the following steps can be followed to complete the testing task:

\begin{small}
\begin{itemize}[leftmargin=4em]
\setlength{\itemsep}{1pt}
\item[\textbf{Step 1:}] \textit{Perform unit tests for lower classes}: This step involves testing the individual lower classes, including \texttt{QModularMultiply}, \texttt{QPhaseEstimation}, \texttt{ContinuedFraction}, and~\texttt{FactorTest}.

\item[\textbf{Step 2:}] \textit{Split the original program into two parts}: Divide the original program into two parts using \\ \texttt{QOrderFinding} as the separation point. These parts are referred to as Part I and Part II, as shown in Figure~\ref{fig:shoruml}(b).

\item[\textbf{Step 3:}] \textit{Utilize test double technology}: Replace the invocation of \texttt{QOrderFinding} in Part II with \texttt{COrderFinding}, which is a classical equivalent of \texttt{QOrderFinding}. This transformation converts Part II into a purely classical component.

\item[\textbf{Step 4:}] \textit{Ensure equivalence between \texttt{COrderFinding} and \texttt{QOrderFinding}}: Verify that \texttt{COrderFinding} exhibits the same external behavior as \texttt{QOrderFinding} through classical equivalence checking.

\item[\textbf{Step 5:}] \textit{Run integration testing for Part II}: Perform integration testing for Part II (purely classical) using a bottom-up order. This step involves testing the interaction between classes such as \texttt{FactorTest}, \texttt{COrderFinding}, \texttt{CFactoring}, and \texttt{Application}.

\item[\textbf{Step 6:}] \textit{Run integration testing for Part I}: Conduct integration testing for Part I, again using a bottom-up order. This step includes testing the interaction between classes like \texttt{QModular}, \texttt{QPhaseEstimation}, \texttt{ContinuedFraction}, and \texttt{QOrderFinding}.

\item[\textbf{Step 7:}] \textit{Final integration}: Now that both Part I and Part II have undergone integration testing, integrate Part I into Part II to complete the final integration.
\end{itemize}
\end{small}

The process of integration testing is shown in Figure~\ref{fig:shoruml}(b), and the bold numbers denote the integration order. In this testing plan, \texttt{COrderFinding} serves as a test double, allowing the testing of upper-level classes to be performed without requiring quantum resources.

\begin{figure}
	\centering
	\includegraphics[scale=0.6]{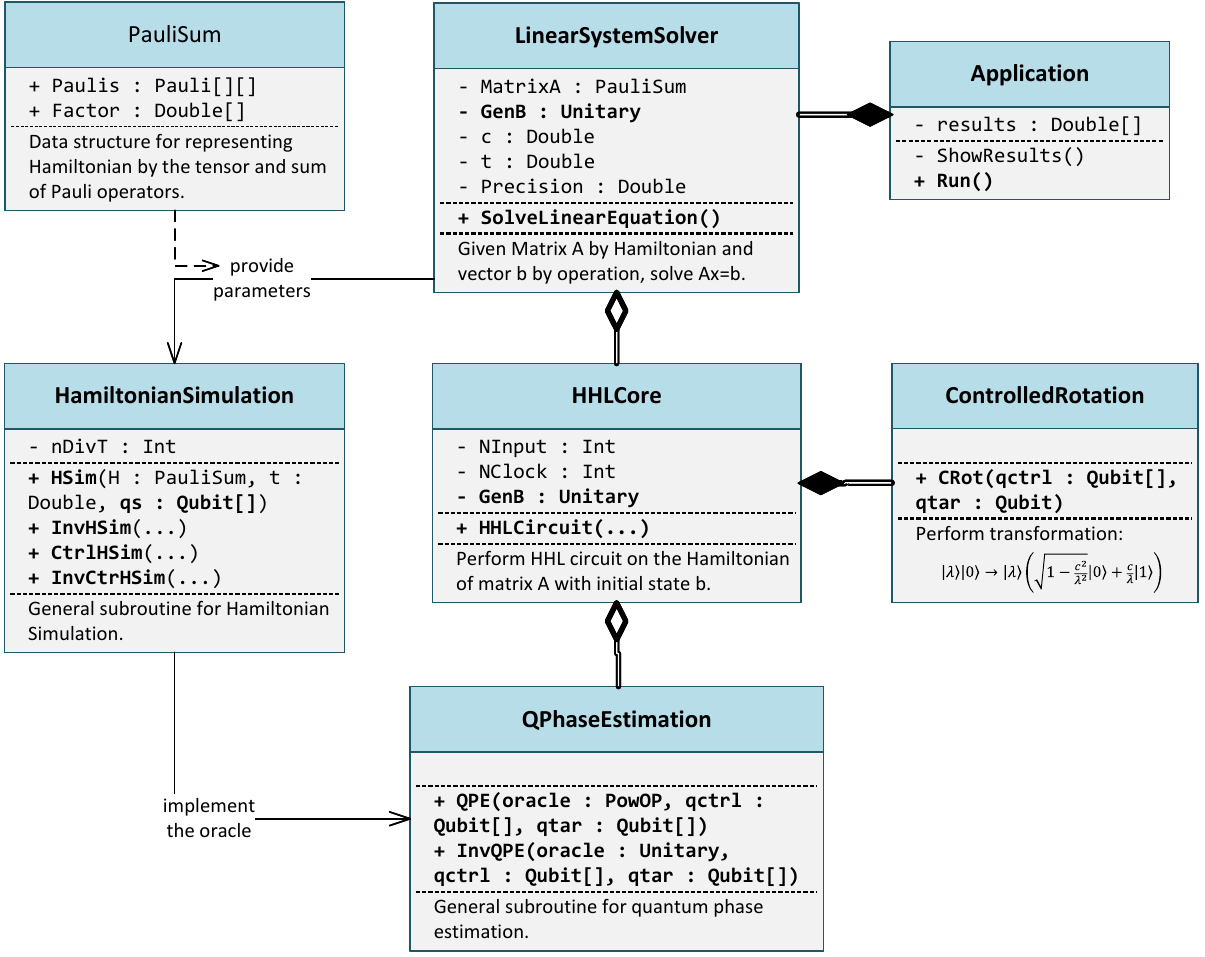}
	\caption{The Q-UML diagram illustrating the structure of the demo program for solving a linear system.}
	\label{fig:hhluml}
\end{figure}

\subsection{Case 3: Linear System Solver}
\label{subsec:CaseHHL}

\subsubsection{Program Structure}
Consider a quantum linear system solver program for $Ax=b$. It allows users to provide the matrix $A$ using a sparse representation (as the sum of several Pauli operators) and $b$ through a state-generation operation. The program utilizes the HHL algorithm to solve the system and performs measurements on the output state a given number of times. The details of the HHL algorithm are given in Appendix~\ref{appendix:algorithms}. The resulting distribution of each measurement outcome is then presented. Figure~\ref{fig:hhluml} illustrates the Q-UML diagram of the program. To represent the matrix $A$, a data-structural class called \texttt{PauliSum} is employed to pass parameters from the \texttt{LinearSystemSolver} class to the \texttt{HamiltonianSimulation} class. In this application, phase estimation is utilized for the Hamiltonian simulation, and the controlled rotation subroutine is implemented in the \texttt{ControlledRotation} class. The \texttt{HHLCore} class implements the HHL circuit depicted in Figure~\ref{fig:hhlcircuit}, while the \texttt{LinearSystemSolver} class encompasses the program's running parameters and calls the \texttt{HHLCore} class.

\subsubsection{Testing Design}
It is important to note that the topmost class, \texttt{Application}, features a display function. This function can be repurposed as a means for result verification, effectively serving as a \textit{testing fixture} for lower-level classes. To achieve thorough testing, we employ a combination of bottom-up and top-down integration strategies, following these outlined steps:

\begin{small}
\begin{itemize}[leftmargin=4em]
\setlength{\itemsep}{1pt}
\item[\textbf{Step 1:}] \textit{Unit testing for lower-level classes}: We initiate the testing process by conducting unit tests for the lower-level classes, including \texttt{ControlledRotation}, \texttt{QPhaseEstimation}, and \texttt{HamiltonianSimulation}.

\item[\textbf{Step 2:}] \textit{Testing the topmost class and creating the testing fixture}: We proceed by evaluating the top-level class, \texttt{Application}, and then reconfigure its display function into a result-checking function. This result-checking function serves as a testing fixture for validating results in the lower-level classes in the subsequent step.

\item[\textbf{Step 3:}] \textit{Integration of \texttt{HHLCore} and \texttt{LinearSystemSolver}}: As we integrate the \texttt{HHLCore} and \\ \texttt{LinearSystemSolver} classes into \texttt{Application}, we continuously examine testing results using the result-checking function created in Step (2).
\end{itemize}
\end{small}

By seamlessly blending bottom-up and top-down approaches while utilizing the result-checking function, we ensure comprehensive test coverage for the entire program.

\begin{tcolorbox}[leftrule=0.5mm, rightrule=0.5mm, toprule=0mm, bottomrule=0mm, colframe=purple, colback=white]
\textbf{Answer to RQ6: } 
Our case studies show that our integration testing design strategies are effective for multi-subroutine quantum programs.
\end{tcolorbox}

\subsection{Insights Learned from the Case Studies}
\label{subsec:learned}

We conclude this section by sharing valuable insights gained from the case studies.


\vspace{2mm}
\noindent $\bullet$ \textbf{Insight 1:} \textit{
If feasible, it is recommended to prioritize the adoption of transform-based methods for checking the running output.}

As discussed in the previous section, there are two main methods for checking the output: (a) the \textit{statistic-based method}, which involves running the target programs multiple times and collecting statistical results; and (b) the \textit{transform-based method}, which applies additional unitary and measurement operations to the output to check for unexpected results. Method (a) requires a large number of repetitions to achieve high accuracy, resulting in decreased efficiency. On the other hand, method (b) allows us to identify errors when unexpected results occur, but its implementation can be challenging in some cases due to the complexity of the required transform. Our case studies also demonstrate that method (b) is more efficient than method (a) whenever it can be successfully implemented. Therefore, we recommend prioritizing the use of transform-based methods to check the running output whenever feasible.

\vspace{2mm}
\noindent $\bullet$ \textbf{Insight 2:} \textit{A recommended testing approach is systematically increasing the testing scale, starting from small-scale scenarios and gradually progressing to larger ones.}

Many quantum programs, including those with variable scales such as QFT (Quantum Fourier Transform), rely on a parameter $n$ representing the number of qubits involved. To ensure efficient testing, we recommend a step-by-step approach, starting from small-scale scenarios and gradually progressing to larger scales. For instance, referring to the Table~\ref{table:qftio} provided, testing QFT would involve examining three cases of $n$: $n=1$, $n=2$, and $n\geq 3$. An effective strategy is, to begin with the smallest scale, testing the $n=1$ case and promptly reporting any discovered bugs to the developers. Once the identified errors have been rectified (i.e., the $n=1$ case passes), the testing can then proceed to the $n=2$ case. Finally, the $n\geq 3$ case can be tested. This sequential order is recommended as smaller-scale programs typically execute more quickly, enabling the timely detection and resolution of errors.

\vspace{2mm}
\noindent $\bullet$ \textbf{Insight 3:} Developers are encouraged to embrace test-driven methods in the development of quantum programs.

Programming quantum systems is inherently challenging, especially in the observation of quantum variable values. To alleviate these challenges, early detection of bugs in the development lifecycle, ideally during the development of each module, proves immensely beneficial. Our research and development practice found that adopting a test-driven approach~\cite{beck2003test} may greatly facilitate early bug detection. This approach entails designing tests before development and subsequently structuring the subroutine to fulfill these tests. Not only does this method enable early bug detection, but it also enhances the reliability of each subroutine. This, in turn, boosts developer confidence, particularly when working on extensive quantum programs encompassing multiple subroutines.

\section{Conclusion And Future Work}
\label{sec:conclusion}

This paper has presented a systematic and comprehensive approach to testing multi-subroutine quantum programs. By investigating the critical properties of quantum programs and considering their unique characteristics, our approach considers the testing for multi-subroutine quantum programs from a whole testing process perspective, which encompasses unit testing and integration testing. The proposed testing principles and criteria provide guidance for designing effective test cases and ensuring the reliability of quantum programs. Through extensive evaluation and case studies on typical quantum subroutines, we have demonstrated the effectiveness of our testing processes, principles, and criteria in detecting bugs and ensuring the correctness of multi-subroutine quantum programs.

However, several open problems require further research:

\begin{itemize}
\item \textit{Identifying additional quantum relations for testing purposes:} Quantum relation checking is a crucial aspect of testing. Beyond the identity, equivalence, and unitarity relations discussed in Section~\ref{subsubsec:SubroutineRelation} and detailed in~\cite{long2023equivalence}, it is advantageous to explore and identify further quantum relations that prove valuable for testing tasks.

\item \textit{Structural testing for general quantum programs:} Quantum programs pose challenges for structural testing due to uncertain execution paths. New methods are needed to accurately describe these paths and measure the coverage rates of quantum programs.

\item \textit{Testing practical quantum software systems}: With the advent of practical quantum software systems, it becomes imperative to assess the effectiveness of current testing methodologies and embark on developing new approaches tailored to the distinctive challenges presented by quantum programming.

\item \textit{Testing processes at higher levels:} While this paper focuses on unit testing and integration testing, testing at higher levels, such as system testing, is crucial for ensuring the overall correctness and reliability of quantum software systems. Future research should explore testing processes at these higher levels.

\item \textit{Testing execution on NISQ quantum devices:} Testing on NISQ devices presents unique challenges due to the substantial influence of quantum noise and decoherence. Therefore, it is essential to explore and advance testing strategies and methodologies specially tailored to these devices through further research.
\end{itemize}

Addressing these open problems will contribute to the advancement of testing techniques for multi-subroutine quantum programs and the overall development of reliable quantum software systems.

\begin{acks}
This work is supported in part by the National Natural Science Foundation of China under Grant 61832015 and JSPS KAKENHI Grant No. JP23H03372.
\end{acks}


\bibliographystyle{acm}
\bibliography{ref}

\section*{Appendix}
\appendix

\section{Related quantum algorithms in this paper}
\label{appendix:algorithms}

Here, we will briefly describe the quantum algorithms involved in this paper, while comprehensive details are available in \cite{nielsen2002quantum} and \cite{harrow2009quantum}.

\begin{figure}[H]
\begin{minipage}[b]{0.49\linewidth}
	\centering
	\subfigure[The overall quantum circuit for Grover search.]{\includegraphics[scale=0.67]{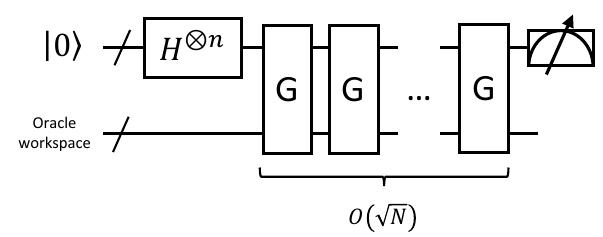}}
	
\end{minipage}\hspace{3mm}
\begin{minipage}[b]{0.46\linewidth}
	\centering
	\subfigure[The quantum circuit for the Grover iteration G.]{\includegraphics[scale=0.67]{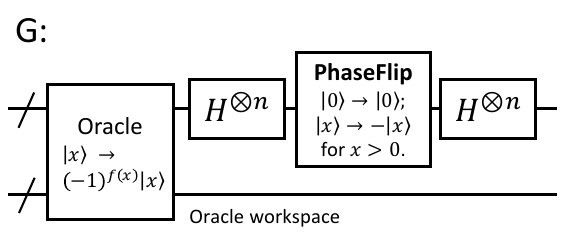}}
\end{minipage}
\caption{The quantum circuits for Grover search algorithm.}
\label{fig:Grover}
\end{figure}

\vspace{1mm}
\noindent $\bullet$ \textbf{Grover Search (\texttt{GS}).}\hspace*{1mm}
Grover's search algorithm (GS)~\cite{grover1996fast} has been proven to outperform classical search algorithms. The circuit diagram of GS is shown in Figure~\ref{fig:Grover}. It comprises approximately $O(\sqrt{N})$ iterations denoted as $G$, where $N = 2^n$ represents the maximum number of elements. Each iteration includes an oracle call and a phase flip subroutine. Remarkably, GS enables the search of a database with $N$ elements using only $O(\sqrt{N})$ oracle queries~\cite{10.5555/870802}.

\vspace{1mm}
\noindent $\bullet$ \textbf{Quantum Fourier Transform (\texttt{QFT}).}\hspace*{1mm}
Consider the following transform:

\begin{equation}
\notag
\left|j\right> \rightarrow \frac{1}{\sqrt{2^n}}\sum_{k=0}^{2^n-1}e^{2\pi i j k / 2^n}\left|k\right>
\end{equation}

\noindent
This transform encodes the Fourier coefficients in the amplitudes of the quantum state, known as the Quantum Fourier Transform (QFT). This algorithm requires only $O(n^2)$ steps for $n$-qubit inputs, compared to the classical fast Fourier transform (FFT), which requires $O(n2^n)$ steps. QFT also has an inverse transform, denoted as IQFT.

\vspace{1mm}
\noindent $\bullet$ \textbf{Quantum Phase Estimation (\texttt{QPE}).}\hspace*{1mm}
QPE is a typical application of QFT to obtain the eigenvalue of a given eigenvector of a target unitary operation. Let $U$ be a unitary operation, and suppose it has an eigenvector $\left|x\right>$ with eigenvalue $e^{2\pi i \theta_x}$ (Note that the length of the eigenvalue of any unitary matrix is 1). The quantum phase estimation (QPE) algorithm estimates $\theta_x$ and encodes its binary representation in a multi-qubit quantum state $\left|\theta_x\right>$. The overall circuit for the QPE algorithm, as shown in Figure~\ref{fig:qpe}, consists of two groups of qubits: the \textit{target qubits} and the \textit{clock qubits}. The target qubits are initialized in the state $\left|x\right>$, and the clock qubits are initialized in an \textit{all-zero} state. The binary representation of $\theta_x$ is then output on the clock qubits. 

\vspace{1mm}
\noindent $\bullet$ \textbf{Quantum Order Finding (\texttt{QOF}).}\hspace*{1mm}
The order finding algorithm takes two positive integers input, $x$, and $N$, and computes the \textit{order} of $x$ modulo $N$, i.e., the minimum value of $r$ satisfying $x^r \equiv1(mod\hspace{1mm}N)$. QOF is a quantum algorithm that achieves order finding with a time complexity of $O(L^3)$, where $L$ represents the length of the binary representation of the input numbers. Currently, no classical order-finding algorithm with polynomial time complexity has been discovered for comparison. The key concept of QOF involves applying QPE on the \textit{modular-multiply} transform:

\begin{equation}
\notag
 \label{equ:OrderFindingU} U\left|y\right> = \left|xy\hspace*{1mm}(mod\hspace*{1mm} N)\right> 
\end{equation}

\vspace*{3mm}
\noindent
Subsequently, the result obtained from QPE undergoes a \textit{continued fraction expansion} to determine the order.

\vspace{1mm}
\noindent $\bullet$ \textbf{Shor's Quantum Factoring (\texttt{Factoring}).}\hspace*{1mm}
An order-finding algorithm can be employed to factorize a positive integer. Shor's Quantum Factoring algorithm~\cite{shor1999polynomial} is built upon QOF. Since QOF is exponentially faster than any known classical order-finding algorithm, it enables the discovery of non-trivial factors for composite numbers $N$ with time complexity of $O((\log N)^3)$. Consequently, Shor's algorithm efficiently factors a given number. This algorithm holds significant importance as it has the potential to break certain cryptosystems, such as RSA, that rely on the complexity of large-number decomposition. Algorithm~\ref{alg:Shor} outlines the process of finding a non-trivial factor for an input composite number $N$. In addition to the QOF subroutine, it involves procedures for testing whether $N$ is in the form of $a^b$ and finding the greatest common divisor (GCD).

\begin{algorithm}
\caption{Quantum\_Factoring}
\label{alg:Shor}
\KwIn{A composite number $N$.}
\KwOut{A non-trivial factor of $N$.}
If $N$ is even, \Return{2}\;
Determine whether $N=a^b$ for integers $a\geq 1$ and $b \geq 2$, and if so \Return{$a$}\;
Randomly choose $x$ in the range $1$ to $N-1$. If \texttt{GCD}$(x,N) > 1$ then \Return{\texttt{GCD}$(x,N)$}\;
Use \texttt{QOF} subroutine to find the order $r$ of $x$ modulo $N$\;
If $r$ is even and $x^{r/2}\neq -1\mod N$, then compute \texttt{GCD}$(x^{r/2}-1,N)$ and \texttt{GCD}$(x^{r/2}+1,N)$. Test to see if one of these is a non-trivial factor, and \Return{the factor} if so. Otherwise, the algorithm fails.
\end{algorithm}

\vspace{1mm}
\noindent $\bullet$ \textbf{Quantum Hamiltonian Simulation (\texttt{QHSim}).}\hspace*{1mm}
The evolution of a closed quantum system satisfies the following equation:

\begin{equation} 
\notag
\left|\psi(t)\right> = e^{-iHt}\left|\psi(0)\right> 
\end{equation}

\vspace*{2mm}
\noindent
In this equation, $H$ represents the Hamiltonian of the system, $\left|\psi(0)\right>$ is the state of the system at time $0$, and $\left|\psi(t)\right>$ is the state of the system at time $t$. QHSim is a technique that enables the implementation of the unitary transformation $e^{-iHt}$ for a given Hamiltonian $H$ and time $t$.

\begin{figure}[h]
	\centering
	\includegraphics[scale=0.65]{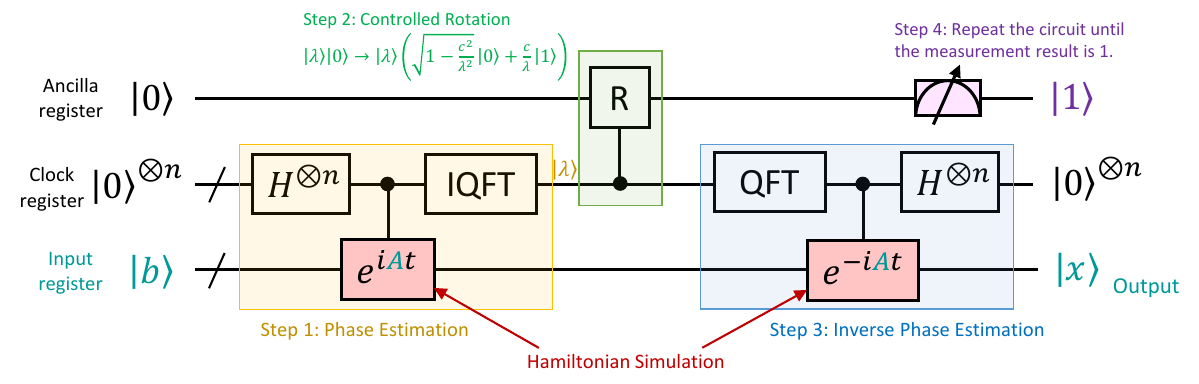}
	\caption{The core quantum circuit of the HHL algorithm.}
	\label{fig:hhlcircuit}
\end{figure}

\vspace{1mm}
\noindent $\bullet$ \textbf{Harrow-Hassidim-Floyd Algorithm (\texttt{HHL}).}\hspace*{1mm}
The HHL algorithm~\cite{harrow2009quantum}, introduced by Harrow, Hassidim, and Lloyd, offers a quantum approach to solving linear systems of equations. This algorithm tackles the problem of determining the unknown vector $x$ in the equation $Ax=b$, where $A$ represents a coefficient matrix of full rank, and $b$ is a constant vector. Figure~\ref{fig:hhlcircuit} illustrates the core circuit of the algorithm. It encodes the matrix $A$ into a Hamiltonian and utilizes QHSim to implement the gate $e^{iAt}$ and its inverse $e^{-iAt}$. The vector $b$ is encoded into the amplitude of the quantum state $\left|b\right>$. The algorithm incorporates a phase estimation step, QPE, for the gate $e^{iAt}$ and the state $\left|b\right>$. The resulting state $\left|\lambda\right>$ is then used to apply a \textit{controlled rotation} gate, \texttt{CRot}, which is a critical operation within the algorithm. Subsequently, an inverse phase estimation is applied for uncomputation, followed by the measurement of the ancilla qubit. The circuit is repeated until the measurement result is 1; at this point, the final output corresponds to the solution state $\left|x\right>$. Ideally, leveraging $n$ qubits, the HHL algorithm has the potential to solve an exponential number of linear systems using polynomial resources, thanks to the $2^n$ amplitudes provided by $n$ qubits.

\end{document}